\documentclass[twoside]{article}

\usepackage[accepted]{aistats2025}
\usepackage[round]{natbib}

\usepackage{fancyhdr}

\usepackage{tikz} 
\usetikzlibrary{positioning,shapes, arrows, math, decorations.text}
\tikzset{events/.style={ellipse, draw, align=center},}

\usepackage[utf8]{inputenc} 
\usepackage[T1]{fontenc}    
\usepackage[colorlinks=true,linkcolor=refcolor,citecolor=refcolor,urlcolor=refcolor]{hyperref}       
\usepackage{url}            
\usepackage{booktabs}       
\usepackage{amsfonts}       
\usepackage{nicefrac}       
\usepackage{microtype}      
\usepackage{xcolor}         
\definecolor{NavyBlueG}{RGB}{28,28,132}
\colorlet{refcolor}{NavyBlueG}
\definecolor{reasonablegreen}{rgb}{0.15, 0.7, 0.15}
\colorlet{reasonableblue}{blue}
\definecolor{purpleG}{RGB}{43,140,190} 
\usepackage{subcaption}
\usepackage{enumitem} 
\usepackage{float}
\usepackage{mathtools}
\usepackage{amsmath}
\usepackage{amssymb}
\usepackage{amsthm}
\usepackage{algorithm}
\usepackage{algorithmicx}
\usepackage{algpseudocode}
\usepackage{bm}
\usepackage{wrapfig}
\usepackage{comment}

\usepackage{amsfonts}

\usetikzlibrary{cd,arrows,calc,shapes,positioning}
\tikzstyle{obs} = [circle,fill=white,draw=black,inner sep=1pt,minimum size=20pt,font=\fontsize{10}{10}\selectfont,node distance=1,thick]
\tikzstyle{latent} = [obs,dotted]
\tikzstyle{pretrained} = [obs,text=NavyBlue,draw=NavyBlue]
\tikzstyle{unfair} = [obs,text=BrickRed,draw=BrickRed]
\tikzstyle{target} = [obs,text=MidnightBlue,draw=MidnightBlue]
\tikzstyle{feature} = [obs,text=ForestGreen,draw=ForestGreen]

\usepackage{cleveref}
\crefname{equation}{Eq.}{Eqs.}
\crefname{align}{Eq.}{Eqs.}
\crefname{section}{Sect.}{Sect.}
\crefname{subsection}{Sect.}{Sect.}
\crefname{subsubsection}{Sect.}{Sect.}
\crefname{aim}{Aim}{Aims}
\crefname{figure}{Fig.}{Figs.}
\crefname{subfigure}{Fig.}{Figs.}
\crefname{principle}{Prin.}{Prin.}
\crefname{definition}{Def.}{Defs.}
\crefname{theorem}{Thm.}{Thms.}
\crefname{appendix}{Appx.}{Appx.}

\usepackage{pdflscape}
\usepackage{booktabs}
\usepackage{pifont}  
\usepackage{makecell} 

\DeclareMathOperator*{\minmax}{min/max}
\DeclareMathOperator{\indep}{\perp\!\!\!\perp}
\DeclareMathOperator{\dep}{\not\! \perp\!\!\!\perp}

\DeclareMathOperator{\cov}{Cov}
\DeclareMathOperator{\var}{Var}
\DeclareMathOperator{\expit}{Expit}

\newcommand{\vect}[1]{\boldsymbol{#1}}
\newcommand{\uvec}[1]{\boldsymbol{#1}}

\theoremstyle{definition}
\newtheorem{theorem}{Theorem}
\newtheorem{definition}{Definition}
\newtheorem{lemma}{Lemma}
\newtheorem{corollary}{Corollary}[theorem]
\newtheorem{principle}{Principle}

\newcommand{\cmark}{{\textcolor{green}{\checkmark}}} 
\newcommand{\xmark}{{\textcolor{red}{\ding{55}}}}    
\newcommand{\partialmark}{{\textcolor{red}{\textbf{-}}}}

\def\Exogenous{\vect{\epsilon}}

\newcommand{\formatDefName}[1]{\textbf{#1}}

\def\CovGZ{\bm\gamma_{\bm g^{ }}}

\begin{document}
%

%

\runningtitle{Partial Identification with Mostly Invalid Instruments}

\twocolumn[

\aistatstitle{BudgetIV: Optimal 
Partial Identification of Causal Effects\\with Mostly Invalid Instruments}

\runningauthor{Penn, \mbox{Bravo-Hermsdorff}, Gunderson, Silva \& Watson}

\aistatsauthor{
    Jordan Penn$^{1}$ \qquad Gecia \mbox{Bravo-Hermsdorff}$^{3}$ \qquad Lee M.~Gunderson$^{2}$ \\ \textbf{Ricardo Silva$^{2}$ \qquad David S.~Watson$^{1}$}

}



\aistatsaddress{$^1$King's College London \qquad $^2$University College London \qquad  $^3$University of Edinburgh} ]

\begin{abstract}
      Instrumental variables (IVs) are widely used to estimate causal effects in the presence of unobserved confounding between an exposure $X$ and outcome $Y$. 
  An IV must affect $Y$ \textit{exclusively} through $X$ and be \textit{unconfounded} with $Y$.
  We present a framework for relaxing 
  these assumptions with tuneable and interpretable "budget constraints". 
  Our algorithm returns a feasible set of causal effects that can be identified exactly given perfect knowledge of observable covariance statistics. 
    This feasible set might contain disconnected sets of possible solutions for the causal effect. 
  We discuss conditions under which this set is \textit{sharp}, i.e., contains all and only effects consistent with the background assumptions and the joint distribution of observable variables. 
  Our method applies to a wide class of semiparametric models, and we demonstrate how its ability to select specific subsets of instruments confers an advantage over convex relaxations in both linear and nonlinear settings. 
  We adapt our algorithm 
  to form confidence sets that are asymptotically valid under a common statistical assumption from the Mendelian randomization literature.  
\end{abstract}

\section{INTRODUCTION}
\label{sec:Introduction}

The causal effect of a ``treatment'' (or ``exposure'') $X$ on an ``outcome'' $Y$ captures the change in the distribution of $Y$ when we intervene on $X$ \citep{pearl2009causality}. 
In a randomized control trial, $X$ is controlled explicitly so that the causal effect can be identified \citep{neyman1923,fisher1935}. 
However, due to financial, ethical or physical constraints, 
confounding factors affecting $X$ and $Y$ often cannot be held fixed or observed and adjusted for \citep{rubin1974,holland1986}.
In these situations, instrumental variables (IVs) are commonly used to infer causal effects. 

IVs are a set of pre-treatment covariates $\vect{Z}$ that are $(\text{A}1)$ associated with $X$; $(\text{A}2)$ unconfounded with $Y$; and $(\text{A}3)$ whose causal effect on $Y$ is exclusively mediated through $X$ \citep{angrist1996,heckman1999} (see \cref{sec:IVassumptions} for formal definition). 
Due to the presence of latent confounders, 
one cannot test whether $(\text{A}2)$ and $(\text{A}3)$ hold for any particular pre-treatment covariate. 
However, with multiple candidate instruments, a simple fact may be exploited: if different sets of covariates predict different values of the causal parameter $\theta$ when treated as IVs, then at most one of the sets contains valid IVs. 

\citet{kang2016} exploit this fact by showing the following in linear models with a scalar treatment: if more than $50\%$ of the pre-treatment covariates are correctly assumed to be valid IVs, then $\theta$ is point-identifiable (i.e., given an infinite number of observations from the model, the value of $\theta$ can be uniquely determined). 
A number of majority rule based approaches to inference with invalid IVs extend this work \citep{Bowden2016WeightedMedian,Hartwig2017,bucur2020,hartford2021}. 

However, in settings where $($a$)$ $50\%$ or fewer candidate IVs are valid, or $($b$)$ more than one causal parameter is needed (e.g., because the treatment is multidimensional), we show that $\theta$ can be at best \textit{partially} identified. 
For example, even with oracle access to the joint distribution $P(X, Y, Z_1, Z_2)$, it may be undecidable whether $Z_1$ is a valid IV and $Z_2$ is not or vice versa. 
Though some values of $\theta$ may be excluded, the parameter will not converge to a single value even in the limit of infinite data.

Rather than assuming that all or a majority of $\bm Z$ are valid IVs, our approach is to set a minimum proportion which are. 
In fact, we consider a more general setting (c), in which ``degrees'' of IV validity 
are set by user-input thresholds. 
It is not assumed \textit{a priori} which thresholds apply to which candidate IVs. 
For each threshold, however, a user-input ``budget'' constrains how many candidates are allowed to violate the IV assumptions up to that degree. 
\citet{bucur2020} and \citet{Xue2023} propose point estimators for $\theta$ under budget-like constraints, restraining fewer than $50\%$ of candidate IVs to be valid. 
As we will see in \cref{sec:problem setup}, however, $\theta$ is only partially identified in this setting. 
The allocation of budgets to $\bm Z$ are also partially identified in general, which can lead to disconnected sets of solutions for $\theta$.





Our discussion applies when IVs are invalid due to violation of $(\text{A}2)$, $(\text{A}3)$ or both.
We show that under the assumption of homogeneous causal effects \citep{holland1986}, which implies an additive separation of the $\bm X$- and $\bm Z$-signals into $Y$
\citep{Newey2003nonparametric}, a latent vector statistic summarizes bias in the estimated $\bm \theta$ due to violation of both assumptions.  
This statistic has a simple interpretation: the residual covariance between pre-treatment covariates $\bm Z$ and the outcome $Y$ not explained by the causal effect of the treatment $X$. 
\citet{vancak_2023} make a similar proposal to quantify the violation of IV assumptions in a single sensitivity parameter. 
However, their approach is limited to a single scalar $Z$, so does not utilize the notion of IV candidates and whether they give consistent estimates of $\theta$.  


We provide an algorithm, \texttt{budgetIV}, that provably finds \textit{sharp} partial identification sets for causal effect parameters under fixed budget constraints.
We show that the set can be inferred from relevant summary statistics, which reduce to the familiar covariance parameters $\cov (X, Z)$ and $\cov (Y, Z)$ for linear models.
In the special case of scalar treatments, we provide a polytime procedure for computing sharp solution sets.
However, under setting (b), we prove that partial identification is NP-hard in the number of instruments, budget thresholds and causal effect parameters.
Finally, we provide a method for handling finite-sample uncertainty. 
In \cref{sec:inference}, we show how \texttt{budgetIV} can be adapted to form (asymptotically) valid confidence sets under the so-called ``no measurement error'' assumption from the Mendelian randomization literature \citep{Bowden2016NOME}.

The remainder of this paper is structured as follows. 
We formalize our problem in \cref{sec:problem setup}.
We introduce the \texttt{budgetIV} algorithm in \cref{sec:BudgetIV} and propose related inference procedures in \cref{sec:inference}.
Experimental results are presented in \cref{sec:experiments}. 
Following a literature review in \cref{sec:related work}, we conclude with a brief discussion in \cref{sec:Discussion}. 
Proofs, pseudocode, and experimental details are given in the appendix. 

\section{PROBLEM SETUP} 
\label{sec:problem setup}

\begin{figure}
    \centering
    \begin{tikzpicture}
    \tikzmath{\PictureWidth=8;\PictureHeight=5;}
    \tikzmath{\PictureLeftX=-2;\PictureBottomY=-1.25;}
    \draw [draw=none,fill=none] (\PictureLeftX,\PictureBottomY) rectangle (\PictureLeftX+\PictureWidth,\PictureBottomY+\PictureHeight);
    \tikzmath{\XXXNodeX=2;\XXXNodeY=0;}
    \tikzmath{\YYYNodeX=5;\YYYNodeY=0;}
    \tikzmath{\ZZZNodeX=-1;\ZZZNodeY=0;}
    \tikzmath{\SubLatentY=2;}
    \tikzmath{\GodLatentY=3.25;}
    \tikzmath{\XYZNodeRadius=0.5;}
    \tikzmath{\NodeFontSize=16;}
    \newcommand{\NodeBorderWidth}{1.5pt}
    \newcommand{\XXYYArrowWidth}{2pt}
    \tikzmath{\ArrowStartX={\XXXNodeX};\ArrowStartY={\XXXNodeY};}
    \tikzmath{\ArrowEndX={\YYYNodeX};\ArrowEndY={\YYYNodeY};}
    \tikzmath{\ArrowBend=0;}
    \tikzmath{\DX=\ArrowEndX-\ArrowStartX;}
    \tikzmath{\DY=\ArrowEndY-\ArrowStartY;}
    \tikzmath{\DD=(\DX*\DX+\DY*\DY)^(0.5);}
    \tikzmath{\Angle=atan(\DY/\DX);}
    \tikzmath{\StartAngle=\Angle+\ArrowBend;\EndAngle=\Angle-\ArrowBend+180;}
    \tikzmath{\ArrowStartXNew=\ArrowStartX+\XYZNodeRadius*cos(\StartAngle);\ArrowStartYNew=\ArrowStartY+\XYZNodeRadius*sin(\StartAngle);}
    \tikzmath{\ArrowEndXNew=\ArrowEndX+\XYZNodeRadius*cos(\EndAngle);\ArrowEndYNew=\ArrowEndY+\XYZNodeRadius*sin(\EndAngle);}
    \draw [black,line width=\XXYYArrowWidth, -latex,
    postaction={decorate, 
    decoration={text along path, 
                text={causal effect},
                text align={center},
                raise=14pt}}] 
    (\ArrowStartXNew,\ArrowStartYNew) to [out=\StartAngle, in=\EndAngle] (\ArrowEndXNew,\ArrowEndYNew);
    \draw [reasonablegreen,line width=\XXYYArrowWidth, -latex,
    postaction={decorate, 
    decoration={text along path, 
                text={of interest},
                text align={center},
                raise=4pt}}] 
    (\ArrowStartXNew,\ArrowStartYNew) to [out=\StartAngle, in=\EndAngle] (\ArrowEndXNew,\ArrowEndYNew);
    \tikzmath{\ArrowStartX={\XXXNodeX};\ArrowStartY={\SubLatentY};}
    \tikzmath{\ArrowEndX={\YYYNodeX};\ArrowEndY={\SubLatentY};}
    \tikzmath{\ArrowBend=30;}
    \tikzmath{\DX=\ArrowEndX-\ArrowStartX;}
    \tikzmath{\DY=\ArrowEndY-\ArrowStartY;}
    \tikzmath{\DD=(\DX*\DX+\DY*\DY)^(0.5);}
    \tikzmath{\Angle=tan(\DY/\DX);}
    \tikzmath{\StartAngle=\Angle+\ArrowBend;\EndAngle=\Angle-\ArrowBend+180;}
    \tikzmath{\ArrowStartXNew=\ArrowStartX+\XYZNodeRadius*cos(\StartAngle);\ArrowStartYNew=\ArrowStartY+\XYZNodeRadius*sin(\StartAngle);}
    \tikzmath{\ArrowEndXNew=\ArrowEndX+\XYZNodeRadius*cos(\EndAngle);\ArrowEndYNew=\ArrowEndY+\XYZNodeRadius*sin(\EndAngle);}
    \draw [dash pattern=on 2pt off 2pt,line width=\XXYYArrowWidth, latex-latex]
    (\ArrowStartXNew,\ArrowStartYNew) to [out=\StartAngle, in=\EndAngle] (\ArrowEndXNew,\ArrowEndYNew);
    \tikzmath{\ArrowStartX={\ZZZNodeX};\ArrowStartY={\ZZZNodeY};}
    \tikzmath{\ArrowEndX={\XXXNodeX};\ArrowEndY={\XXXNodeY};}
    \tikzmath{\ArrowBend=0;}
    \tikzmath{\DX=\ArrowEndX-\ArrowStartX;}
    \tikzmath{\DY=\ArrowEndY-\ArrowStartY;}
    \tikzmath{\DD=(\DX*\DX+\DY*\DY)^(0.5);}
    \tikzmath{\Angle=atan(\DY/\DX);}
    \tikzmath{\StartAngle=\Angle+\ArrowBend;\EndAngle=\Angle-\ArrowBend+180;}
    \tikzmath{\ArrowStartXNew=\ArrowStartX+\XYZNodeRadius*cos(\StartAngle);\ArrowStartYNew=\ArrowStartY+\XYZNodeRadius*sin(\StartAngle);}
    \tikzmath{\ArrowEndXNew=\ArrowEndX+\XYZNodeRadius*cos(\EndAngle);\ArrowEndYNew=\ArrowEndY+\XYZNodeRadius*sin(\EndAngle);}
    \draw [blue,line width=\XXYYArrowWidth, -latex,
    postaction={decorate, 
    decoration={text along path, 
                text={instrumental},
                text align={center},
                raise=14pt}}] 
    (\ArrowStartXNew,\ArrowStartYNew) to [out=\StartAngle, in=\EndAngle] (\ArrowEndXNew,\ArrowEndYNew);
    \draw [reasonableblue,line width=\XXYYArrowWidth, -latex,
    postaction={decorate, 
    decoration={text along path, 
                text={effect},
                text align={center},
                raise=4pt}}] 
    (\ArrowStartXNew,\ArrowStartYNew) to [out=\StartAngle, in=\EndAngle] (\ArrowEndXNew,\ArrowEndYNew);
    \tikzmath{\ArrowStartX={\ZZZNodeX};\ArrowStartY={\ZZZNodeY};}
    \tikzmath{\ArrowEndX={\YYYNodeX};\ArrowEndY={\YYYNodeY};}
    \tikzmath{\ArrowBend=-35;}
    \tikzmath{\DX=\ArrowEndX-\ArrowStartX;}
    \tikzmath{\DY=\ArrowEndY-\ArrowStartY;}
    \tikzmath{\DD=(\DX*\DX+\DY*\DY)^(0.5);}
    \tikzmath{\Angle=atan(\DY/\DX);}
    \tikzmath{\StartAngle=\Angle+\ArrowBend;\EndAngle=\Angle-\ArrowBend+180;}
    \tikzmath{\ArrowStartXNew=\ArrowStartX+\XYZNodeRadius*cos(\StartAngle);\ArrowStartYNew=\ArrowStartY+\XYZNodeRadius*sin(\StartAngle);}
    \tikzmath{\ArrowEndXNew=\ArrowEndX+\XYZNodeRadius*cos(\EndAngle);\ArrowEndYNew=\ArrowEndY+\XYZNodeRadius*sin(\EndAngle);}
    \draw [red,dashed,dash pattern=on 8pt off 4pt,line width=\XXYYArrowWidth, -latex,
    postaction={decorate, 
    decoration={text along path, 
                text={instrumental leakage},
                text align={center},
                raise=4pt}}] 
    (\ArrowStartXNew,\ArrowStartYNew) to [out=\StartAngle, in=\EndAngle] (\ArrowEndXNew,\ArrowEndYNew);
    \tikzmath{\ArrowStartX={\ZZZNodeX};\ArrowStartY={\SubLatentY};}
    \tikzmath{\ArrowEndX={\YYYNodeX};\ArrowEndY={\SubLatentY};}
    \tikzmath{\ArrowBend=50;}
    \tikzmath{\DX=\ArrowEndX-\ArrowStartX;}
    \tikzmath{\DY=\ArrowEndY-\ArrowStartY;}
    \tikzmath{\DD=(\DX*\DX+\DY*\DY)^(0.5);}
    \tikzmath{\Angle=atan(\DY/\DX);}
    \tikzmath{\StartAngle=\Angle+\ArrowBend;\EndAngle=\Angle-\ArrowBend+180;}
    \tikzmath{\ArrowStartXNew=\ArrowStartX+\XYZNodeRadius*cos(\StartAngle);\ArrowStartYNew=\ArrowStartY+\XYZNodeRadius*sin(\StartAngle);}
    \tikzmath{\ArrowEndXNew=\ArrowEndX+\XYZNodeRadius*cos(\EndAngle);\ArrowEndYNew=\ArrowEndY+\XYZNodeRadius*sin(\EndAngle);}
    \draw [red, dash pattern=on 2pt off 2pt,line width=\XXYYArrowWidth, latex-latex,
    postaction={decorate, 
    decoration={text along path, 
                text={latent confounding},
                text align={center},
                raise=-14pt}}] 
    (\ArrowStartXNew,\ArrowStartYNew) to [out=\StartAngle, in=\EndAngle] (\ArrowEndXNew,\ArrowEndYNew);
    \tikzmath{\ArrowStartX={\ZZZNodeX};\ArrowStartY={\SubLatentY};}
    \tikzmath{\ArrowEndX={\XXXNodeX};\ArrowEndY={\SubLatentY};}
    \tikzmath{\ArrowBend=30;}
    \tikzmath{\DX=\ArrowEndX-\ArrowStartX;}
    \tikzmath{\DY=\ArrowEndY-\ArrowStartY;}
    \tikzmath{\DD=(\DX*\DX+\DY*\DY)^(0.5);}
    \tikzmath{\Angle=atan(\DY/\DX);}
    \tikzmath{\StartAngle=\Angle+\ArrowBend;\EndAngle=\Angle-\ArrowBend+180;}
    \tikzmath{\ArrowStartXNew=\ArrowStartX+\XYZNodeRadius*cos(\StartAngle);\ArrowStartYNew=\ArrowStartY+\XYZNodeRadius*sin(\StartAngle);}
    \tikzmath{\ArrowEndXNew=\ArrowEndX+\XYZNodeRadius*cos(\EndAngle);\ArrowEndYNew=\ArrowEndY+\XYZNodeRadius*sin(\EndAngle);}
    \draw [reasonableblue, dash pattern=on 2pt off 2pt,line width=\XXYYArrowWidth, latex-latex]
    (\ArrowStartXNew,\ArrowStartYNew) to [out=\StartAngle, in=\EndAngle] (\ArrowEndXNew,\ArrowEndYNew);
    \tikzmath{\ArrowStartX={\ZZZNodeX};\ArrowStartY={\SubLatentY};}
    \tikzmath{\ArrowEndX={\ZZZNodeX};\ArrowEndY={\ZZZNodeY};}
    \tikzmath{\ArrowBend=0;}
    \tikzmath{\DX=\ArrowEndX-\ArrowStartX;}
    \tikzmath{\DY=\ArrowEndY-\ArrowStartY;}
    \tikzmath{\DD=(\DX*\DX+\DY*\DY)^(0.5);}
    \tikzmath{\Angle=270;}
    \tikzmath{\StartAngle=\Angle+\ArrowBend;\EndAngle=\Angle-\ArrowBend+180;}
    \tikzmath{\ArrowStartXNew=\ArrowStartX+\XYZNodeRadius*cos(\StartAngle);\ArrowStartYNew=\ArrowStartY+\XYZNodeRadius*sin(\StartAngle);}
    \tikzmath{\ArrowEndXNew=\ArrowEndX+\XYZNodeRadius*cos(\EndAngle);\ArrowEndYNew=\ArrowEndY+\XYZNodeRadius*sin(\EndAngle);}
    \draw [black,line width=\XXYYArrowWidth, -latex] (\ArrowStartXNew,\ArrowStartYNew) to [out=\StartAngle, in=\EndAngle] (\ArrowEndXNew,\ArrowEndYNew);
    \tikzmath{\ArrowStartX={\XXXNodeX};\ArrowStartY={\SubLatentY};}
    \tikzmath{\ArrowEndX={\XXXNodeX};\ArrowEndY={\XXXNodeY};}
    \tikzmath{\ArrowBend=0;}
    \tikzmath{\DX=\ArrowEndX-\ArrowStartX;}
    \tikzmath{\DY=\ArrowEndY-\ArrowStartY;}
    \tikzmath{\DD=(\DX*\DX+\DY*\DY)^(0.5);}
    \tikzmath{\Angle=270;}
    \tikzmath{\StartAngle=\Angle+\ArrowBend;\EndAngle=\Angle-\ArrowBend+180;}
    \tikzmath{\ArrowStartXNew=\ArrowStartX+\XYZNodeRadius*cos(\StartAngle);\ArrowStartYNew=\ArrowStartY+\XYZNodeRadius*sin(\StartAngle);}
    \tikzmath{\ArrowEndXNew=\ArrowEndX+\XYZNodeRadius*cos(\EndAngle);\ArrowEndYNew=\ArrowEndY+\XYZNodeRadius*sin(\EndAngle);}
    \draw [black,line width=\XXYYArrowWidth, -latex] (\ArrowStartXNew,\ArrowStartYNew) to [out=\StartAngle, in=\EndAngle] (\ArrowEndXNew,\ArrowEndYNew);
    \tikzmath{\ArrowStartX={\YYYNodeX};\ArrowStartY={\SubLatentY};}
    \tikzmath{\ArrowEndX={\YYYNodeX};\ArrowEndY={\YYYNodeY};}
    \tikzmath{\ArrowBend=0;}
    \tikzmath{\DX=\ArrowEndX-\ArrowStartX;}
    \tikzmath{\DY=\ArrowEndY-\ArrowStartY;}
    \tikzmath{\DD=(\DX*\DX+\DY*\DY)^(0.5);}
    \tikzmath{\Angle=270;}
    \tikzmath{\StartAngle=\Angle+\ArrowBend;\EndAngle=\Angle-\ArrowBend+180;}
    \tikzmath{\ArrowStartXNew=\ArrowStartX+\XYZNodeRadius*cos(\StartAngle);\ArrowStartYNew=\ArrowStartY+\XYZNodeRadius*sin(\StartAngle);}
    \tikzmath{\ArrowEndXNew=\ArrowEndX+\XYZNodeRadius*cos(\EndAngle);\ArrowEndYNew=\ArrowEndY+\XYZNodeRadius*sin(\EndAngle);}
    \draw [black,line width=\XXYYArrowWidth, -latex] (\ArrowStartXNew,\ArrowStartYNew) to [out=\StartAngle, in=\EndAngle] (\ArrowEndXNew,\ArrowEndYNew);
    %
    %
    \draw[black,line width=\NodeBorderWidth] (\XXXNodeX,\XXXNodeY) circle[radius=\XYZNodeRadius];
    \draw[black,line width=\NodeBorderWidth] (\YYYNodeX,\YYYNodeY) circle[radius=\XYZNodeRadius];
    \draw[black,line width=\NodeBorderWidth] (\ZZZNodeX,\ZZZNodeY) circle[radius=\XYZNodeRadius];
    \node[circle,draw=none,font=\fontsize{\NodeFontSize}{0}\selectfont] at (\XXXNodeX,\XXXNodeY) {$\bm X$};
    \node[circle,draw=none,font=\fontsize{\NodeFontSize}{0}\selectfont] at (\YYYNodeX,\YYYNodeY) {$Y$};
    \node[circle,draw=none,font=\fontsize{\NodeFontSize}{0}\selectfont] at (\ZZZNodeX,\ZZZNodeY) {$\bm Z$};
    \draw[black,line width=\NodeBorderWidth, dash pattern=on 2pt off 2pt] (\XXXNodeX,\SubLatentY) circle[radius=\XYZNodeRadius];
    \draw[black,line width=\NodeBorderWidth, dash pattern=on 2pt off 2pt] (\YYYNodeX,\SubLatentY) circle[radius=\XYZNodeRadius];
    \draw[black,line width=\NodeBorderWidth, dash pattern=on 2pt off 2pt] (\ZZZNodeX,\SubLatentY) circle[radius=\XYZNodeRadius];
    \node[circle,draw=none,font=\fontsize{\NodeFontSize}{0}\selectfont] at (\XXXNodeX,\SubLatentY) {$\Exogenous_{\vect{X}}$};
    \node[circle,draw=none,font=\fontsize{\NodeFontSize}{0}\selectfont] at (\YYYNodeX,\SubLatentY) {$\Exogenous_{Y}$};
    \node[circle,draw=none,font=\fontsize{\NodeFontSize}{0}\selectfont] at (\ZZZNodeX,\SubLatentY) {$\Exogenous_{\vect{Z}}$};
    \end{tikzpicture}
    \caption{
    \textbf{Acyclic directed mixed graph for our problem setup.}
     Solid circles represent observable variables and dashed circles latent variables. 
    Bidirected arrows are interpreted as any mutual dependence between noise residuals. 
    The dotted black arrow indicates the unobserved confounding between $\vect{X}$ and $Y$.  
    The relevance assumption (A$1$) requires at least one of the blue arrows. 
    The red arrows contribute to violations of the exogeneity conditions (A2) and (A3).
    The green arrow indicates the causal effect of interest.  
    }
    \vspace{-3mm}
    \label{fig:Causal diagram}
\end{figure}
Our observable variables include a set of candidate instruments $\bm{Z} \in \Omega_{\bm Z} \subseteq \mathbb{R}^{d_{\bm{Z}}}$; treatments $\bm{X} \in \Omega_{\bm X} \subseteq \mathbb{R}^{d_{\bm{X}}}$; and a univariate outcome $Y \in \Omega_Y \subseteq \mathbb{R}$. 
We assume that the ground truth structural equation model (SEM) between these variables takes the following form:
\begin{align}
    \vect{Z} &:= f_{\vect{z}}\left(\Exogenous_{\vect{z}}\right), \label{eqn:Z} \\
    \vect{X} &:= f_{\vect{x}} \left( \vect{Z},  \Exogenous_{\vect{x}} \right), \label{eqn:X} \\
    Y &:= \vect{\theta^*} \cdot \vect{\Phi} ( \vect{X} ) + g_y (\vect{Z}, \Exogenous_{\vect{y}}), \label{eqn:Y}
\end{align}
where $\Exogenous_{\vect{z}}$, $\Exogenous_{\vect{x}}$, and $\Exogenous_{\vect{y}}$ are noise residuals that capture any and all effects from latent variables, including unobserved confounding. 
The additive separability of $\bm X$ and $\bm Z$ in $Y$ is guaranteed by assuming that the treatment effect is homogeneous, i.e., $P\big(y \mid \mathrm{do} (\bm x)\big) - P\big(y \mid \mathrm{do} (\bm x_0)\big)$ is independent of $\bm Z$ and unobserved confounding between $\bm X$ and $Y$.

The function $\vect{\Phi}: \Omega_{\bm X} \mapsto \Omega_{\bm \Phi} \subseteq \mathbb R^{d_{\bm \Phi}}$ may provide a basis-expansion of a nonlinear treatment effect or a representation of a high-dimensional treatment $\bm X$. 
Though some authors have studied the underspecified regime in which $d_{\vect \Phi} > d_{\vect{Z}}$ \citep{pfister22a, ailer2023sequential}, we restrict attention to the more common case where $d_{\bm \Phi} \leq d_{\vect{Z}}$.
This guarantees the identification result of \cref{thm:identifiability_completeness}, first shown in linear models by \citet{koopmans1949identification}.

The vector $\bm \theta^{\bm *} \in \mathbb{R}^{d_{\bm \Phi}}$ is the causal parameter of interest. 
Together with $\vect \Phi$, it defines the average treatment effect:
\begin{align*}
    \mathrm{ATE} (\vect{x}; \vect{x}_0) = \vect{\theta^*} \cdot \big( \vect{\Phi}(\vect{x}) - \vect{\Phi}(\vect{x}_0)\big),
\end{align*}
which represents the expected change in $Y$ if we were to replace the intervention $do(\vect X=\vect{x}_0)$ with $do(\vect X=\vect{x})$.




\subsection{A Sensitivity Parameter for IV Violation}
\label{sec:IVassumptions}

We call $\bm Z$ a set of valid IVs if they satisfy the following properties: 
\begin{description}[noitemsep, itemindent=0.2em, font=\normalfont]
    \item[(A1)] \textit{Association}: $\vect{Z} \dep \vect{X},$
    \item[(A2)] \textit{Unconfoundedness}: $\vect{Z} \indep \Exogenous_{\uvec{y}},$
    \item[(A3)] \textit{Exclusion}: $\vect{Z} \indep Y \mid \{\vect{X}, \Exogenous_{\uvec{y}} \}.$
\end{description}
In several well-studied regimes \citep{koopmans1949identification,imbens1994,Newey2003nonparametric,Heckman2005,sorowit2022}, the assumptions above allow for point identification of the target parameter $\bm \theta^*$.
For our SEM (\cref{eqn:Z,eqn:X,eqn:Y}), we prove point identification under a stronger version of (A1) and an assumption that combines aspects of (A2) and (A3), despite being strictly weaker than either. 
To show this, we introduce the following parameters:
\begin{align*}
    \CovGZ &:= \text{Cov}\big(g_y(\vect{Z}, \vect{\epsilon_y}),\vect{Z}\big), \\
    \bm{\beta_{\Phi}} &:= \text{Cov}\big(\bm \Phi(\bm X), \bm Z \big),\\
    \bm{\beta_y} &:= \text{Cov}\big(Y, \bm Z\big).
\end{align*}

By taking the covariance between each term in \cref{eqn:Y} with $\bm Z$, we see that target parameter $\bm \theta^*$ is related to $\bm \gamma_{\bm g}$ by $\bm \gamma_{\bm g} = \bm \beta_{\bm y} - \bm \theta^* \cdot \bm \beta_{\bm \Phi}$. 
Thus, $\bm \gamma_{\bm g}$ is the residual covariance between $Y$ and $\bm Z$ not explained by the ground truth causal effect of $\bm X$ on $Y$. 
Notice that if both (A2) and (A3) are satisfied, then $\bm \gamma_{\bm g} = \bm 0$.

Given that $d_{\bm \Phi} \leq d_{\bm Z}$, the following constraints are sufficient for identification: 
\begin{description}[noitemsep, itemindent=0.2em, font=\normalfont]
    \item[(B$1^*$)] \textit{Association (strong)}: $\text{rank}(\bm{\beta_\Phi}) = d_{\bm \Phi}$\\and  
    $(\bm \beta_{\bm \Phi})_i \neq 0$  
    ($\forall Z_i  \in \bm{Z}$);
    \item[(B$2^*$)] \textit{Exogeneity (strong)}: 
    $\bm \gamma_{\bm g} = \bm 0,$
\end{description}
where $\left(\bm \beta_{\bm \Phi}\right)_i := \cov \left( \bm \Phi (\bm X), Z_i \right)$ and ``exogeneity'' refers to the conjunction of (A$2$) and (A$3$).

\begin{theorem}[\formatDefName{Identifiability}]\label{thm:identifiability_completeness}
    Assume \cref{eqn:Z,eqn:X,eqn:Y} and claims (B$1^*$), (B$2^*$) hold for some $d_{\bm \Phi} \leq d_{\bm Z}$. 
    Assume the existence of a ground truth joint distribution $P(\bm X, Y, \bm Z)$ with finite  
    covariance 
    parameters $\bm{\beta_\Phi}^*, \bm{\beta_y}^*$.
    Then the causal parameter $\bm \theta^*$ can be identified exactly as the unique solution to $\bm \beta_{\bm y}^* - \bm \theta^* \cdot \bm \beta_{\bm \Phi}^* = \bm 0$. 
\end{theorem}
We can relax (B$1^*$) and (B$2^*$) further, in particular using $\bm \gamma_{\bm g}$ to model violations of exogeneity:

\begin{description}[noitemsep, itemindent=0.2em, font=\normalfont]
    \item[(B1)] ~\textit{Association (relaxed)}: 
    $\bm{\beta_\Phi} \neq \bm 0$;
    \item[(B2)] ~\textit{Exogeneity (relaxed)}: $\bm \gamma_{\bm g} \in \bm \Gamma$.
\end{description}
Such a relaxation may still allow for partial identification, which we will see in the following section.
In \cref{sec:Budget constraints} we introduce budget constraints as tuneable and interpretable choices for $\bm \Gamma$.


\subsection{Formalizing Optimal Partial Identification}

We say that the causal parameter $\bm \theta^*$ is \textit{partially identified} when more than one (but not all) of its possible values are consistent with the data and our structural assumptions. 
In this section, we define a certain notion of \textit{optimality} for general partial identification problems, and establish the ingredients for an optimal solution in our setting.


The following definitions allow us to state and prove these results.
Let $\mathcal M$ be a class of SEMs and $m^* \in \mathcal{M}$ the ground truth model. 
Each $m \in \mathcal M$ implies a joint distribution $P_m(\bm V)$ over the observables $\bm V \in \Omega_{\bm V}$. (In our setting $\bm V = \{\bm X, Y, \bm Z\}$ and $\Omega_{\bm V} = \Omega_{\bm X} \times \Omega_Y \times \Omega_{\bm Z}$.)

A constraint $c: \mathcal M \mapsto \{0,1\}$ is a logical formula that either does or does not hold for any given model.
For instance, $c$ may bound the range of some parameter(s) in $m$ or impose conditional independence on certain variables in $\bm V$.
Let $\mathcal{C}$ be 
a 
set of such constraints, with $\mathcal{C}^* := \{ c \in \mathcal{C} : c(m^*) = 1 \}$ denoting ground truth. 
(In our setting, $\mathcal C$ includes the relaxed association and exogeneity assumptions (B1) and (B2).)

An \textit{observable} statistic $s$ is a functional of the joint distribution, $s : \{ P_m (\bm V) : m \in \mathcal{M} \} \mapsto \Omega_s$, with ground truth value $s^* := s\left( P_{m^*} (\bm V)\right)$.
Examples include (conditional) moments or correlations between variables. (In our setting, $s$ comprises the cross-covariance parameters $\bm{\beta_y}, \bm{\beta_\Phi}$.)
The target parameter $q^* := q(m^*)$ is the ground truth for some 
 \textit{latent} statistic 
$q : \mathcal{M} \mapsto \Omega_q$, which cannot be determined by $P_m (\bm V)$ alone. (In our setting, $q^*$ is the causal parameter $\bm \theta^*$.)

Given a constraint $c \in \mathcal{C}$ and statistic $s$, the ``plausible'' values of $q^*$ form a \textit{solution set} $\mathcal{T}(c, s) \subseteq \Omega_q$. Such sets are the image of a \textit{solution map} for $q$, $\mathcal{T} : \mathcal{C} \times \Omega_s \mapsto \mathcal P(\Omega_q)$, where $\mathcal P$ denotes the power set.


We define an \textit{optimal} solution map in terms of soundness, completeness, and minimality criteria. We defer discussion of computational complexity and finite sample inference to \cref{sec:BudgetIV,sec:inference}, respectively.

\begin{definition}[\formatDefName{Soundness}] \label{def:Soundness}
    A solution map $\mathcal{T}$ is \textit{sound} if, for any ground truth model $m^* \in \mathcal{M}$, given statistic $s^* := s\left(P_{m^*} (\bm V)\right)$ and constraint $c^* \in \mathcal{C}^*$, we have $q^* \in \mathcal{T} (c^*, s^*)$. 
\end{definition}
This condition ensures that our solution map cannot exclude the target $q^*$ when provided with ground truth inputs.

\begin{definition}[\formatDefName{Completeness}] \label{def:Completeness}
    A solution map $\mathcal{T}$ is \textit{complete} if, for any ground truth model $m^* \in \mathcal{M}$, given $s^*$ and any $c^* \in \mathcal{C}^*$, the following holds. 
    For all $q \in \mathcal{T} (c^*, s^*)$, there is at least one model $m_{q} \in \mathcal{M}$ for which $q = q(m_{q})$ and $s\left(P_{m_{q}} (\bm V)\right) = s^*$. 
\end{definition}
This condition ensures that no sound map can exclude more values of $q \in \Omega_q \backslash  \{q^*\}$. 
Solution sets that are both sound and complete are said to be \textit{sharp}, as originally defined by \citet{Manski1990, manski2003partial}. 


For the next definition, we introduce a partial order on observable statistics with respect to their information content. 
We say that $s'$ is at least as informative as $s$ if there exists a deterministic function $f: \Omega_{s'} \mapsto \Omega_s$ such that, for any $m \in \mathcal M$, $s \left( P_m (\bm V) \right) = f \left( s' \left( P_m (\bm V) \right) \right)$. In this case, we write $s' \preceq s$.

\begin{definition}[\formatDefName{Minimality}]
\label{def:minmality}
    A sharp solution map $\mathcal{T} : \mathcal{C} \times \Omega_s \mapsto \Omega_q$ is \textit{minimal} if, for any other sharp solution map $\mathcal{T}' : \mathcal{C} \times \Omega_{s'} \mapsto \Omega_q$ that takes a different input statistic $s'$, we have $s' \preceq s$.
\end{definition}
Minimality ensures that no sharp map could be constructed using strictly less information.
With \cref{def:Soundness,def:Completeness,def:minmality} in place, we can show the existence of an optimal solution map for our problem.  
\begin{theorem}[\formatDefName{Optimal solution map}]
\label{thm:Optimal solution map}
    Let the model class $\mathcal{M}$ consist of all SEMs consistent with \cref{eqn:Z,eqn:X,eqn:Y} for some given $\bm \Phi$ and $d_{\bm \Phi} \leq d_{\bm Z}$. Assume the existence of a ground truth joint distribution $P(\bm X, Y, \bm Z)$ with finite  
    covariance 
    parameters $\bm{\beta_\Phi}^*, \bm{\beta_y}^*$.
    Define the affine function $h (\bm \theta) := \bm \beta_{\bm y} - \bm \theta \cdot \bm \beta_{\bm \Phi}$, for some $\bm \beta_{\bm y}$, $\bm \beta_{\bm \Phi}$. 
    Then the solution map $\mathcal{T}$ defined by: 
    \begin{align}
        \mathcal{T} \left(c, s = (\bm \beta_{\bm \Phi}, \bm \beta_{\bm y}) \right)
        = \{ \bm \theta \in \mathbb{R}^{d_{\bm \Phi}} : h(\bm \theta) \in \bm \Gamma \},
    \end{align}
    is sound, complete, and minimal with respect to background constraint set $\mathcal{C} = \{ \mathbb{I} [ \bm \gamma_{\bm g} \in \bm \Gamma] : \bm \Gamma \subseteq \mathbb{R}^{d_{\bm Z}} \}$,
    where $\mathbb I[\cdot]$ denotes the indicator function.
\end{theorem}
In \cref{app:conditions for optimality} we examine how extra assumptions
can result in smaller solution sets.
In particular, we show that this is the case for models with categorical \citep{Balke1997} or bounded \citep{Manski1990} outcomes. 

\subsection{Budget Background Constraints}
\label{sec:Budget constraints}

We introduce sensible choices for the search space $\bm \Gamma$. 
For any $\CovGZ \in \mathbb{R}^{d_{\bm Z}}$ and any $\alpha \in [0, 1)$, if $\CovGZ' := \alpha \CovGZ$, the degree of IV violation attributed to each candidate and the combined violation is strictly weaker for $\CovGZ'$ than for $\CovGZ$. 
Below, we formalize the notion that if $\CovGZ$ is plausible, then so too is $\CovGZ'$.
\begin{definition}[\formatDefName{Star domain}]
    A set of points $A \subseteq \mathbb{R}^d$ is a \textit{star domain} if there exists a point $\vect{a} \in A$ such that the line between $\vect{a}$ and any other point $\vect{a'} \in A$ is contained within $A$.
    Equivalently, a star domain is a space $A \subseteq \mathbb{R}^d$ with a nonempty convex kernel: 
    \begin{align*}
        \mathrm{ck} (A) := 
         \{ &\vect{a} \in A \mid \forall \vect{a'} \in A, \eta \in [0, 1] : \\
         &~\eta\vect{a} + (1 - \eta)\vect{a'} \in A \}.
    \end{align*}
\end{definition}

\begin{principle}[\formatDefName{Starfish principle}]
\label{prin:starFISH}
    Relax a structural assumption by adding a latent sensitivity parameter, $\bm \delta$, whose direction is related to the mechanisms of violation and whose magnitude increases with the degree of violation by these mechanisms. 
    Bound the support of this parameter within a star domain for the FeasIble SearcH space---i.e., a starFISH---whose convex kernel includes the point $\vect{\delta} = \vect{0}$.
\end{principle}
This principle encapsulates current literature restricting the number\footnote{Notice that the $L_0$ norm constraint $\lVert h(\bm \theta)\rVert_0 \leq b$ is a star domain with a convex kernel including the point $\bm \theta = \bm 0$.} \citep{kang2016,Hartwig2017,silva_shimizu_2017,hartford2021} or total effect \citep{ramsahai2012, Conley2012, silva_evans:16, watson2024bounding,jiang2024} 
of invalid instruments. 
Convex restrictions lead to convex identified sets when the sensitivity parameter is an affine function of the causal effect parameters.
If this affine function intersects the convex kernel, the intersection is connected (by definition). 
In the general case, the identified set for the quantities of interest may be disconnected (see \cref{fig:gamma-space illustrations}).
However, each disjoint subset might also provide a distinct mechanistic interpretation for the plausible violations.

\begin{figure*}
    \centering
    \vspace{-0.15cm}
    \input{IndividualSections/2aFig2Tikz}
    \vspace{-0.05cm}
    \caption{
    \textbf{The topology of a feasible set depends on the shape of the background constraints.}\\
    Plots of constrained search spaces $\bm \Gamma$ (shaded) and lines $h (\theta)$ corresponding to $d_{\vect{Z}} = 2, d_{\Phi} = 1$.
    The intersection between a line and shaded region determines a feasible set of $\theta$. 
    The constraints are a subspace of $\mathbb{R}^{d_{\vect{Z}}}$ while $h (\theta)$ are $d_{\Phi}$-dimensional affine subspaces. (Left) Convex $\vect{\Gamma}$ entails convex feasible sets. 
    (Right) Budget constraints $\bm \Gamma (\bm \tau, \bm b)$ form a star domain. 
    They are non-convex in general and are unbounded for $b_K < d_{\bm Z}$ (i.e., some $\gamma_{g_i}$ may be unconstrained).
    This can lead to disconnected or even unidentifiable causal effect. In \cref{app:Unidentifiability} we show that unidenfitiability occurs only under violation of (B$1^*$) and can be tested in polytime.
    }
    \label{fig:gamma-space illustrations}
\end{figure*}

Here, we introduce budget constraints, which are based on two user-specified components: thresholds, $\bm \tau := (\tau_1, \tau_2, \ldots, \tau_K)$, which describe degrees of IV invalidity; and budgets, $\bm b := (b_1, b_2, \ldots, b_K)$, which determine the minimum number of IVs assumed to be at least as valid as the corresponding threshold.
The thresholds must be nonnegative and increasing: 
$0 \leq \tau_1 < \tau_2 < \ldots < \tau_K < \infty$.
Integer budgets are increasing and strictly positive: 
$0 < b_1 < b_2 < \dots < b_K \leq d_{\bm Z}$.
Thresholds and budgets with the same index $i \in [K]$ form pairs.   
The number of thresholds--budget pairs $K$ is chosen by the user but we require $K \leq d_{\bm Z}$.


We define the $d_{\bm Z}$-dimensional latent statistic $\bm U (\bm \gamma_{\bm g})$ through the following relationship:
\begin{align*}
    U_i = \ell \Longleftrightarrow \tau_{\ell-1} \leq | \gamma_{g_i} | \leq \tau_\ell.
\end{align*}
Let $U_{i1}, U_{i2}, \dots, U_{iK}$ represent the encoding of $U_i$ such that $U_{i\ell} = \mathbb{I}[U_i \leq \ell]$ for all $\ell \in [K]$, where $\mathbb{I}[\cdot]$ is the indicator function.
We define the budget constraint restriction $\vect{\Gamma} (\vect{\tau}, \vect{b})$ as the set of vectors $\CovGZ$ satisfying, for all $\ell \in [K], \sum_{i = 1}^{d_{\vect Z}} U_{i\ell} \geq b_\ell$ for some choice of budgets and thresholds. 
We denote the set of $\bm U$ with encodings satisfying the inequality above by $\bm \Sigma_{\bm b}$. 
In particular, if equality holds, we say $\bm U \in \bm \Sigma_{\bm b}^{(\mathrm{max})}$ is a maximally relaxed assignment. 
These definitions will be useful in the next section. 



Since $\bm U (\bm \gamma_{\bm g})$ depends on $\bm \theta^*$ it is only partially identifiable in general. 
We define the set of plausible budget assignments:
\begin{align*}
    \{ \bm U (h(\bm \theta)) : \bm \theta \in \mathbb{R}^{d_{\bm Z}}, h(\bm \theta) \in \bm \Gamma (\bm b, \bm \tau) \}.
\end{align*}
For instance, if $K=1$ and $\tau_1 = 0$, we assume at least $b_1$ candidate IVs are valid (having $\gamma_{g_i} = 0$), but allow $d_{\bm Z} - b_1$ to be unrestricted. 
This extends the $L_0$-norm constraint from \citet{kang2016} to violation of (A$2$) and/or (A$3$).

\section{THE ALGORITHM}
\label{sec:BudgetIV}

The objective of \texttt{budgetIV} is twofold: 
$(\text{O}1)$ discover plausible budget assignments; and
$(\text{O}2)$ partially identify the ATE.
We simplify (O$1$), which for general $ d_{\bm Z} > d_{\bm \Phi} > 1$ allows the solution set for $\bm \theta$ to be calculated using subset selection over $\bm \Sigma_{\bm b}^{(\mathrm{max})}$.
We conjecture that subset selection cannot be avoided, which means finding any $\bm \theta$ in the solution set is NP-hard (w.r.t. $d_{\bm Z}$ and $d_{\bm \Phi}$), while finding the entire solution set is \#P-hard.
As it stands, $\texttt{budgetIV}$ should not be run for $d_{\bm Z}, d_{\bm \Phi} > 10$.  
When $d_{\Phi} = 1$, however, we show the solution set can be found in polynomial time using our algorithm \texttt{budgetIV\_scalar}.
Pseudocode for both algorithms is given in \cref{app:Algorithms}; executable code is available online.\footnote{\url{https://github.com/jpenn2023/budgetIVr}.}


\paragraph{Simplifying $(\mathrm{O}1)$ for $d_{\bm \Phi} > 1$.}
Since $\bm U (\bm \gamma)$ is uniquely defined for each $\bm \gamma \in \mathbb{R}^{d_{\bm Z}}$, the feasible region $\bm \Gamma (\bm \tau, \bm b)$ decomposes into subsets $\bm \Gamma_{\bm U}$ for which $\bm \gamma \in \bm \Gamma_{\bm U} \implies \bm U(\bm \gamma) = \bm U$.
However, in \cref{app:Further technical details}, we show that $\bm \Gamma (\bm \tau, \bm b)$ can also be thought of as a union of overlapping cuboids. 
Indeed, in \cref{fig:gamma-space illustrations} we see a feasible region made up of two overlapping rectangles. 
Each cuboid $\bm{\Tilde{\Gamma}}_{\bm{\Tilde{U}}}$ is indexed by a maximal budget assignment $\bm{\Tilde{U}} \in \bm \Sigma_{\bm b}^{(\mathrm{max})}$. 

\subsection{\texttt{budgetIV} with Oracle \texorpdfstring{$\bm \beta$}{TEXT} Parameters}
\label{sec:budgetIV oracle beta parameters}

We summarize the $\texttt{budgetIV}$ algorithm below.


(1) For each $\Tilde{\bm U} \in \bm \Sigma^{(\mathrm{max})}$, test for the intersection between $h(\bm \theta) = \bm \beta_{\bm y} - \bm \theta \cdot \bm \beta_{\bm \Phi}$ and $\Tilde{\bm \Gamma}_{\Tilde{\bm U}}$. 

(2) If there is an intersection, solve the following linear program to find the bounds on $\mathrm{ATE} (\bm x; \bm x_0)$ for each $\bm x \in \Omega_{\bm X}$ of interest and baseline treatment $\bm x_{\bm 0}$:
\begin{align*}
    \minmax_{\bm \theta: ~h(\bm \theta) \in \Tilde{\bm \Gamma}_{\Tilde{\bm U}}} ~\bm \theta \cdot \big(\bm \Phi (\bm x) - \bm \Phi (\bm x_0)\big).
\end{align*}

(3) Let $\bm \theta^{-/+}(\bm x)$ denote the argmin/argmax solution (respectively) to the optimization problem in step (2) at point $\bm x$. The function $U_i (h(\bm \theta))$ returns the value of the latent variable $U_i$ at the point $h(\bm \theta)$. 
Compute $\check{\bm U}$, defined by its components:
\begin{align*}
    \check{U}_i := \max_{\bm x \in \Omega_{\bm X}} \max_{\diamond \in \{ -, +\}} U_i \big(\bm \theta^{\diamond} (\bm x)\big).
\end{align*}

(4) Return all unique $\check{\bm U}$ (O$2$) along with the corresponding ATE bounds (O$1$).

\paragraph{Polytime \texttt{budgetIV\_scalar}.}
The algorithm above describes an ILP 
that requires a linear search over $\lvert \Sigma_{\bm b}^{(\mathrm{max})}\rvert$, which itself is bounded above by $d_{\bm Z}^{d_{\bm Z}}$, corresponding to the case where $K=d_Z$.
Though standard solvers are highly optimized, this task can quickly become intractable with many candidate instruments. 
When $d_{\Phi} = 1$, however, exact partial identification of the ATE can be computed in $\Tilde{\mathcal{O}} (d_{\bm Z} K)$ time. 
This is done by noticing that the value of $\bm U$ associated with $h(\theta)$ can only change as a function of $\theta$ at the $2 d_{\bm Z} K$ points: 
\begin{align*}
    \Theta_{ik}^{\pm} := \frac{\pm \tau_k - (\beta_y)_i}{(\beta_{\Phi})_i},
\end{align*}
where we ignore any $i$ for which $(\beta_{\Phi})_i =0$.
Similarly, the full set of plausible budget assignments can be calculated in polynomial time. 
See \cref{app:Algorithms} for further details.

\subsection{The $L_0$-norm Constraint} \label{sec:L_0 norm constraint}


\citet{kang2016} show that point identification is possible in the linear setting with univariate exposure $X$ and some invalid instruments---provided at least half are valid. 
We show that this result cannot be generalized to multidimensional exposures.
However, we place tight bounds on the cardinality of the feasible set with respect to the minimum number of valid IVs $b$. 
We use the following shorthand: 
\begin{align*}
    \mathcal{T}_{L_0} (b) &:= \mathcal{T} \left(c = \mathbb{I} \left[ \bm \gamma_{\bm g} \in \bm \Gamma (0,b) \right], s = \{\bm \beta_{\bm \Phi}, \bm \beta_{\bm y}\} \right). 
\end{align*}

\begin{theorem}[\formatDefName{$t$-point identification}] \label{thm:n-point identifiability}
    Assume Eqs. \ref{eqn:Z}, \ref{eqn:X}, \ref{eqn:Y} and claims (B$1^*$), (B2) hold for some $\bm \Gamma (0, b)$. 
    Then, for all $b < d_{\bm Z} - d_{\bm \Phi} + 1$, the cardinality of the feasible set $t := \lvert \mathcal{T}_{L_0} (b) \rvert \in \mathbb N$ is bounded above by:
    \begin{align*}
        t \leq \frac{d_{\bm Z}!}{(d_{\bm \Phi} - 1)!(d_{\bm Z} - d_{\bm \Phi} + 1)!} \left\lfloor \frac{d_{\bm Z} - d_{\bm \Phi} + 1}{d_{\bm Z} - b - d_{\bm \Phi} + 1} \right\rfloor.
    \end{align*}
    This bound is tight in the sense that equality holds for some values of $\bm \beta_{\bm \Phi}$, $\bm \beta_{\bm y}$. 
    We extend these results to incorporate violation of (B$1^*$) in the proof (\cref{app: Proof of n point identifiability etc}).
\end{theorem}

\begin{corollary}[\formatDefName{No point identification for $d_{\bm \Phi} > 1$}]
    There is no value of $b$ for which $t \leq 1$ is guaranteed for all $\lvert \mathcal{T}_{L_0} \rvert$ when $d_{\bm \Phi} > 1$. 
\end{corollary}

\begin{corollary}[\formatDefName{$t$-point identification for $d_{\Phi} = 1$}]
    \label{coly:n-point identification scalar phi}
    In the case of $d_{\Phi} =1$, we have: 
    \begin{align*}
        \left\lvert \mathcal{T}_{L_0} (b) \right\rvert =: t \leq \left\lfloor \frac{d_{\bm Z}}{d_{\bm Z} - b} \right\rfloor \leq d_{\bm Z}.
    \end{align*}
    This reduces to point identifiability when $b < d_{\bm Z}/2$.
\end{corollary}

\section{INFERENCE}
\label{sec:inference}

In this section we consider finite-sample uncertainty. 
We seek confidence sets with (asymptotically) valid coverage over the solution set $\{ \bm \theta \in \mathbb{R}^{d_{\bm Z}} : h(\bm \theta) \in \bm \Gamma (\bm \tau, \bm b)\}$. 

Since $\mathcal{T}$ is a deterministic map, it is possible to use a confidence set $(\bm B_{\bm \Phi}, \bm B_{\bm y})$ over the covariance parameters $\bm \beta_{\bm \Phi}$, $\bm \beta_{\bm y}$.
The confidence set over the causal parameter is then all $\bm \theta \in \mathbb{R}^{d_{\bm \Phi}}$ satisfying:
\begin{align*}
	\forall (\bm \beta_{\bm \Phi}, \bm \beta_{\bm y}) \in (\bm B_{\bm \Phi}, \bm B_{\bm y}) : \left( \bm \beta_{\bm y} - \bm \theta \cdot \bm \beta_{\bm \Phi} \right) \in \bm \Gamma (\bm \tau, \bm b).
\end{align*}
Provided the estimators $\bm{\hat \beta}_{\bm \Phi}$, $\bm{\hat \beta}_{\bm y}$ have finite variance, an asymptotically valid confidence set can be constructed by modeling these estimators as multivariate normal (see \cref{app:Proof of coverage}).  
However, we choose to make the following simplifications taken from the applied IV literature. 


\subsection{Coverage with Summary Statistics under the NOME Assumption}

Candidate IVs are often selected from a pool of covariates, with inclusion based solely on marginal association with the exposure.
For instance,
Mendelian randomization (MR) is a popular approach in genetic epidemiology whereby genetic variants $\bm Z$ are used as IVs to determine the causal effect of phenotype(s) $\bm X$ on a health outcome $Y$.
The $\bm Z \rightarrow \bm X$ link is usually established by a genome-wide association study (GWAS).
Empirically, the chosen $\bm Z$ tend to be less strongly associated with $Y$ than with $\bm X$.
This can occur for various reasons: strong latent confounding between $\bm X$ and $Y$; a weak causal effect of $\bm X$ on $Y$; low variation in $Y$ (e.g., rare diseases); and/or smaller sample sizes for evaluating $P(\bm Z, Y)$ than $P(\bm Z, \bm X)$ \citep{Pierce2013}.

In such cases, finite-sample error in $\hat{\bm{\theta}}$ is mostly explained by finite-sample error in $\hat{\bm{\beta}}_{\bm y}$.
This has led to the introduction of a \textit{no measurement error} (NOME) assumption in MR studies \citep{Bowden2016NOME, Bowden2019BeyondNOME}, under which one assumes finite sample error is only due to error in $\hat{\bm{\beta}}_{\bm y}$.
For complex choices of $\bm \Phi$, it may also be the case that $p$-values for testing $(\hat{\beta}_{\bm \Phi})_{ij} = 0$ are lower than those for testing $({\hat{\beta}}_{y})_i = 0$, which would justify the NOME assumption. 
We have adapted \texttt{budgetIV} to construct confidence sets for $\bm \theta$ with (asymptotically) valid probabilities to cover the entire solution set $\{ \bm \theta \in \mathbb{R}^{d_{\bm \Phi}} : h(\bm \theta) \in \bm \Gamma (\bm \tau, \bm b) \}$ under the NOME assumption.

In particular, we use a Bonferroni adjustment to construct a box-shaped confidence set over $\bm{\beta}_{\bm y}$.
For a target coverage $(1-\alpha) \times 100\%$, we take a union over $(1-\alpha/d_{\bm Z}) \times 100\%$ confidence intervals corresponding to each  $(\beta_y)_j$.
Let $(\delta \beta_y)_i$ denote the half width of the confidence interval over $(\beta_y)_i$. 
While box-shaped confidence sets are conservative, they neatly superimpose with the $\bm \tau$-thresholds: 
\begin{multline*}
     \Big| (\hat{\beta}_y)_i \pm (\delta \beta_y)_i - (\bm \theta \cdot \bm{\hat{\beta}}_{\bm \Phi})_i \Big| \leq \tau \\
    \Longleftrightarrow \Big| (\hat{\beta}_y)_i - (\bm \theta \cdot \bm{\hat{\beta}}_{\bm \Phi})_i \Big| \leq \tau + (\delta \beta_y)_i. \qquad
\end{multline*}
We use this relationship to, we add slack $( \pm \delta \beta_{y})_i$ to the corresponding face of each $\Tilde{\bm \Gamma}_{\Tilde{\bm U}}$ for $\Tilde{\bm U} \in \bm \Sigma^{(\mathrm{max})}$ to form $\hat{\bm \Gamma}_{\Tilde{\bm U}}$, and optimize over these domains to construct corresponding confidence intervals for the $\mathrm{ATE}(\bm x; \bm x_{\bm 0})$ (see \cref{sec:budgetIV oracle beta parameters}). 
Our approach has the following guarantee.

\begin{theorem}[\formatDefName{Coverage}] \label{thm:Coverage}
    Fix the target level $\alpha \in (0,1)$. Let $\bm{\hat \Gamma}_\alpha := \bigcup_{\Tilde{\bm U} \in \bm \Sigma^{(\mathrm{max})}} \bm{\hat \Gamma}_{\Tilde{\bm U}}$ be formed from the $(1 - \alpha/d_{\bm Z}) \times 100\%$ confidence intervals for each component of $\bm{\hat \beta}_{\bm y}$ as described above, where the intervals are estimated from a dataset of $n$ samples drawn $\mathrm{iid}$ from $P(\bm X, Y, \bm Z)$. 
    Using the shorthand:
    \begin{align*}
        \hat{\mathcal{T}}_{\alpha} := \mathcal{T} \left(c = \mathbb{I} [ \bm \gamma_{\bm g} \in \bm{\hat \Gamma}_\alpha], s = (\bm \beta_{\bm \Phi}^{\bm *}, \bm{\hat \beta}_{\bm y}) \right),
    \end{align*}
    we have, as $n \rightarrow \infty$:
    \begin{align*}
        P\big(\bm \theta^* \in \hat{\mathcal{T}}_{\alpha} \big) \geq 1 - \alpha.
    \end{align*}
\end{theorem}

\section{EXPERIMENTS} \label{sec:experiments}

\begin{figure}
    \centering
    \begin{tikzpicture}
        \node[draw=none] at (0,0) {\includegraphics[trim=0 1.75cm 0 0, clip, width=0.85\linewidth]{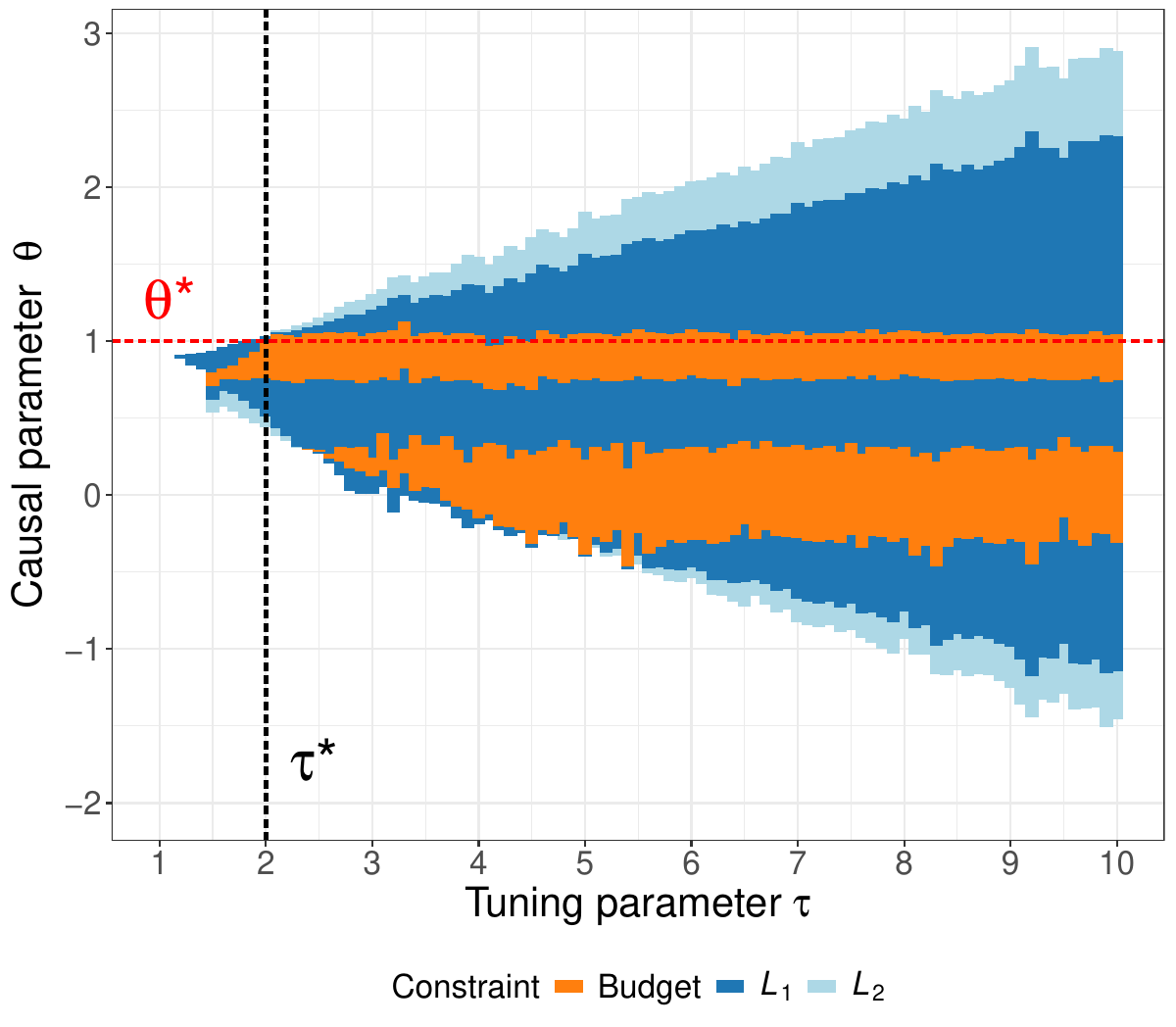}};
        \node[draw=none,anchor=west] at (1.5,0.63) {\fontsize{8}{0}\selectfont $\texttt{budgetIV}$};
        \node[draw=none,anchor=west] at (1.5,-0.2) {\fontsize{8}{0}\selectfont $\texttt{budgetIV}$};
        \draw[draw=black,fill=white] (-2.65,2.1) rectangle (1.85,2.55);
        \draw[draw=black,fill=white] (-2.65,1.80) rectangle (0.45,2.55);
        \draw[draw=none,fill=white] (-2.65,2.1) rectangle (1.85,2.55);
        \draw[draw=none,fill=white] (-2.65,1.80) rectangle (0.45,2.55);
        \node[draw=none,anchor=west] at (-2.7,2.3) {\fontsize{8}{0}\selectfont \textsf{Our method yields sharper bounds}};
        \node[draw=none,anchor=west] at (-2.7,2.0) {\fontsize{8}{0}\selectfont \textsf{than convex relaxations}};
        \node[draw=none,anchor=west,rotate=8,text=black] at (1.5,1.2) {\fontsize{8}{0}\selectfont $L_1$-norm};
        \node[draw=none,anchor=west,rotate=17,text=black] at (1.5,1.8) {\fontsize{8}{0}\selectfont $L_2$-norm};
    \end{tikzpicture}
    \caption{
    \textbf{Our method yields sharper bounds than convex relaxations in linear models.}
    Bounds on $\theta$ for a series of linear models with scalar exposure $X$ and $\bm \gamma_{\bm g}^* = (-2, -0.4)$. Plug-in estimators are used throughout. Orange bounds come from $\texttt{budgetIV}$ with $\bm \tau = (0.6, \tau)$; dark blue from the $L_1$-norm constraint $\lVert \bm \gamma_{\bm g}\rVert_1 \leq \tau + 0.6$; and light blue from the $L_2$-norm constraint $\lVert \bm \gamma_{\bm g} \rVert_2 \leq \sqrt{\tau^2 + 0.6^2}$. We vary $\tau$ linearly from $0$ to $10$---each bound is an experiment.}
    \label{fig:linear_exp}
\end{figure}

\begin{figure}
    \centering
    \begin{tikzpicture}
        \node[draw=none] at (0,0) {\includegraphics[trim=0 1.75cm 0 0, clip, width=0.85\linewidth]{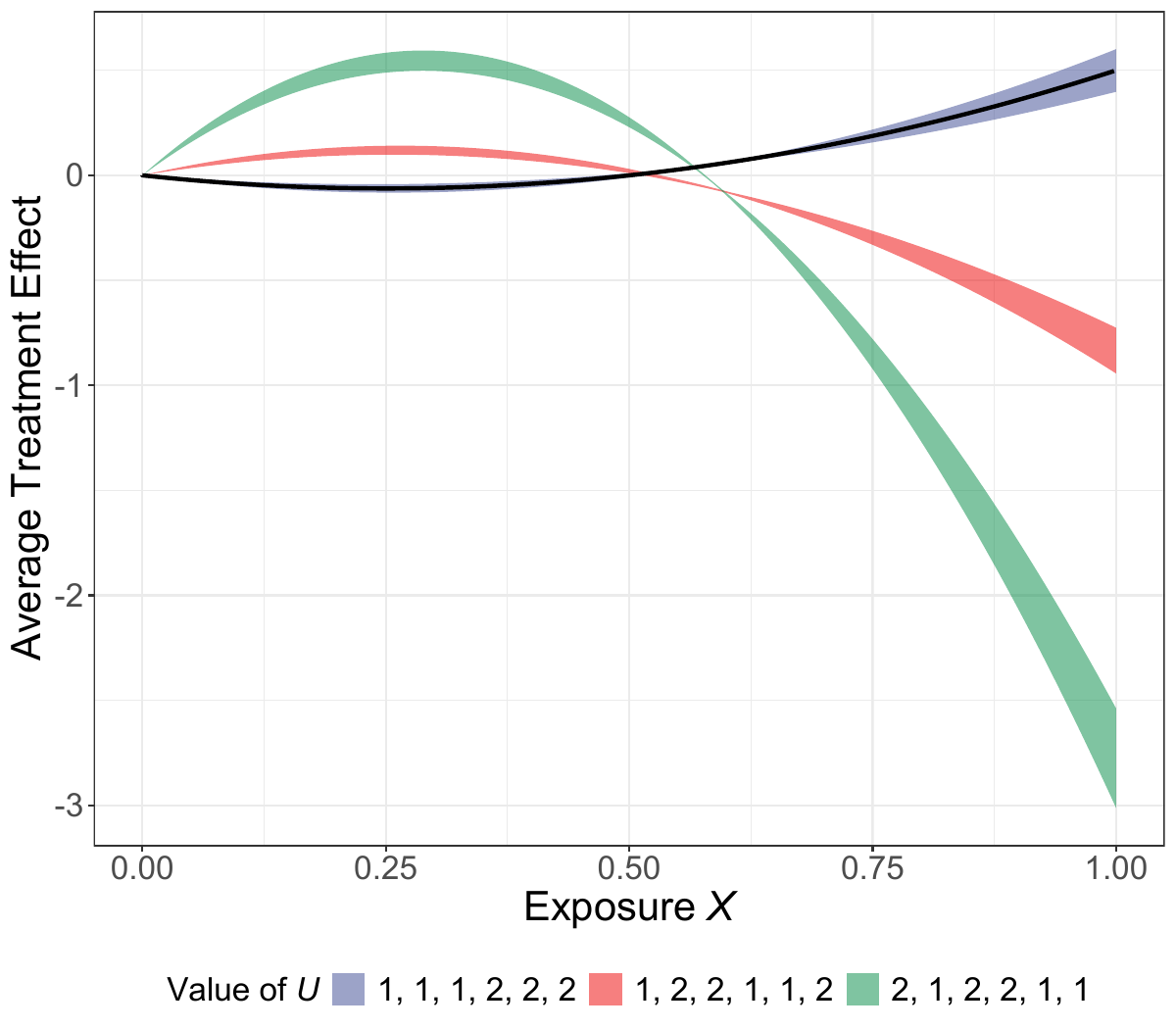}};
        \draw[overlay, decorate, decoration={text along path, text={|\fontsize{8}{0}\selectfont|U=[1 1 1 2 2 2]}, text align={left}}] 
        (0.75,1.93) arc[start angle=270+6, end angle=270+45, radius=7cm];
        \draw[overlay, decorate, decoration={text along path, text={|\fontsize{8}{0}\selectfont|U=[1 2 2 1 1 2]}, text align={left}}] 
        (1.3,1.57) arc[start angle=90-10, end angle=90-30, radius=6cm];
        \draw[overlay, decorate, decoration={text along path, text={|\fontsize{8}{0}\selectfont|U=[2 1 2 2 1 1]}, text align={left}}] 
        (1.65,0.95) arc[start angle=42, end angle=15, radius=10.0cm];
        \draw[draw=black,fill=white] (-2.35,-1.75) rectangle (1.85,-0.65);
        \draw[draw=none,fill=white] (-2.35,-1.75) rectangle (1.85,-0.65);
        \node[draw=none,anchor=west] at (-2.4,-0.85) {\fontsize{8}{0}\selectfont \textsf{Different choices of}};
        \node[draw=none,anchor=west] at (-2.4,-1.20) {\fontsize{8}{0}\selectfont \textsf{plausible budget assignments} U};
        \node[draw=none,anchor=west] at (-2.4,-1.55) {\fontsize{8}{0}\selectfont \textsf{offer different plausible ATEs}};

    \end{tikzpicture}
    \caption{
    \textbf{Budget constraints provide information about the structure of the problem.}
    Feasible values of the ATE relative to a baseline of $x_0=0$ as exposure $X$ varies in a nonlinear SEM with $d_{\bm Z}=6$. The true ATE is given by the solid black curve. 
    Each colored region corresponds to a unique intersection of $\bm \gamma_{\bm g}$ and the star domain $\bm \Gamma$. The union of such intersections at each value of $X$ produces a disconnected feasible set.}
    \label{fig:nonlinear_exp}
\end{figure}

The full simulation studies are detailed in \cref{app:Experiments}, along with additional results. 

\paragraph{Sharper than convex relaxations.}
In \cref{fig:linear_exp} we study the effect of varying the thresholds $\bm \tau$ on the feasible set. 
We consider a linear Gaussian model with $d_X = 1$ and $d_{\bm Z} = 2$ where (A2) is violated through correlation between $\bm Z$ and $\epsilon_y$, while (A3) is satisfied.
We fix the ground truth parameters $\theta^* = 1$, $\bm \gamma_{\bm g}^* = (-2, 0.4)$ and $\bm \beta_{\Phi}^* = (2, -4)$. 
The remaining parameters are randomized while $\bm \Gamma (\tau, b = 1)$ is varied.

The results illustrate that if the degree of validity for some instruments can be bounded, \texttt{budgetIV} may be insensitive to weak constraints on other instruments.
By contrast, when used for partial identification, the convex relaxations suffer from these bottlenecks of limited background knowledge, and the ATE bounds grow linearly as a result. 

\paragraph{Budget constraints highlight possible mechanisms.}
\cref{fig:nonlinear_exp} shows bounds on the ATE for feasible values of $\bm U$ with a quadratic ground truth $\Phi^* (X)$ under the presence of (A2) and (A3) violations.
We study a collection of violations in \cref{app:Experiments}.

Feasible sets returned by $\texttt{budgetIV}$ represent a union of convex bounds, each of which corresponds to a unique causal hypothesis that cannot be determined by the data or budget constraints alone (see Fig.~\ref{fig:nonlinear_exp}).
This drastically reduces the search space of possible causal mechanisms for the IV candidates.
If one or more of these solutions can be ruled out by expert knowledge---e.g., if a monotonicity assumption is justified \citep{angrist1996}---then the method can be rerun with added constraints, pruning the search space still further.
In this way, $\texttt{budgetIV}$ can help practitioners evaluate causal systems in a dynamic and principled manner, aiding in hypothesis generation and experimental design.

In \cref{fig:nonlinear_exp} the ground truth corresponds to the only plausible ATE with positive convexity. 
With additional expert knowledge, one might infer that $Z_1$, $Z_2$, and $Z_3$ are the true set of valid instruments. 

\begin{table}

\begin{minipage}{\columnwidth}
    \centering
    \includegraphics[width=\linewidth]{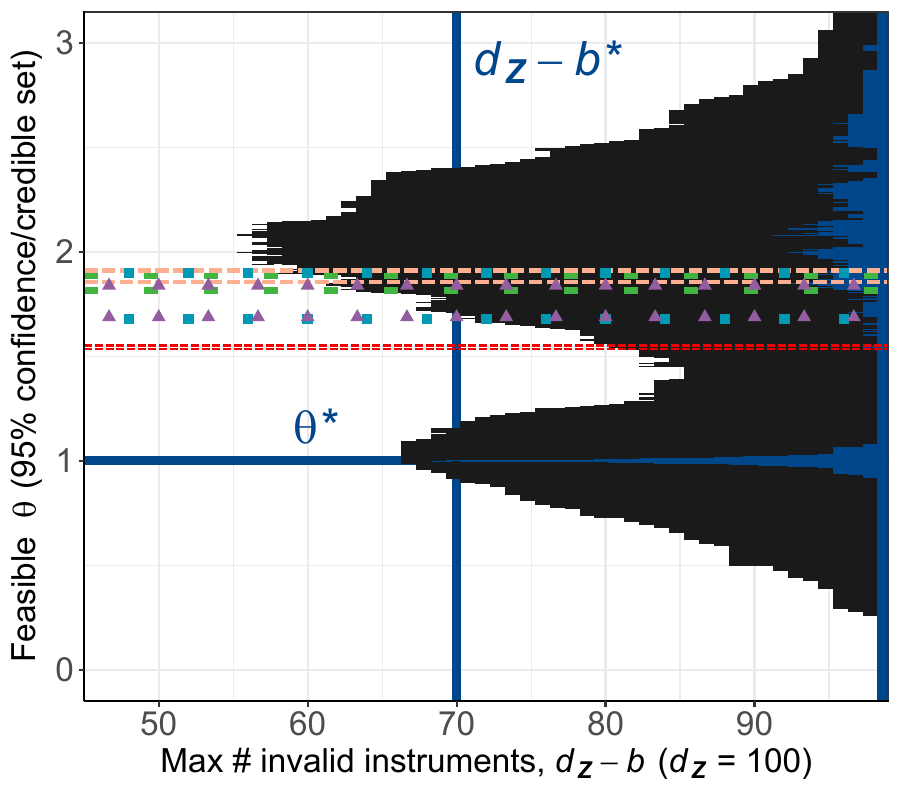}
\end{minipage}

\vspace{1em}

\resizebox{\columnwidth}{!}{
\begin{tabular}{|l|c|c|c}
\hline
{\textbf{Method}} & \textbf{Confidence/Credible Set} & $\ni \theta^*$ \\
\hline
\textcolor[HTML]{0099B4}{\textbf{MR-Egger}} & $[1.680, 1.898]$ & \xmark \\
\textcolor[HTML]{42B540}{\textbf{MR-Median}} & $[1.817, 1.884]$ & \xmark \\
\textcolor[HTML]{FDAF91}{\textbf{MBE}} & $[1.857, 1.911]$ & \xmark \\
\textcolor[HTML]{925E9F}{\textbf{IVW}} & $[1.688, 1.839]$ & \xmark\\
\textcolor[HTML]{ED0000}{\textbf{MASSIVE}} & $[1.536, 1.552]$ & \xmark\\
\textcolor[HTML]{333333}{\textbf{\texttt{budgetIV}}} & $[0.914, 1.160] \cup [1.705, 2.403]$ & \cmark\\
\textbf{\textcolor[HTML]{00468B}{Oracle}} & $[0.997, 1.003]$ & \cmark\\
\hline
\end{tabular}
}

\captionof{figure}{\textbf{\texttt{budgetIV} captures the true causal effect when most candidates are invalid IVs.} Results from a simulation study with $d_{\bm Z} = 100$ candidate IVs, $70$ of which violate (A$3$), benchmarking 
(\textcolor[HTML]{333333}{Black}) $95\%$ coverage of the feasible set according \texttt{budgetIV\_scalar}, where the budget constraints $\bm \Gamma (\tau = 0, b)$ are varied along the $x$-axis. 
(\textcolor[HTML]{00468B}{Blue}) The optimal solution set relative to the constraint $\bm \Gamma (\tau = 0.001, b)$ (for visibility) captures the true causal effect if and only if the choice of $b$ doesn't exclude the ground truth $\bm \gamma_{\bm g}^{\bm *}$.
(Others) Confidence intervals for benchmark methods produce do not include $\theta^*$. 
\label{fig:benchmark main}
}

\end{table}

\paragraph{Consistent inference under relaxed assumptions.} \cref{fig:benchmark main} shows a benchmarking experiment with $d_{\bm Z} = 100$ candidate IVs, $30$ of which are valid and the remaining $70$ violate (A$3$). 
The data is simulated via a linear with a scalar exposure $X$ and multivariate Gaussian $(\epsilon_{x}, \epsilon_y, \bm Z := \bm \epsilon_{\bm z})$.  
Linearity, (A$3$) violation, scalar $X$ and Gaussian exogenous variables reflect the modeling assumptions in the benchmarking methods. 
We apply \texttt{budgetIV\_scalar} for computational efficiency.

We apply a series of budget constraints $\bm \Gamma (\tau = 0, b)$ where $b$ takes each value from $1$ to $100$ ($b=0$ corresponds to no constraint).  
We calculate $\bm \delta \bm \beta_{\bm y}$ from a $95\%$ confidence set using $100,000$ samples, which is a typical GWAS sample size.

The resulting confidence sets decompose into three kinds: 
($d_{\bm Z} - b < 56$) empty, falsifying the budget constraints; ($56 \leq d_{\bm{Z}} - b < 67$) nonempty but excluding $\theta^*$; and ($d_{\bm Z} - b \geq 67$)  containing $\theta^*$. 
Whenever $b^* < b$ (so that $\bm \gamma_{\bm g}^{\bm *} \in \bm{\Gamma} (0, b)$) our confidence set captures $\theta^*$.   
Similarly, the solution set, calculated using $\bm \beta_{\bm y}^{\bm *}$ and $\bm \beta_{\bm \Phi}^{\bm *}$, is shown to contain $\theta^*$ if and only if $b < b^*$.


The methods we benchmark against give biased results. 
Inverse variance weighting (IVW) is a classical method that relies on all candidates being valid IVs. 
MR-median \citep{Bowden2016WeightedMedian} assumes $\mathrm{median} ( \gamma_{g_i}^* )_{i \in [d_{\bm Z}]} = 0$.
MR-Egger assumes candidates are invalid through independent mechanisms (e.g., not affecting $Y$ via shared mediators or being confounded with $Y$ via shared confounders).
These assumptions do not hold in our simulation study so we expect the results to be biased.  
Intriguingly, the stated assumptions for the Bayesian approach MASSIVE \citep{bucur2020} and the mode based estimate MBE \citep{Hartwig2017} hold in our experiment.

The MASSIVE estimator assumes an $L_0$-norm constraint $\bm \Gamma(\tau = 0, b)$ with a tuneable value of $b$, which we set to $b^*$. 
We have shown $\theta^*$ is only partially identified under such constraints. 
MASSIVE applies Bayesian model averaging over plausible sets of valid IVs. 
The resulting posterior distribution sits between the two disjoint confidence intervals returned by \texttt{budgetIV\_scalar}, which themselves correspond to different sets of valid IVs.

MBE assumes $\mathrm{mode} ( \gamma_{g_i}^* )_{i \in [d_{\bm Z}]} = 0$. 
\cref{fig:benchmark main} shows that this assumption holds because the optimal solution set peaks at $\theta = \theta^* := 1$.  
They estimate the corresponding modal causal effect $\theta$ using the summary statistics $\bm{\hat \beta_{\bm y}}$, $\bm{\hat \beta_{\bm \Phi}}$, standard errors and the bandwidth selection rule of \citet{bickel2008}.
Performing statistical inference on an estimator of the mode is not straightforward \citep{Genovese2015}.
\citet{Hartwig2017} base their confidence intervals a normal approximation bootstrap. 
However, mode estimators generally do not converge to a normal distribution, which may explain the underestimated uncertainty.
Indeed \cref{fig app:Linear high dZ simulation grid} in \cref{app:Linear high dZ experiment} shows an expanded grid of experiments under which the estimates from MBE are highly variable.   



\section{RELATED WORK} 
\label{sec:related work}

There is a substantial literature on partial identification for IV models with restricted outcome domains. 
Early work includes seminal papers by \citet{Manski1990} and \citet{Balke1997}. 
Later work considered bounded violations of the IV assumptions \citep{ramsahai2012, silva_evans:16, jiang2024}. 
Others have proposed 
generic methods for bounding counterfactual probabilities in discrete settings---either via polynomial programs \citep{duarte2023, sachs2023} or Markov chain Monte Carlo \citep{zhang2022}---with applications to IV models.

In the continuous setting, partial identification for pseudo-IV models has typically been formulated with respect to some convex relaxation of either (A2) or (A3), in both linear \citep{Conley2012, watson2024bounding} and nonlinear SEMs \citep{Newey2003nonparametric, Gunsilius2019}. 
In recent years, several authors have described more generic solutions based on stochastic gradient descent \citep{kilbertus2020class, hu2021, padh_stochastic2023}. 
Unlike the linear programming approach of \texttt{budgetIV}, these methods are not guaranteed to converge on global optima.

In the MR literature, various methods are designed to try to handle linkage disequilibrium and/or pleiotropy. 
One strategy is to include a large number of candidate instruments and assume that biases will tend to cancel out in the limit \citep{Kolesar2015, Bowden2015}. Others take a feature selection approach in which $Z$'s may be rejected on the basis of statistical tests \citep{chu_semi_instrumental, kang2020} or $L_1$-penalized regression \citep{kang2016, Guo2018, Windmeijer2019, Xue2023}. Alternatively, instruments may be pooled into a single feature \citep{kuang2020}.
An especially popular choice is the modal validity assumption \citep{Hartwig2017, hartford2021}, which asserts that the most common causal effect estimate is consistent. 
The goal in these works is point identification, which may be unrealistic if underlying assumptions fail.

Bayesian methods for causal inference in IV models are well-established. 
Priors can be used to encode uncertainty with regard to latent parameters that track either (A2) \citep{Shapland2019} or (A3) violations \citep{bucur2020, Gkatzionis2021}.
%

In recent work, \citet{vancak_2023} define a single sensitivity parameter for violation of (A2) and/or (A3). 
We extend this approach by using a latent statistic of a nonparametric function. 
Star-domain restrictions for partial identification have also been proposed \citep{molinari200881}, though not as a guiding principle. 

For a summary of the constraints and affordances of various relaxed IV methods, see Appx. \ref{appx:lit}, Table \ref{tab:literature}.

\section{DISCUSSION} 
\label{sec:Discussion}

The $\texttt{budgetIV}$ optimization problem is NP-hard for any $d_{\bm \Phi} > 1$. 
Therefore, with multivariate $\bm \Phi$, the method becomes impractical for large $d_{\bm Z}$. Approximate solutions that rely on grid search may be preferable in such cases.
Many MR studies are built on just one or a handful or genetic variants, so a cap on $d_{\bm Z}$ may not be overly restrictive in such settings. 


The additive separability of causal effects in pseudo-IV models has been assumed by various authors \citep{Newey2003nonparametric, sorowit2022, Christiansen2021}.
We have considered the special case of a homogeneous treatment effect, though our method can be further generalized if $\bm \Phi$ is promoted to known functions of $\bm X$, $\bm Z$ and/or $\bm \epsilon_{\bm z}$. 
Extending our approach to compute feasible sets of \textit{conditional} average treatment effects may result in more informative outputs \citep{cai2007, hartford2021, kennedy2023}.

We rely on the assumption $d_{\bm \Phi} \leq d_{\bm Z}$ and assume our choice of $\bm \Phi$ is sufficient to describe the ground truth causal effect of $\bm X$ on $Y$ exactly.
Future work could could investigate error arising from misspecification of $\bm \Phi$ or incorporate thresholds on this kind of error into the background assumptions.
In particular, basis expansions that have been truncated to satisfy the bound on $d_{\bm \Phi}$ may be of interest. 


While we have presented our method as inferring the ATE from the distribution of instrument validity, the correspondence can be thought of as bidirectional.  
One promising direction for future work is to invert such methods 
for the purpose of instrument discovery \citep{silva_shimizu_2017}. 
Another direction could be to use background knowledge about the functional form of the ATE (when  $d_{\bm \Phi} > 1$) to restrict the solution set.

There are several other extensions to \texttt{budgetIV} that could be of interest.  
Sharper partial identification may be achieved in models with restricted outcome domains. 
The finite-sample properties of our method might also be improved. 
We used a Bonferroni correction to construct the confidence set over $\bm \beta_{\bm y}$. 
Since this is conservative, a future direction may involve adaptively selecting the confidence thresholds for each $(\beta_y)_i$ to minimize the width of the bounds on the ATE.
On the other hand, coverage under finite sample uncertainty without the NOME assumption remains an open problem. 








\bibliography{references.bib}


\newpage
\appendix
\onecolumn
\section{PROOFS}

\subsection{Proof of \cref{thm:identifiability_completeness}}

This is a special case of a standard result from the classical IV literature, which goes back to \citet{koopmans1949identification}: $\bm \theta$ has a unique solution under the assumptions (B$1^*$) and (B$2^*$) because $\bm \beta_{\bm y} - \bm \theta \cdot \bm \beta_{\bm \Phi} = \bm 0$ is a complete system of simultaneous equations.
With finite samples, it is common to estimate $\bm \theta$ solving this equation via two-stage least squares (2SLS) or (for $d_{\Phi} = 1$) inverse variance weighting (IVW). 
The 2SLS estimator is given by:
\begin{align*}
    \bm \theta_{\mathrm{(2SLS)}} := (\bm{\beta_\Phi} \cdot \Sigma_{\bm{ZZ}}^{-1} \cdot \bm{\beta_\Phi}^{\top})^{-1} \cdot \bm{\beta_\Phi} \cdot \Sigma_{\bm{ZZ}}^{-1} \cdot \bm \beta_{\bm y},
\end{align*}
where $\Sigma_{\bm{ZZ}}$ denotes the $d_{\bm Z} \times d_{\bm Z}$ covariance matrix between the instruments. For details, see \citet{wright1928, bt_iv, angrist2009}.

\subsection{Proof of \cref{thm:Optimal solution map}}\label{app:Proof of Optimality}

Let the structural \cref{eqn:Z,eqn:X,eqn:Y}
hold for some ground truth functions $f_{\bm z}^*$, $f_{\bm x}^*$, $\vect{\Phi}^*$, $g_y^*$, and causal parameter $\vect{\theta}^* \in \mathbb{R}^{d_{\vect{\Phi}}}$.

\paragraph{Soundness} 
We prove soundness of the feasible map $\mathcal{T}$ for which:
\begin{align*}
    \mathcal{T} \left(c = \mathbb{I} [ \bm \gamma_{\bm g} \in \bm{\Gamma}], s = (\bm \beta_{\bm \Phi}, \bm \beta_{\bm y}) \right) = \{ \bm \theta \in \mathbb{R}^{d_{\bm \Phi}} : h(\bm \theta) \in \bm \Gamma \},
\end{align*} 
where $h(\bm \theta) = \bm \beta_{\bm y} - \bm \theta \cdot \bm \beta_{\bm \Phi}$.
For ease of notation, we use the shorthand $\mathcal{T} (\bm \Gamma, \bm \beta_{\bm \Phi}, \bm \beta_{\bm y})$ hereinafter.

It follows immediately from \cref{eqn:Y} and the left-linearity of the covariance operator that:
\begin{align*}
    \bm \gamma_{\bm g}^{\bm *} := \cov (g_y(\vect{Z}, \vect{\epsilon_y}), \vect{Z}) & = \cov (Y, \vect{Z}) - \cov (\vect{\theta^*} \cdot \vect{\Phi} (\vect{X}),\vect{Z}) \\
    & = \bm \beta_{\bm y}^* - \bm \theta^* \cdot \bm \beta_{\bm \Phi}^*.
\end{align*}
Therefore, $\bm \theta^* \in \mathcal{T} (\bm \Gamma, \bm \beta_{\bm \Phi}^{\bm *}, \bm \beta_{\bm y}^{\bm *})$ whenever $\bm \gamma_{\bm g}^{\bm *} = (\bm \beta_{\bm y}^* - \bm \theta^* \cdot \bm \beta_{\bm \Phi}^*) \in \bm \Gamma$ as required. 

\paragraph{Completeness} To prove completeness of $\mathcal{T}$ we have to show that any $\bm \theta \in \mathcal{T} (\vect{\Gamma}, \bm \beta_{\bm \Phi}, \bm \beta_{\bm y})$ cannot be rejected by any statistic of the observed joint distribution $P (\bm Z, \bm X, Y)$. 
Notice that in specifying \cref{eqn:Z,eqn:X,eqn:Y}, we have not made any a priori assumptions about the joint distribution $P_{\bm \epsilon}(\bm \epsilon_{\bm z}, \bm \epsilon_{\bm x}, \bm \epsilon_{\bm y})$ or the function classes to which $f_{\bm z}^*$, $f_{\bm x}^*$, $g_{y}^*$ belong. 
In \cref{app:conditions for optimality} we extend this proof to find conditions under which $\mathcal{T}$ remains complete when further structural assumptions are made.

Consider any $\vect{\theta}^{\dagger} \in \mathbb{R}^{d_{\vect{\Phi}}}$. Then the following holds:

(Z) There exists at least one function $f_{\bm z}^{\dagger}$ with the following property. 
Given any $\vect{z} \in \Omega_{\bm Z}$, either: (i) $P (\vect{x}, y, \vect{z}) = 0$ for all $\bm x \in \Omega_{\bm X}$ and $y \in \Omega_{Y}$; or (ii) there exists at least one value $\vect{\epsilon}_{\uvec{z}}^{\dagger}$ that solves the equation:
\begin{align*}
    f_{\bm z}^\dagger (\vect{\epsilon}_{\uvec{z}}^{\dagger}) = \vect{z}.
\end{align*}  
Note that the ground truth $f_{\bm z}^*$ is one valid choice of $f_{\bm z}^\dagger$.  

We can, therefore, define a function $\vect{\epsilon}_{\bm z}^{\dagger} (\vect{z})$ that satisfies the above equation for all $\vect{z}$ for which case (i) is false. 
We have not demanded either $f_{\bm z}^{\dagger}$ or $\bm \epsilon_{\bm z}^{\dagger} (\vect{z})$ to be unique.

(X) Likewise, there exists at least one function $f_{\bm x}^{\dagger}$ with the following property. 
Given any $\vect{x} \in \Omega_{\bm X}, \vect{z} \in \Omega_{\bm Z}$, either: (i) $P (\vect{x}, y, \vect{z}) = 0$ for all $y \in \Omega_{Y}$, or (ii) there exists at least one value $\vect{\epsilon}_{\uvec{x}}^{\dagger}$ that solves the equation:
\begin{align*}
    f_{\bm x}^\dagger (\vect{z}, \vect{\epsilon}_{\uvec{x}}^{\dagger}) = \vect{x},
\end{align*}  
and $f_{\bm x}^*$ is one valid choice of $f_{\bm x}^\dagger$.  
We can, therefore, define at least one function $\vect{\epsilon}_{\bm x}^{\dagger} (\vect{z}, \vect{x})$ that satisfies the above equation for all $\vect{x}$ and $\vect{z}$ for which case (i) is false. 

(Y) Likewise, for any $\vect{x} \in \Omega_{\bm X}$, $y \in \Omega_{Y}$, $\vect{z} \in \Omega_{\bm Z}$, either (i) $P(\bm x, y, \bm z) = 0$, or (ii) for any $\bm \theta^{\dagger} \in \mathbb{R}^{d_{\bm \Phi}}$ there exists at least function $g_y^{\dagger}$ and one value $\vect{\epsilon}_{\uvec{y}}^{\dagger}$ that solves the equation:
\begin{align*}
    g_y^{\dagger} (\vect{z}, \vect{\epsilon}_{\uvec{y}}^{\dagger}) = y - \vect{\theta}^{\dagger} \cdot \vect{\Phi} (\vect{x}).
\end{align*}
For instance we could have $g_y (\bm z, \bm \epsilon_{\bm y}) = \bm \epsilon_{\bm y}$ and $\bm \epsilon_{\bm y}^{\dagger} := \bm y - \bm \theta^{\dagger} \cdot \bm \Phi (\bm x)$ is one such choice. 
We can therefore define a function $\vect{\epsilon}_{\uvec{y}}^{\dagger} (\vect{z}, \vect{x}, y; \vect{\theta}^{\dagger})$ for each $\vect{\theta}^{\dagger} \in \mathbb{R}^{d_{\bm \Phi}}$ that solves the above equation for any $\vect{x}$, $y$ and $\vect{z}$. 

The full joint distribution that generates $P (\vect{X}, Y, \vect{Z})$ from the structural equations can be factorized as follows:
\begin{equation*}
    P(\vect{X}, Y, \vect{Z}, \vect{\epsilon}_{\uvec{x}}, \vect{\epsilon}_{\uvec{y}}) = P(\vect{\epsilon}_{\uvec{x}}, \vect{\epsilon}_{\uvec{y}} \mid \vect{X}, Y, \vect{Z}) ~P(\vect{X}, Y, \vect{Z}).
\end{equation*}
We can therefore define (for any $\vect{\theta}^{\dagger}$) at least one joint distribution consistent with the structural assumptions, $\vect{\theta}^{\dagger}$, and the observed joint distribution $P (\vect{X}, Y, \vect{Z})$:
\begin{equation*}
    P^{\dagger} (\vect{X}, Y, \vect{Z}, \vect{\epsilon}_{\uvec{x}}, \vect{\epsilon}_{\uvec{y}}) := D^{\dagger} (\vect{\epsilon}_{\uvec{x}} \mid \vect{X}, \vect{Z}) ~\delta (\vect{\epsilon}_{\uvec{y}} - \vect{\epsilon}_{\uvec{y}}^{\dagger} (\vect{X}, Y, \vect{Z}; \vect{\theta})) ~P (\vect{X}, Y, \vect{Z}),
\end{equation*}
where:
\begin{equation*}
    D^{\dagger} (\vect{\epsilon}_{\uvec{x}} \mid \vect{x}, \vect{z}) = \begin{cases}
        \delta (\vect{\epsilon}_{\uvec{x}} - \vect{\epsilon}_{\uvec{x}}^{\dagger} (\vect{x}, \vect{z}) ) & \exists y \in \Omega_{Y}: P(\vect{x}, y, \vect{z}) \neq 0 \\
        0 & \text{otherwise},
    \end{cases}
\end{equation*}
and $\delta$'s are Dirac delta measures over the domains of their arguments. 

Therefore, we cannot reject any $\vect{\theta} \in \mathbb{R}^{d_{\vect{\Phi}}}$ based on the structural equations and the ground truth $P (\vect{X}, Y, \vect{Z})$ alone, since there exists some joint distribution consistent with both.

Assume the validity of a background constraint $\vect{\Gamma}$, so that $\cov (\vect{Z}, g_y) \equiv (\bm \beta_{\bm y}^{\bm *} - \vect{\theta^*} \cdot \bm \beta_{\bm \Phi}^{\bm *} ) \in \vect{\Gamma}$.
We can only use $\vect{\Gamma}$ to exclude $\vect{\theta}$ for which $(\bm \beta_{\bm y}^{\bm *} - \vect{\theta} \cdot \bm \beta_{\bm \Phi}^{\bm *}) \notin \vect{\Gamma}$. 
Therefore, it follows immediately that the feasible map $\mathcal{T}$ for which $\mathcal{T} (\vect{\Gamma}, \bm \beta_{\bm \Phi}^{\bm *}, \bm \beta_{\bm y}^{\bm *}) = \{ \vect{\theta} : (\bm \beta_{\bm y}^{\bm *} - \vect{\theta} \cdot \bm \beta_{\bm \Phi}^{\bm *}) \in \vect{\Gamma} \}$ is complete.

\paragraph{Minimality} We have proven that any and all $\bm \theta \in \mathcal{T} (\bm \Gamma, \bm \beta_{\bm \Phi}, \bm \beta_{\bm y})$ satisfy $(\bm \beta_{\bm y} - \bm \theta \cdot \bm \beta_{\bm \Phi}) \in \bm \Gamma$. 
Since $\bm \Gamma$, $\bm \beta_{\bm \Phi}$ and $\bm \beta_{\bm y}$ can be varied arbitrarily and independently of each other, it is clear the $d_{\bm Z} d_{\bm \Phi} + d_{\bm Z}$ independent, real parameters required to specify $\bm \beta_{\bm \Phi}$ and $\bm \beta_{\bm y}$ are needed to specify the feasible map $\mathcal{T}$. 
This completes the proof of optimality.

\subsection{Proof of \cref{thm:n-point identifiability}}
\label{app: Proof of n point identifiability etc}

We prove a slight generalization of the theorem that accounts for violation of (B$\bm 1^{\bm *}$).
As we will see, the the tight bound depends on a quantity $B$ that equals $0$ when (B$\bm 1^{\bm *}$) is satisfied. 

\textbf{We can account for violation of (B$\bm 1^{\bm *}$) by defining a reduced problem}

We extend the theorem as stated by allowing for (B$1^*$) violation. 
By simple linear algebra, the left and right null spaces and accounting for (B$1^*$) violation can be identified polynomial time. 

If $\bm \beta_{\bm \Phi}$ has a nonempty left null space, then each point $h (\bm \theta) \in h := h[\bm \theta]$ (i.e., the affine space that is the image of the affine map $h(\bm \theta)$) corresponds to a continuum of possible $\bm \theta \in \Theta \subseteq \mathbb{R}^{d_{\bm \Phi}}$, where $\Theta$ is unbounded. 
This means any nonempty feasible set $\mathcal{T}_{L_0}$ will also be unbounded, and depending on the choice of $\bm \Phi$, the ATE may be vacuous. 
We choose to ignore this kind of violation of (B$1^{\bm *}$), which corresponds to $\bm \beta_{\bm \Phi}$ not being full rank. 



We decompose the set of covariance-irrelevant candidate instruments $I = \{ i \in [d_{\vect{Z}}] : (\bm \beta_{\bm \Phi})_i = \bm 0\}$ into the following subsets:
\begin{align*}
    & I_{=0} := \{ i \in [d_{\vect{Z}}] : (\bm \beta_{\bm \Phi})_i = \bm 0 \wedge (\beta_{y})_i = 0 \}, \\
    & I_{\neq 0} := \{ i \in [d_{\vect{Z}}] : (\bm \beta_{\bm \Phi})_i = \bm 0 \wedge (\beta_{y})_i \neq 0 \},
\end{align*}
where the former corresponds to irrelevant candidate instruments that are uncorrelated with the outcome while the latter corresponds to those correlated with the outcome. 
Any $\bm \gamma_{\bm g} \in h := h[\bm \theta]$ will satisfy $(\gamma_{g})_{i \in I_{=0}} = 0$ and $(\gamma_{g})_{i \in I_{\neq 0}} \neq 0$. Therefore, we can count whether these components are always or never $0$ irrespective of $\bm \theta$. 

This motivates the following definitions:
\begin{align*}
    D_{\bm Z} &:= d_{\bm Z} - \lvert I_{=0} \rvert - \lvert I_{\neq 0} \rvert, \\
    B &:= b - \lvert I_{\neq 0} \rvert,
\end{align*}
through which we define the reduced problem of finding $\bm \theta$ for which:
\begin{align*}
    \lVert H(\bm \theta) \rVert_{0} \leq B,
\end{align*}
where:
\begin{align*}
    H (\bm \theta) = \sum_{j \in [d_{\bm Z}] \backslash I} \big( (\beta_y)_i - \bm \theta \cdot (\bm \beta_{\bm \Phi})_i \big) \bm e_i.
\end{align*}
The solutions for $\bm \theta$ are exactly the same solutions as those to $\lVert h(\bm \theta) \rVert_0 \leq b$. 
Notice that $H := H[\bm \theta]$ is a $d_{\bm \Phi}$-dimensional affine subspace of $\mathbb{R}^{D_{\bm Z}}$ (we can ignore $i \in I$). 
Defining $J = [d_{\bm Z}] \backslash I$, we see that none of the basis vectors $\bm e_j$ for $j \in J$ are orthogonal to $H$.

Notice that $D_{\bm Z} = d_{\bm Z}$, $B = b$ and $H(\bm \theta) = h (\bm \theta)$ iff (B$1^*$) holds. 


\newcommand{\scalesim}[1]{\overset{#1}{\scalebox{1.5}[1.0]{$\sim$}}}

\textbf{We begin with $d_{\Phi} = 1$}


If $d_{\Phi} = 1$ then $H$ represents a line embedded in the $D_{\bm Z}$-dimensional Euclidean space. 
Consider a point $\bm \gamma := H(\theta) \in H$ for some $\theta \in \mathbb{R}$. 
This point satisfies $\lVert \bm \gamma \rVert_0 \leq B$ iff there are at least $D_{\bm Z} - B$ many values of $j \in J$ for which $\gamma_j = 0$. 

Since there are no $j \in J$ for which $H$ is orthogonal to $J$, we know there is exactly one solution $\theta_{(j)}$ to the equation $(\beta_y)_j - \theta (\beta_{\Phi})_j = 0$. 
Thus the equivalence relation $\sim$ defined by:
\begin{align*}
    \forall i,j \in J : i \sim j \Longleftrightarrow \theta_{(i)} = \theta_{(j)},
\end{align*}
forms a partition over $[J]$. 
Each equivalence class $\langle j \rangle$ of this partition must have cardinality at least $D_{\bm Z} - B$ for $\theta_{(j)}$ to be in the feasible set.
Since $\lvert J \rvert = D_{\bm Z}$, the number of unique $\theta_{(j)}$ solving the constraint, $n$, is bounded above by: 
\begin{align}
    \label{eqn app:Scalar phi l0 bound}
    \left\lfloor \frac{D_{\bm Z}}{D_{\bm Z} - B} \right\rfloor.
\end{align} 
This proves Cor. \ref{coly:n-point identification scalar phi}, including the special case of point identification when $B > D_{\bm Z} / 2$ first shown by \citet{kang2016}.

\textbf{For $d_{\bm \Phi} > 1$ we enumerate the plausible unique points satisfying the budget constraints}

\paragraph{Consider demanding a single instrument is valid}

In general, $H$ corresponds to a $d_{\bm \Phi}$-dimensional affine space, not orthogonal to $\bm e_j$ for any $j \in J$. 
For some choice of $j \in J$, $H$ will intersect the $(D_{\bm Z} - 1)$-dimensional affine space $\{\bm \gamma \in \mathbb{R}^{D_{\bm Z}} : \gamma_j = 0 \}$ to form $H_{\backslash j} := \{ \bm \gamma' \in H : \gamma'_j = 0 \}$.
$H_{\backslash j}$ is necessarily a $(d_{\bm \Phi} - 1)$-dimensional affine space by the rule for the dimension of the intersection of two affine spaces and the fact that $\bm e_j \not\perp H$ (where $\perp$ means orthogonal).
Notice that $\bm e_j \perp H_{\backslash j}$.

It is possible that $H_{\backslash j} = H_{\backslash k}$ for some $k \in J \backslash j$, which motivates introducing another equivalence relation $\scalesim{d_{\bm \Phi}}$:
\begin{align*}
    \forall j, k \in J : j \, \scalesim{d_{\bm \Phi}} \, k \Longleftrightarrow H_{\backslash j} = H_{\backslash k},
\end{align*}
which induces a partition over $J$. 
The equivalence classes $\langle j \rangle_{d_{\bm \Phi}}$ represent a single $(d_{\bm \Phi} - 1)$-dimensional affine space that consists of points $\bm \gamma$ that simultaneously satisfy the constraints $\gamma_k = 0$ for all $k$ in the equivalence class. 
In other terms:
\begin{align*}
    H_{\backslash j} = H_{\backslash \langle j \rangle_{d_{\bm \Phi}}} := \{ \bm \gamma \in H : \forall k \in \langle j \rangle_{d_{\bm \Phi}} (\gamma_k = 0) \}.
\end{align*}
The number of unique equivalence classes with respect to $\scalesim{d_{\bm \Phi}}$, $n_{d_{\bm \Phi}}$, is anywhere between $1$ and $D_{\bm Z}$.
However, if any of the equivalence classes has cardinality at least $D_{\bm Z} - B$, then there is at least one $(d_{\bm \Phi}-1)$-dimensional solution to the $L_0$-norm constraint.

\paragraph{What if we demand more instruments are valid?}
We could equally ask how many sets of constraints $(\gamma_q)_{q \in Q} = 0$ for $q \in Q \subseteq J$ lead to unique $(d_{\bm \Phi} - 2)$-dimensional affine spaces $H_{\backslash Q} := \{ \bm \gamma \in H : \forall k \in Q (\gamma_k = 0) \}$.
The $Q$ such that $H_{\backslash Q}$ is a $(d_{\bm \Phi} - 2)$-dimensional affine space is the union of at least two equivalence classes $\langle j \rangle_{d_{\bm \Phi}}$ and $\langle k \rangle_{d_{\bm \Phi}} \neq \langle j \rangle_{d_{\bm \Phi}}$. 
Therefore the number of unique $(d_{\bm \Phi} - 2)$-dimensional affine spaces $H_{\backslash Q}$, denoted $n_{d_{\bm \Phi} -1}$, is anywhere between $1$ and ${n_{d_{\bm \Phi}} \choose 2}$.

Likewise, the number of sets $Q$ generating unique $1$-dimensional affine spaces (lines) $H_{\backslash Q}$, denoted $n_1$, is anywhere between $1$ and ${n_{d_{\bm \Phi}} \choose d_{\bm \Phi} - 1}$. 
This implies the upper bound:
\begin{align*}
    n_1 \leq {D_{\bm Z} \choose d_{\bm \Phi} - 1} = \frac{D_{\bm Z}!}{(d_{\bm \Phi} -1)! (D_{\bm Z} - d_{\bm \Phi} + 1)!},
\end{align*}
where this upper bound corresponds to all such $Q \subset J$ having length $d_{\bm \Phi} - 1$, which is the minimum length $Q$ can take.

\paragraph{Bounding the number of unique points for a given line} 
Consider some $Q$ for which $H_{\backslash Q} := \{ \bm \gamma \in H : \forall k \in Q (\gamma_k = 0) \}$ is a line. 
Then define $Q' := \{ i \in J : \forall \bm \gamma \in H_{\backslash Q} (\gamma_k = 0) \}$, which expresses every constraint satisfied along the line.
We ask how many unique points along $H_{\backslash Q}$ can satisfy the $L_0$-norm constraint $\lVert H_{\backslash Q} \rVert_0 \leq b$. 
Along the entire line, there are $\lvert Q' \rvert$ many components for which $\gamma_k = 0$.
For a point to satisfy the constraint, we must have at least $D_{\bm Z} - \lvert Q' \rvert$ many $\gamma_k = 0$ for $k \in J \backslash Q'$.
Therefore, by the same arguments that lead to \cref{eqn app:Scalar phi l0 bound}, the number of unique points solving the constraint along $H_Q$, denoted $p_Q$, is bounded by:
\begin{align*}
    p_{Q} \leq \left\lfloor \frac{D_{\bm Z} - \lvert Q'\rvert}{D_{\bm Z} - \lvert Q' \rvert - B} \right\rfloor.
\end{align*}

\paragraph{Putting the pieces together}
As discussed earlier, the minimum length of $Q'$ is $d_{\bm \Phi} - 1$. 
This value maximizes $p_Q$. 
The maximum number of unique lines $n_1$ also increases as the length $\lvert Q' \rvert$ for each line decreases.
Therefore, the upper bound:
\begin{align*}
    n \leq \frac{D_{\bm Z}!}{(d_{\bm \Phi} -1)! (D_{\bm Z} - d_{\bm \Phi} + 1)!} \left\lfloor \frac{D_{\bm Z} - d_{\bm \Phi} + 1}{D_{\bm Z} - d_{\bm \Phi} + 1 - B} \right\rfloor,
\end{align*}
which corresponds to all unique lines $H_Q$ having $Q' = d_{\bm \Phi} - 1$, is valid and tight. 

\subsection{Proof of \cref{thm:Coverage}}
\label{app:Proof of coverage}
We begin with the assumption that the estimated statistic $\hat{\beta}_{\Phi}$ is exactly equal to the observable ground truth $\beta_{\Phi}^*$.
Furthermore, since elliptical confidence sets require estimating $\cov \left((\hat{\beta}_y)_i, (\hat{\beta}_y)_j \right)$ for pairs $i \neq j$---estimates for these are not openly available with GWAS summary statistics---we construct box constraints using the Bonferroni correction.
Our $(1-\alpha) \times 100\%$ confidence set over $\bm \beta_{\bm y}$, consists of all $\bm \beta_{\bm y}$ for which each component $(\beta_y)_i$ is in the corresponding $(1-\alpha / d_{\bm Z}) \times 100\%$ confidence interval, defined below. 

Calculate the estimator $\bm {\hat \beta}_{\bm y} := \widehat{\cov}(Y, \bm Z)$ using $N$ $\mathrm{i.i.d.}$ samples from $P(\bm X, Y, \bm Z)$.
It is well known from the central limit theorem \citep{Vaart_1998} that if this estimator has finite marginal standard errors $\mathrm{Se} \,(\hat{\beta}_{y})_i = \sqrt{\mathrm{Var} (\hat{\beta}_{y})_i}$,then the following convergence in distribution holds as our choice of $N$ approaches $\infty$:
\begin{align*}
    \frac{\sqrt{N} \left((\hat{\beta}_y)_i - (\beta_y^*)_i\right)}{\mathrm{Se}(\hat{\beta}_y)_i} \overset{d}{\longrightarrow} \mathcal{N} (0, 1),
\end{align*}
where $\bm \beta_{\bm y}^{\bm *}$ denotes the ground truth.
A similar statement can be made about $\bm \beta_{\bm \Phi}$ under the assumption that $\mathrm{Se} \, (\beta_{\Phi})_{ij}$ are finite for all $i \in [d_{\bm \Phi}], j \in [d_{\bm Z}]$.
However, by the NOME assumption, we choose to neglect any variation in $\bm{\hat{\beta}}_{\bm \Phi}$.

We model $\mathrm{Se}(\hat{\beta}_y)_i$ with the plug in estimator $\widehat{\mathrm{Se}}(\hat{\beta}_y)_i$ and construct the required confidence intervals using quantiles of $\mathcal{N} (0, 1)$.
The confidence intervals are symmetric about $(\hat{\beta}_y)_i$ and have width $2 (\delta \beta_y)_i$ and we denote the resulting confidence set $\bm B_{\bm y}^{\alpha}$. 

Since $\mathcal{T} (C = \{ \bm \gamma_{\bm g} \in \bm \Gamma \}, s = \{ \bm \beta_{\bm \Phi}^*, \bm{\hat \beta}_{\bm y} \}) $ is a deterministic functional of $\bm{\hat \beta}_{\bm y}$, we can compute a confidence set over $\mathcal{T}$ with asymptotically valid coverage explicitly: 
\begin{align*}
    \mathcal{T}^{\alpha} \left( c = \mathbb{I} ( \bm \gamma_{\bm g} \in \bm \Gamma ), s = \{ \bm \beta_{\bm \Phi}^*, \bm B_{\bm y}^{\alpha} \} \right) = \left\{ \bm \theta \in \mathbb{R}^{d_{\bm \Phi}} : \exists \bm \beta_{\bm y} \in \bm B_{\bm y}^{\alpha} \big( (\bm \beta_{\bm y} - \bm \theta \cdot \bm \beta_{\bm \Phi}^*) \in \bm \Gamma \big) \right\}.
\end{align*}
Confidence sets over a functional that are constructed in this way appear, for example, in \citet{duarte2023} and \citet{malloy2021optimalconfidenceregionsmultinomial}.

We can rewrite $\mathcal{T}^{\alpha} (c, s)$ defined above as the following set: 
\begin{align*}
    \left\{ \bm \theta \in \mathbb{R}^{d_{\bm \Phi}} : \exists \bm \beta_{\bm y} \in \bm B_{\bm y}^{\alpha}, \Tilde{\bm U} \in \bm \Sigma_{\bm b}^{\mathrm{(max)}} \left( (\bm \beta_{\bm y} - \bm \theta \cdot \bm \beta_{\bm \Phi}^*) \in \Tilde{\bm \Gamma}_{\Tilde{\bm U}} \right) \right\}.
\end{align*}

For any particular $\Tilde{\bm U} \in \bm \Sigma_{\bm b}$, define:
\begin{align*}
    \tau_i^{\Tilde{\bm U}} = \tau_{\ell} \Longleftrightarrow \Tilde{U}_i = \ell.
\end{align*}
Then $\bm \theta \in \mathcal{T}^{\alpha} (c, s)$, iff there exists a $\Tilde{\bm U} \in \bm \Sigma_{\bm b}$ for which: 
\begin{align*}
     \Big| (\hat{\beta}_y)_i \pm (\delta \beta_y)_i - \bm \theta \cdot \bm{\hat \beta}_{\bm X} \Big| \leq \tau_i^{\Tilde{\bm U}} \Longleftrightarrow \Big| (\hat{\beta}_y)_i - \bm \theta \cdot \bm{\hat \beta}_{\bm X} \Big| \leq \tau_i^{\Tilde{\bm U}} + (\delta \beta_y)_i.
\end{align*}
We can therefore define $\bm{\hat \Gamma}_{\alpha} := \bigcup_{\Tilde{\bm U} \in \bm \Sigma_{\bm b}} \bm{\hat \Gamma}_{\Tilde{\bm U}}^{\alpha}$, where: 
\begin{align*}
    \bm{\hat \Gamma}_{\Tilde{\bm U}}^{\alpha} := \left\{ \bm \gamma \in \mathbb{R}^{d_{\bm Z}} : \forall i \in [d_{\bm Z}] \left( \lvert \gamma_i \rvert \leq \tau_i^{\Tilde{\bm U}} + (\delta \beta_y)_i \right) \right\},
\end{align*}
and the confidence set can be rewritten as:
\begin{align*}
    \mathcal{T}^{\alpha} \left( c = \mathbb{I} [ \bm{\gamma}_{\bm g} \in \bm{\Gamma}], s = \{ \bm{\beta}_{\bm \Phi}^*, \bm{\hat{B}}_{\bm y}^{\alpha} \} \right) = \mathcal{T} \left( c' = \mathbb{I} [ \bm \gamma_{\bm g} \in \bm{\hat{\Gamma}}_{\alpha}], s = \{ \bm \beta_{\bm \Phi}^*, \bm{\hat \beta}_{\bm y} \}\right).
\end{align*}
Because $\mathcal{T}^{\alpha}$ includes all $\bm \theta \in \mathcal{T} (c, s = \{ \bm \beta_{\bm \Phi}^*, \bm{\beta}_{\bm y}^*\})$ with probability at least $(1-\alpha)\times 100\%$, we have the following guarantee: 
\begin{align*}
    P\Big(\bm \theta^* \in \mathcal{T} \left(c', s = \{ \bm \beta_{\bm \Phi}^*, \bm{\hat{\beta}_{\bm y}}\}) \right) \Big) \geq 1 - \alpha,
\end{align*}
provided $\bm \gamma_{\bm g}^* \in \bm \Gamma$ (i.e., for all $c \in \mathcal{C}^*$).

\section{ADDITIONAL THEOREMS} \label{app:additional theorems}

\subsection{A Condition for the Optimality of the Feasible Map under Stronger Structural Assumptions} \label{app:conditions for optimality}

Here we show that with stricter structural assumptions, sharper feasible maps can be obtained. 
In particular, we focus on the case in which the model class $\mathcal{M}'$ consists of all SCMs satisfying \cref{eqn:Z,eqn:X,eqn:Y} for some predetermined restricted function classes $\mathcal{F}_{\bm z} \ni f_{\bm z}$, $\mathcal{F}_{\bm x} \ni f_{\bm x}$ and $\mathcal{G}_{y} \ni g_{y}$.

We show that restrictions on $\mathcal{F}_{\bm z}$ and $\mathcal{F}_{\bm x}$ alone do not affect the sharpness of $\mathcal{T}$ and we construct necessary and sufficient conditions for restrictions over $\mathcal{G}_{y}$ to lead to the existence of sharper feasible maps.  

The fact that $\mathcal{G}_{y}$ is the important restriction mirrors results in the 
(valid) IV literature, in which nonparameteric bounds can be put on treatment effects based solely on outcomes $Y$ being categorical \citep{Balke1997} or bounded continuous \citep{Manski1990}.
In fact, the following result shows that if $\Omega_Y$ is any proper subset of the real line (e.g., positive but not absolutely bounded), then a sharper feasible map exists.

\begin{theorem}[Necessary and sufficient condition for optimality under stricter assumptions]
\label{thm:n and c condition for optimality}
    Let $\mathcal{M}'$ be all SCMs consistent with \cref{eqn:Z,eqn:X,eqn:Y} under restricted function classes as described above.
    The feasible map $\mathcal{T}$ described in \cref{thm:Optimal solution map} remains sound against the constraint set $\mathcal{C} := \{ \mathbb{I} (\bm \gamma_{\bm g} \in \bm \Gamma) : \bm \Gamma \in \mathbb{R}^{d_{\bm Z}} \}$ under any such $\mathcal{M}'$.
    Furthermore, the map remains complete if and only if for all $\bm z \in \Omega_{\bm Z}$, the \text{section} of $g_y$ at that $\bm z$ is \textit{onto} the full real line $\mathbb{R}$. 
\end{theorem}

\begin{corollary}(Restricted outcome domain)
    \label{app coly: restricted outcome domain}
    In particular, $\mathcal{T}$ is not complete if $\Omega_Y$ is a bounded subset of $\mathbb{R}$.  
    This includes if $\Omega_{Y}$ is bounded and categorical, e.g., $\Omega = \{ 0, 1\}$.
\end{corollary}

For any $\bm z \in \bm \Omega_{\bm Z}$, the "section" of $g_y$ at $\bm Z = \bm z$ is defined as the function $g_{y \mid \bm z}$ defined over the domain $\bm \Omega_{\bm \epsilon_{\bm y}}$ for which $g_{y \mid \bm z} (\bm \epsilon_{\bm y}) = g_y(\bm z, \bm \epsilon_{\bm y})$.
The necessary and sufficient condition for completeness is that the image of each $g_{y \mid \bm z}$, denoted $g_{y \mid \bm z}[\epsilon_{\bm y}]$, is the full real line $\mathbb{R}$. 
Before proving the theorem, we introduce an intuitive example when this image restriction does not hold.

\subsubsection{An Example in which the Map is Incomplete} \label{app:counterexample codomain}

Suppose $\mathcal{M}'$ requires that at some point $\vect{z} \in \Omega_{\bm Z}$, the section:
\begin{equation*}
    g_{y \mid \bm z} (\epsilon_y) := g_y (\vect{Z} = \vect{z}, \epsilon_y) = \frac{a}{1+e^{-\epsilon_y}},
\end{equation*}
for some constant $a$. 
Intuitively, at this value $\bm z$, the proportion of $\var (Y)$ due to confounding is restricted: there is a value of $\bm z$ for which we know the confounding between $\bm X$ and $Y$ is restricted. 

Then, for any $\vect{x} \in \Omega_{\bm X}$ and $y \in \Omega_{Y}$ for which $P (\vect{X} = \vect{x}, Y = y \mid \vect{Z} = \vect{z}) \neq 0$, we may write:
\begin{equation}
    y = \vect{\theta} \cdot \vect{\Phi (x)} + \frac{a}{1+e^{\epsilon_y}}.
    \label{eqn:app bounding theta with troublesome image}
\end{equation}
Assuming $\vect{\Phi} (\vect{x}) \neq 0$, the above equation imposes a restriction on $\vect{\theta}$.
To see this, consider the case whereby $d_{\Phi} = 1$ and notice that: 
\begin{equation*}
    \frac{1}{1+e^{-\epsilon_y}} \in [0, 1].
\end{equation*}
Then, the scalar $\theta$ is bounded by \cref{eqn:app bounding theta with troublesome image}:
\begin{equation*}
    \theta \in \left[ \frac{y}{\Phi (\vect{x})}, \frac{y}{\Phi (\vect{x})} + \frac{a}{\Phi (\vect{x})} \right].
\end{equation*}
Therefore, with only a single value of $\vect{z}$, a strong restriction on $\theta$ has been imposed through the structural equations and $P(\vect{X}, Y \mid \vect{Z} = \vect{z})$ alone.

\subsubsection{Proof of \cref{thm:n and c condition for optimality}} \label{app:proof of necessary and sufficient conditions} 

This proof mirrors that of the original optimality result \cref{thm:Optimal solution map} in \cref{app:Proof of Optimality}. 
We explicitly describe the effect of the restrictions $f_{\bm z} \in \mathcal{F}_{\bm z}$, $f_{\bm x} \in \mathcal{F}_{\bm x}$ and $g_y \in \mathcal{G}_y$ at each stage. 

Let the structural \cref{eqn:Z,eqn:X,eqn:Y}
hold for some ground truth functions $f_{\bm z} \in \mathcal{F}_{\bm z}$, $f_{\bm x} \in \mathcal{F}_{\bm x}$, $\vect{\Phi}^*$, $g_y \in \mathcal{G}_y$ and causal parameter $\vect{\theta}^* \in \mathbb{R}^{d_{\vect{\Phi}}}$.

\paragraph{Soundness} 
The proof of soundness is not affected by the functional restrictions. 

It follows immediately from \cref{eqn:Y} and the left-linearity of the covariance operator that:
\begin{align*}
    \bm \gamma_{\bm g}^{\bm *} := \cov (g_y(\vect{Z}, \vect{\epsilon_y}), \vect{Z}) & = \cov (Y, \vect{Z}) - \cov (\vect{\theta^*} \cdot \vect{\Phi} (\vect{X}),\vect{Z}) \\
    & = \bm \beta_{\bm y}^* - \bm \theta^* \cdot \bm \beta_{\bm \Phi}^*.
\end{align*}
Therefore, $\bm \theta^* \in \mathcal{T} (\bm \Gamma, \bm \beta_{\bm \Phi}^{\bm *}, \bm \beta_{\bm y}^{\bm *})$ whenever $\bm \gamma_{\bm g}^{\bm *} = (\bm \beta_{\bm y}^* - \bm \theta^* \cdot \bm \beta_{\bm \Phi}^*) \in \bm \Gamma$ as required. 

\paragraph{Completeness} Consider any $\vect{\theta}^{\dagger} \in \mathbb{R}^{d_{\vect{\Phi}}}$. Then the following holds:

(Z) There exists at least one function $f_{\bm z}^{\dagger}$ with the following property. 
Given any $\vect{z} \in \Omega_{\bm Z}$, either: (i) $P (\vect{x}, y, \vect{z}) = 0$ for all $\bm x \in \Omega_{\bm X}$ and $y \in \Omega_{Y}$; or (ii) there exists at least one value $\vect{\epsilon}_{\uvec{z}}^{\dagger}$ that solves the equation:
\begin{align*}
    f_{\bm z}^\dagger (\vect{\epsilon}_{\uvec{z}}^{\dagger}) = \vect{z}.
\end{align*}  
We know this must hold because the ground truth $f_{\bm z}^* \in \mathcal{F}_{\bm z}$ is a valid choice of $f_{\bm z}^\dagger$.  

We can, therefore, define a function $\vect{\epsilon}_{\bm z}^{\dagger} (\vect{z})$ that satisfies the above equation for all $\vect{z}$ for which case (i) is false. 
We have not demanded either $f_{\bm z}^{\dagger}$ or $\bm \epsilon_{\bm z}^{\dagger} (\vect{z})$ to be unique.

(X) Likewise, there exists at least one function $f_{\bm x}^{\dagger}$ with the following property. 
Given any $\vect{x} \in \Omega_{\bm X}, \vect{z} \in \Omega_{\bm Z}$, either: (i) $P (\vect{x}, y, \vect{z}) = 0$ for all $y \in \Omega_{Y}$, or (ii) there exists at least one value $\vect{\epsilon}_{\uvec{x}}^{\dagger}$ that solves the equation:
\begin{align*}
    f_{\bm x}^\dagger (\vect{z}, \vect{\epsilon}_{\uvec{x}}^{\dagger}) = \vect{x},
\end{align*}  
and, again, $f_{\bm x}^* \in \mathcal{F}_{\bm x}$ is a valid choice of $f_{\bm x}^\dagger$.  
We can, therefore, define at least one function $\vect{\epsilon}_{\bm x}^{\dagger} (\vect{z}, \vect{x})$ that satisfies the above equation for all $\vect{x}$ and $\vect{z}$ for which case (i) is false. 

(Y) In the proof for \cref{thm:Optimal solution map}, we showed that for any $\vect{x} \in \Omega_{\bm X}$, $y \in \Omega_{Y}$, $\vect{z} \in \Omega_{\bm Z}$, either (i) $P(\bm x, y, \bm z) = 0$, or (ii) for any $\bm \theta^{\dagger} \in \mathbb{R}^{d_{\bm \Phi}}$ there exists at least one value $\vect{\epsilon}_{\uvec{y}}^{\dagger}$ that solves the equation:
\begin{align*}
    g_y^{\dagger} (\vect{z}, \vect{\epsilon}_{\uvec{y}}^{\dagger}) = y - \vect{\theta}^{\dagger} \cdot \vect{\Phi} (\vect{x}).
\end{align*}
However, for some choices of $\mathcal{G}_y$ this is not true. 
The statement can only be true if, whenever $\bm \Phi (\bm x) \neq \bm 0$, we have $g_{y \mid \bm z}[\epsilon_{\bm y}] = \mathbb{R}$.
Otherwise, there is a choice of $\bm \theta^{\dagger}$ and a value $g \in \mathbb{R} \backslash g_{y \mid \bm z}[\bm \epsilon_{\bm y}]$ such that $y - \bm \theta^{\dagger} \cdot \bm \Phi (\bm x)$.   

We define the set of potentially problematic samples: 
\begin{align}
    \bm \Omega_{\mathfrak{V}} = \{ (\bm x, y, \bm z) \in \bm \Omega_{\bm x} \times \Omega_Y \times \bm \Omega_{\bm Z} : P(\bm X, Y, \bm Z) \neq 0 \wedge \bm \Phi (\bm x) \neq \bm 0 \},
\end{align} 
and then the values that are "missing" in at least one section of $g_y$:
\begin{align}
    \mathfrak{G} = \bigcup_{(\bm x, y, \bm z) \in \Omega_{\mathfrak{V}}} \left\{ \mathfrak{g} \in \mathbb{R}\, \backslash \, g_{y \mid \bm z}[\epsilon_{\bm y}] \right\}.
\end{align}
For each $\mathfrak{g} \in \mathfrak{G}$ there are a set of disallowed values for $\bm \theta$:
\begin{align*}
    \Theta_{\mathfrak{g}} := \left\{ \bm \theta \in \mathbb{R}^{d_{\bm \Phi}} : \exists (\bm x, y, \bm z) \in \Omega_{\mathfrak{V}} \left( \bm \theta \cdot \bm \Phi (\bm x) \right) = y - \mathfrak{g} \right\}, 
\end{align*}
from which we define $\Theta_{\mathfrak{G}} := \bigcup_{\mathfrak{g} \in \mathfrak{G}} \Theta_{\mathfrak{g}}$.

We can therefore construct the sharper feasible set than $\mathcal{T} = \{ \bm \theta : h(\bm \theta) \in \bm \Gamma \}$ by excluding every $\bm \theta$ which is disallowed: 
\begin{align*}
    \mathcal{T}' (c = \mathbb{I} [\bm \gamma_{\bm g} \in \bm \Gamma], s' = P(\bm X, Y, \bm Z)) = \mathcal{T} \, \backslash \, \Theta_{\mathfrak{G}},
\end{align*}
with $\mathcal{T}' = \mathcal{T}$ iff for all $\bm z \in \Omega_{\bm z}$, $g_{y \mid \bm z} [\bm \epsilon_{\bm y}] = \mathbb{R}$ (equivalently, $\mathfrak{G} = \emptyset$).

We know the corresponding feasible map $\mathcal{T}'$ is sound because it only excludes $\bm \theta \in \mathbb{R}^{d_{\bm Z}}$ if either (i) the sound feasible map $\mathcal{T}$ excludes $\bm \theta$, or (ii) there does not exist a $g_y \in \mathcal{G}_y$ (and thus an $m \in \mathcal{M}'$) which is consistent with $P(\bm x, y, \bm z)$. 
We have therefore proven the necessity of the assumption that the image of each section of $g_y$ is the full real line for sharpness of $\mathcal{T}$. 

Sufficiency is simple to prove by following step (Y) in the proof of \cref{thm:Optimal solution map} in \cref{app:Proof of Optimality} to construct a full joint distribution consistent with each $P(\bm X, Y, \bm Z)$ and the SCM:
\begin{equation*}
    P^{\dagger} (\vect{X}, Y, \vect{Z}, \vect{\epsilon}_{\uvec{x}}, \vect{\epsilon}_{\uvec{y}}) := D^{\dagger} (\vect{\epsilon}_{\uvec{x}} \mid \vect{X}, \vect{Z}) ~\delta (\vect{\epsilon}_{\uvec{y}} - \vect{\epsilon}_{\uvec{y}}^{\dagger} (\vect{X}, Y, \vect{Z}; \vect{\theta})) ~P (\vect{X}, Y, \vect{Z}).
\end{equation*}
For definitions of the $\delta$'s and $D$, visit this part of the appendix. 

This concludes the proof of \cref{thm:n and c condition for optimality}.

\paragraph{Minimality} We note that $\mathcal{T}$ is not necessarily minimal for general $\mathcal{F}_{\bm z}$, $\mathcal{F}_{\bm x}$ and $\mathcal{G}_{y}$. 
If $\mathcal{T}$ is sharp, it may not be minimal because underlying symmetries in these function classes may constrain which $\bm \beta_{\bm y}$, $\bm \beta_{\bm \Phi}$ can arise, which allow one to describe the solution set $\{ \bm \theta : (\bm \beta_{\bm y}^{\bm *} - \bm \theta \cdot \bm \beta_{\bm \Phi}^{\bm *}) \in \bm \Gamma \}$ with fewer parameters. 
On the other hand, $\bm \beta_{\bm y}$, $\bm \beta_{\bm \Phi}$ may be insufficient for specifying $\mathcal{T}'$ when $g_{y \mid \bm z} [\bm \epsilon_{\bm y}] \subset \mathbb{R}$. 

\subsection{Polytime Testability and Necessary Condition for Unidentifiability} 
\label{app:Unidentifiability}

We say the ground truth causal effect $\bm \theta^{\bm *}$ is \textit{unidentifiable} when the feasible set $\mathcal{T} (\bm \Gamma, \bm \beta_{\bm \Phi}^{\bm *}, \bm \beta_{\bm y}^{\bm *}) = \mathbb{R}^{d_{\bm \Phi}}$. 
In plain English, this means nothing can be learned about $\bm \theta^{\bm *}$ from the data under the current assumptions. 

In this section we show that budget constraints enable an efficient test for unidentifiable causal effects even when $d_{\bm \Phi} > 1$.
We also show unidentifiability can only occur under budget constraints if assumption (B$1^*$) is violated (recall the definition from \cref{sec:IVassumptions}). 

\begin{theorem}[\formatDefName{Unidentifiability}] \label{thm:Unidentifiability}
    Assume Eqs. \ref{eqn:Z}, \ref{eqn:X}, \ref{eqn:Y} and budget constraints according to some $\bm \Gamma(\bm \tau, \bm b)$.
    Then unidentifiability of $\bm \theta^{\bm *}$ can be decided in $\mathcal O(K d_{\bm Z} d_{\bm \Phi})$ time. 
    Moreover, unidentifiability never occurs under (B$1^*$).
\end{theorem}


\subsubsection{Proof of \cref{thm:Unidentifiability}}

We prove necessary and sufficient conditions for $\mathcal{T} (\vect{\Gamma}(\vect{\tau}, \vect{b}), \bm \beta_{\bm \Phi}, \bm \beta_{\bm y}) = \mathbb{R}^{d_{\vect{\Phi}}}$.

Defining the function $\bm V : \mathbb{R}^{d_{\bm Z}} \mapsto [K+1]^{d_{\bm Z}}$ through its components:
\begin{align*}
    V_i (\bm \gamma_{\bm g}) = \begin{cases}
        1 & \lvert ({\gamma_g})_i \rvert \leq \tau_1 \\
        l \in \{ 2, 3, \dots, K \} & \tau_{l-1} \leq \lvert ({\gamma_g})_i \rvert \leq \tau_l \\
        K+1 & \lvert ({\gamma_g})_i \rvert > \tau_K,
    \end{cases}
\end{align*}
we have $\bm V (\bm \gamma_{\bm g}^*) = \bm U^*$ is the ground truth for the latent variable $\bm U$. 
We can represent $\bm V$ by its one-hot encoding:
\begin{align*}
    V_{il} (\bm \gamma_{\bm g}) = \begin{cases}
        1 & V_i (\bm \gamma_{\bm g}) \leq l \\
        0 & \text{otherwise}.
    \end{cases}
\end{align*}
Then budget background search space is explicitly written as:
\begin{equation*}
    \vect{\Gamma} (\vect{\tau}, \vect{b}) := \left\{ \bm \gamma_{\bm g} \in \mathbb{R}^{d_{\vect{Z}}} : \forall l \in [K] \left( \sum_{i=1}^{d_{\bm Z}} V_{il} (\bm \gamma_{\bm g}) \geq b_l \right) \right\}.
\end{equation*}
Unidentifiability occurs iff $h(\bm \theta) = \bm \beta_{\bm y} - \bm \theta \cdot \bm \beta_{\bm \Phi}$ is contained within $\vect{\Gamma} (\vect{\tau}, \vect{b})$ for any $\bm \theta \in \mathbb{R}^{d_{\bm \Phi}}$.
Equivalently, unidentifiability occurs iff the affine space $h := h[\bm \theta]$ formed by the image of $h(\bm \theta)$ is contained within $\vect{\Gamma} (\vect{\tau}, \vect{b})$.

Consider the standard orthonormal basis $\{ \vect{e}_i \}_{i=1}^{d_{\vect{Z}}}$ for which $\bm \gamma_{\bm g} = \sum_{i =1}^{d_{\bm Z}} (\gamma_{g})_i \vect{e}_i$. 
For any direction $\bm e_{i}$ we have that either (i) $(\beta_y)_i - (\bm \theta \cdot \bm \beta_{ \bm \Phi})_i = (\beta_y)_i$ is constant and $h$ is orthogonal to $\bm e_{i}$, or (ii) $(\beta_y)_i - (\bm \theta \cdot \bm \beta_{ \bm \Phi})_i$ is unbounded. 
If (i) holds, it must be the case that $(\bm \beta_{\bm \Phi})_i = \bm 0$ and thus (B$1^*$) is violated, and we say candidate instrument $Z_i$ is covariance-irrelevant to $\bm \Phi (\bm X)$. 

We can define the index set for all covariance-irrelevant candidate instruments:
\begin{equation*}
    I := \{ i \in [d_{\vect{\Phi}}] : (\bm \beta_{\bm \Phi})_i = \bm 0 \}.
\end{equation*}
Then, $h$ is unbounded in all directions $[d_{\bm Z}] \backslash I$ but fixed at a single value for each direction in $I$. 
Since $h$ is affine it is also unbounded in the direction $\sum_{i\in [d_{\bm Z}] \backslash I} \bm e_{i}$.
Moreover, since points in $h$ are parameterized by $\bm \theta$, there exists values of $\bm \theta$ for which $V_i \big(h(\bm \theta)\big) = K+1$ for all $i \in [d_{\bm Z}] \backslash I$.

At such a value for $\bm \theta$, we have $\bm V \big(h(\bm \theta)\big) = \bm V (\bm P)$, where $\bm P \in \mathbb{R}^{d_{\bm Z}}$ is defined by:
\begin{equation*}
    P_i = \begin{cases}
        (\beta_y)_i & i \in I \\
        \tau_{K} + 1 & \text{otherwise}.
    \end{cases}
\end{equation*}
Therefore, the necessary and sufficient condition for unidentifiability is: 
\begin{equation*}
    \sum_{i =1}^{d_{\bm Z}} V_{il} (\bm P) \geq b_l,
\end{equation*}
where $I$ and $\bm P$ can be computed and the condition evaluated by straightforward linear algebra in $\mathcal{O} (K d_{\bm Z} d_{\bm \Phi})$ time. 

\section{ALGORITHMS} \label{app:Algorithms}

\begin{algorithm}
    \caption{$\texttt{budgetIV}$ general case.}
    \label{alg:general budget constraints}
    \begin{algorithmic}[1]
    \For{$\Tilde{\bm U} \in \Sigma_{\bm b}^{\mathrm{(max)}}$} \Comment{Iterate through search space (combinatorial)}
        \If{$h(\bm \theta) \in \Tilde{\bm \Gamma}_{\Tilde{\bm U}}$ for some $\bm \theta$} \Comment{Linear CSP with convex constraints}
            \State $\check{\bm U} (\Tilde{\bm U} ) \gets \bm 0$
            \For{$\vect{x} \in \Omega_{\bm X}$} \Comment{Grid evaluation (for $d_{\bm X} \ll d_{\bm \Phi}$)}
                \State Calculate $\mathrm{ATE}^{+/-}_{\Tilde{\bm U}} (\bm x; \bm x_{\bm 0})$ \Comment{Linear program with convex constraints} 
                \State \hphantom{Calculate} $\bm \theta^{+/-}_{\Tilde{\bm U}} (\bm x; \bm x_{\bm 0})$ \Comment{Arguments for above LP} 
                \For{$i \in [d_{\bm Z}]$}
                    \State $\check{U}_i (\Tilde{\bm U}) \gets \max \{ \check{U}_i (\Tilde{\bm U}), U_i (\bm \theta^+), U_i (\bm \theta^-)\}$
                \EndFor
            \EndFor
        \EndIf
    \EndFor
    \State \Return $\left\{ \left( \mathrm{ATE}_{\bm \Tilde{\bm U}}^{+/-}, \check{\bm U} (\Tilde{\bm U}) \right) : \Tilde{\bm U} \in \Sigma_{\bm b}^{\mathrm{(max)}} \right\}$
    \end{algorithmic}
\end{algorithm}

\begin{algorithm}
    \caption{Polytime $\texttt{budgetIV}$ with $d_{\Phi} = 1$.} \label{alg:lines in budget}
\begin{algorithmic}[1]
    \State Let $c' (\theta)$ be the indicator function for $h(\theta) \in \bm \Gamma (\bm \tau, \bm b)$ and $\bm U (\theta)$ be the unique value of $\bm U$ for which $h(\theta) \in \bm \Gamma_{\bm U} (\bm \tau, \bm b)$ provided $c' (\theta)$. 
    \For{$(i,j) \in [d_{\uvec{Z}}] \times [K]$}
        \If{$(\beta_{\Phi})_i = 0 \land (\beta_{y})_i \leq \tau_j$}
            \State $b_j \gets b_j  - 1$
        \ElsIf{$(\beta_{\Phi})_i \neq 0$}
            \State $\theta_{ij}^{\pm} \gets \frac{(\beta_{y})_i}{(\beta_{\Phi})_i} \mp \frac{\tau_j}{(\beta_{\Phi})_i}$
        \EndIf
    \EndFor
    \If{$\forall i \in [K] (b_i \leq 0)$}
        \State \Return \emph{unidentifiable}
    \EndIf
    \State $\bm \Theta \gets$ sort($\theta_{ij}^{\pm}$)
    \State $\bm \Theta' \gets \left(\Theta_1 - 1, \frac{\Theta_2 + \Theta_1}{2}, \frac{\Theta_3 + \Theta_2}{2}, \ldots, \frac{\Theta_{\lvert \bm \Theta \rvert} + \Theta_{\lvert \bm \Theta \rvert - 1}}{2}, \Theta_{\lvert \bm \Theta \rvert} + 1 \right)$
    \State $\mathrm{intervals} \gets \left\{ \left\langle [\Theta_i, \Theta_{i+1} ], \bm U (\Theta'_i) \right\rangle : c'(\Theta_i') \right\}$ 
    \State $\mathrm{points} \gets \left\{ \left\langle \Theta_i, \bm U (\Theta_i) \right\rangle  : c'(\Theta_i) \wedge \neg c'(\Theta'_i) \neg c'(\Theta'_{i+1}) \right\}$
    \If{$\mathrm{intervals} \cup \mathrm{points} == \emptyset$}
        \State \Return \emph{infeasible}
    \Else
        \State \Return $\mathrm{intervals} \cup \mathrm{points}$
    \EndIf
\end{algorithmic}
\end{algorithm}


Alg. \ref{alg:general budget constraints} depicts a  combinatorial search algorithm that finds the exact feasible set of $\bm \theta$ and the resulting feasible set of $\mathrm{ATE} (\bm x; \bm x_{\bm 0})$ subject to budget background constraints. 
The ATE is recovered using a grid search and program over the linear weights $\bm \theta$ with plug-in values of $\bm x \in \Omega_{\bm X}$. 

Alg. \ref{alg:lines in budget} depicts a polynomial time algorithm for the case of $d_{\Phi} = 1$. 
This method utilises the single solution to the equation $A_i - B_i \theta = \tau$ for $B_i = 0$, and our ability to order such solutions along $\mathbb{R}$.





\section{EXPERIMENTS}
\label{app:Experiments}

\subsection{Linear Simulation Study}

\cref{fig:linear_exp} is the result of a simple experiment to understand the differences between the nonconvex budget background constraints and any convex relaxation thereof. 
We consider violation of (A2) in a linear model where association between $X$ and $\bm Z$ is wholly due to unobserved confounding:
\begin{align*}
    \bm Z & := \bm \epsilon_{\bm z}, \\
    X & := \epsilon_x, \\
    Y & := \theta^* X + \epsilon_y.
\end{align*}
We consider $d_X = 1$ and $d_{\bm Z} = 2$, take the exogenous variables to have a joint distribution $(\bm \epsilon_{\bm z}, \epsilon_x, \epsilon_y) \sim \mathcal{N} (\bm 0, \bm \Sigma_{\bm \epsilon \bm \epsilon})$, where the covariance is given by the terms:
\begin{align*}
    \vect{\Sigma}_{\bm \epsilon \bm \epsilon} := \begin{pmatrix} 
        \eta_{Z_1}^2 & \rho_{Z_1 Z_2} \eta_{Z_1} \eta_{Z_2} & \rho_{Z_1 \epsilon_x} \eta_{Z_1} \eta_{\epsilon_x} & \rho_{Z_1 \epsilon_y} \eta_{Z_1} \eta_{\epsilon_y} \\
         & \eta_{Z_2}^2 & \rho_{Z_2 \epsilon_x} \eta_{Z_2} \eta_{\epsilon_x} & \rho_{Z_2 \epsilon_y} \eta_{Z_2} \eta_{\epsilon_y} \\
         &  & \eta_{\epsilon_x}^2 & \rho_{\epsilon_x \epsilon_y} \eta_{\epsilon_x} \eta_{\epsilon_y} \\
         &  &  & \eta_{\epsilon_y}^2 \\
    \end{pmatrix}.
\end{align*}
We fix the following throughout the study: 
\begin{align*}
    \theta^* &:= 1, \\
    \bm \beta_{\bm x}^* &:= \cov (X, \bm Z) = ( 2, -4 ), \\
    \bm \gamma_{\bm g}^* &:= \cov (\epsilon_y, \bm Z) = ( -2, 0.4 ),
\end{align*}
which, in turn, imply $\bm \beta_{\bm y}^* = (0, 4.4 )$.

\subsubsection{Sweep Through Increasingly Uncertain Background Constraints}

We consider three kinds of background constraints: (a) budget constraints, (b) an $L_2$-norm relaxation (a similar setting to \citet{watson2024bounding}), and (c) an $L_1$-norm relaxation (which is motivated by the relaxation in \citet{kang2016}, under the setting where point identification is not guaranteed): 
\begin{itemize}[noitemsep, leftmargin=3cm]
    \item[(a)] $\bm \gamma_{\bm g} \in \vect{\Gamma} (\vect{\tau} = (\tau, 0.6), \vect{b} = (1,2))$,
    \item[(b)] $\lVert \bm \gamma_{\bm g} \rVert_2 \leq \tau$,
    \item[(c)] $\lVert \bm \gamma_{\bm g} \rVert_1 \leq \tau$.
\end{itemize}
We adjust $\tau$ from $0$ to $10$ linearly across a grid of $101$ simulation studies. 

\subsubsection{Randomized Parameters for the Covariance Matrix}

We select random values for $\bm \Sigma_{\bm \epsilon \bm \epsilon}$ between simulations. 
The data generating process is constrained by two requirements: (i) the matrix $\vect{\Sigma}_{\bm \epsilon \bm \epsilon}$ must be positive definite, and (ii) the fixed values of $\vect{g^*}$ and $\vect{\beta_x^*}$, which demand: 
\begin{align}
    & (\gamma^*_g)_1 = \rho_{Z_1 \epsilon_y} \eta_{Z_1} \eta_{\epsilon_y}, \label{eqn app:rho z_1 e_y} \\
    & (\gamma^*_g)_2 = \rho_{Z_2 \epsilon_y} \eta_{Z_2} \eta_{\epsilon_y}, \label{eqn app:rho z_2 e_y} \\
    & (\beta_x^*)_1 = \rho_{Z_1 \epsilon_x} \eta_{Z_1} \eta_{\epsilon_x}, \label{eqn app:rho z_1 e_x} \\
    & (\beta_x^*)_2 = \rho_{Z_2 \epsilon_x} \eta_{Z_2} \eta_{\epsilon_x}. \label{eqn app:rho z_2 e_x}
\end{align}

We choose to generate $\bm \Sigma_{\bm \epsilon \bm \epsilon}$ by a rejection method.  
We sample the marginal variances according to: 
\begin{align*}
    \eta_{Z_1}, \eta_{Z_2}, \eta_{\epsilon_x}, \eta_{\epsilon_{y}} \overset{\mathrm{i.i.d}}{\sim} \mathrm{Exp} (1),
\end{align*}
and calculate the would-be $\rho$'s from \cref{eqn app:rho z_1 e_y,eqn app:rho z_2 e_y,eqn app:rho z_1 e_x,eqn app:rho z_2 e_x}.
This enforces condition (ii). 
We then test whether the resultant $\bm \Sigma_{\bm \epsilon \bm \epsilon}$ is positive definite. 
If it is, (i) is satisfied and we perform the experiment; otherwise we resample the $\eta$'s from $\mathrm{Exp} (1)$.
\cref{fig:linear_exp} shows the resulting feasible sets with plug-in $\hat{\beta}_{\Phi}$ and $\hat{\beta}_{y}$, calculated from a dataset of $N = 10000$ samples.  

The feasible sets over $\theta$ in \cref{fig:linear_exp} were found using the efficient $\texttt{budgetIV}$ algorithm for $d_{\Phi} = 1$, an implementation of which is provided in the supplement. 

\subsection{Nonlinear Simulation Study} 
\label{app:Nonlinear simulation study}

In this experiment, we show another advantage of the budget background constraint approach over convex relaxations.
Different values of the decision variables $\Tilde{\bm U}$ lead to different functions $\mathrm{ATE} (\bm x; \bm x_{\bm 0})$. 


\subsubsection{Endogenous Equations} \label{app:nonlinear structural parameters}

The candidate instruments are in the domain $\Omega_{\bm Z} = \{ 0, 1\}^{d_{\vect{Z}}}$, where $d_{\bm Z} = 6$; the exposure is a scalar in $\Omega_{\bm X} = [0, 1]$; and the outcome is a scalar in the full real line $\Omega_{Y} = \mathbb{R}$.

The ground truth structural equations are: 
\begin{align*}
     \bm Z &:=  \bm \epsilon_{\bm z} \\
     X &:= f_{x} (\bm Z, \bm \epsilon_{\bm x}), \\
     Y &:= \bm \theta^* \cdot \bm \Phi^* (\bm X) + g_y (\bm Z, \bm \epsilon_{\bm y}),
\end{align*} 
where the functions take the form: 
\begin{align*}
   f_x (\bm Z, \epsilon_x) &= \expit (\epsilon_x - \bm m \cdot \bm Z), \\
    g_y (\bm Z, \epsilon_{y}) &= \lambda^{\text{(A$3$)}} \vect{Z} \cdot \vect{\Lambda} \cdot \vect{Z} + \lambda^{\text{(A$2$)}} \epsilon_y,\\
    \Phi^* (X) &= s_{\Phi} \left( (X - 1/4)^2 - 1/16 \right),
\end{align*}
and $\bm \theta^* = 1$.

The real vector $\bm m \in \mathbb{R}^{d_{\bm Z}}$ and the binary matrix $\bm \Lambda \in \{ 0,1\}^{d_{\bm Z} \times d_{\bm Z}}$ are sampled randomly for each experiment. 
The components of $\bm m$ are $\mathrm{i.i.d}$ normal and the components of $\bm \Lambda$ are independent Bernoulli trials according to: 
\begin{align*}
    & m_1, \ldots, m_{d_{\bm Z}} \overset{\mathrm{i.i.d}}{\sim} \mathcal{N} (1,4), \\
    & \Lambda_{ij} \sim \begin{cases}
        \mathcal{N} (0.9,0.9) & i=j>3, \\
        \mathcal{N} (0.3,0.3) & i \neq j \text{ and } i, j >3, \\
        0 & \text{otherwise}.
    \end{cases}
\end{align*}
Notice that candidate instruments $Z_1, Z_2, Z_3$ all do not violate (A$3$). 

The remaining parameters $s_{\Phi}, \lambda^{\text{(A$2$)}}, \lambda^{\text{(A$3$)}} \geq 0$ are restricted to be positive but are otherwise varied accross simulation settings (see \cref{app:nonlinear varying}).

\subsubsection{Generating the Exogenous Variables} \label{app:nonlinear exogenous distribution}

We generate $(\bm Z, \epsilon_x, \epsilon_y)$ to have a block diagonal covariance structure where so that $Z_1$, $Z_2$ and $Z_3$ are valid instruments.

This is achieved by a Markov chain approach over binary $Z$ with: 
\begin{align*}
    Z_1 & \sim \mathrm{Ber} (0.05), \\
    Z_4 & \sim \mathrm{Ber} (0.05), \\
    Z_{j+1} \mid Z_{j} = 1 & \sim \mathrm{Ber} (0.9), \\
    Z_{j+1} \mid Z_{j} = 0 & \sim \mathrm{Ber} (0.05),
\end{align*}
where $j \in \{ 2, 3, 5, 6\}$. 
Then the noise residuals are given by: 
\begin{align*}
    \epsilon_x & = \bm \gamma_{\bm \epsilon_{\bm x}} \cdot \bm Z + \epsilon',\\
    \epsilon_y & = \bm \gamma_{\bm \epsilon_{\bm y}} \cdot \bm Z + \epsilon',
\end{align*}
where:
\begin{align*}
    (\gamma_{\epsilon_{x}})_1, \ldots, (\gamma_{\epsilon_{x}})_{6} & \overset{\mathrm{i.i.d}}{\sim} \mathcal{N} (0, 1), \\
    (\gamma_{\epsilon_{y}})_1, \ldots, (\gamma_{\epsilon_{y}})_{3} & = 0, \\
    (\gamma_{\epsilon_{x}})_4, \ldots, (\gamma_{\epsilon_{x}})_{6} & \overset{\mathrm{i.i.d}}{\sim} \mathcal{N} (0, 1), \\
    \epsilon' & \sim \mathcal{N} (0, 1).
\end{align*}

\begin{figure}
    \centering 
    \begin{subfigure}[t]{0.3\textwidth}
    \centering
    \includegraphics[width=\textwidth]{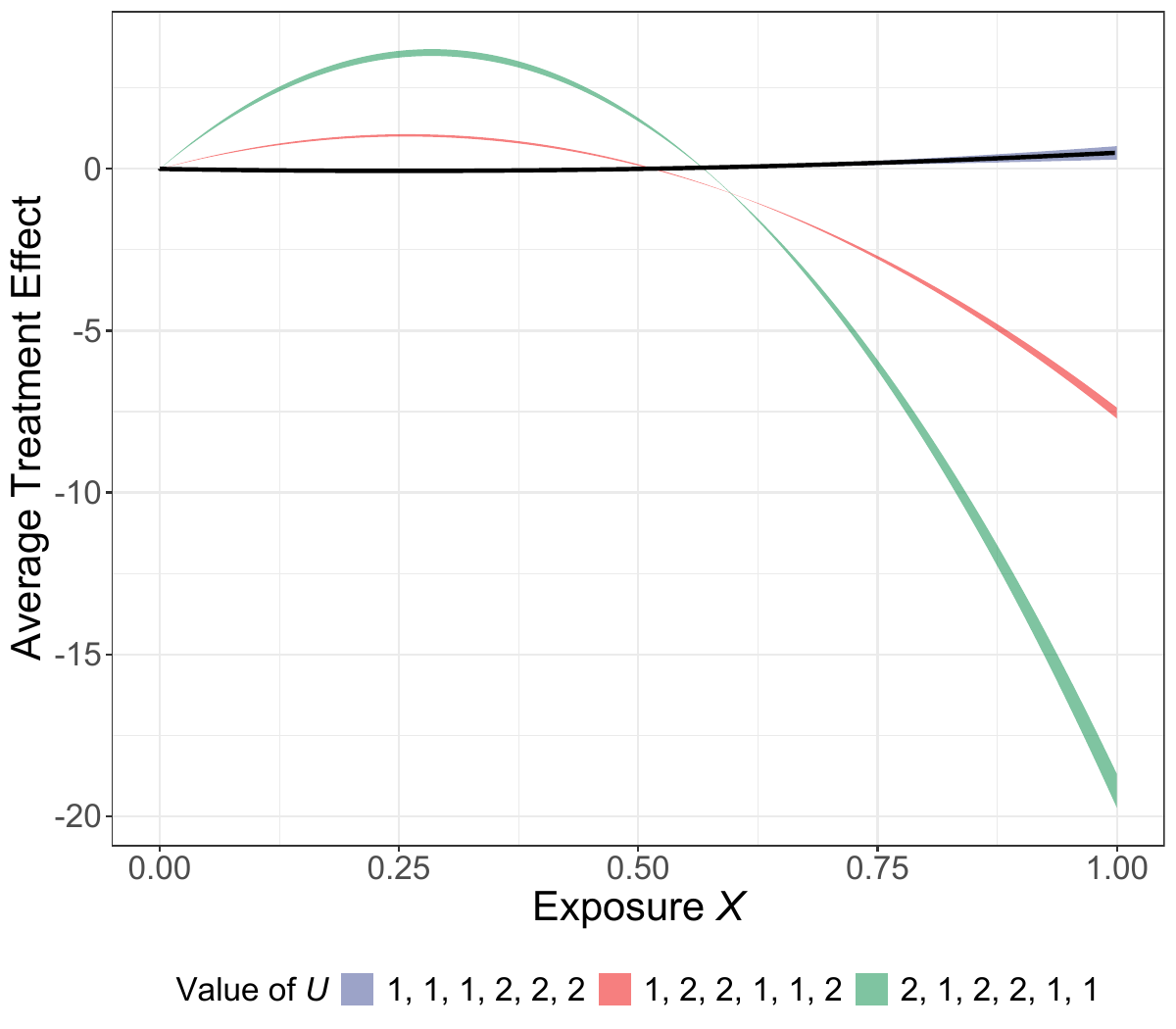}
    \subcaption{$R = 1/2$, $\mathrm{SNR}_y = 1/10$}
    \end{subfigure}
    \begin{subfigure}[t]{0.3\textwidth}
    \centering
    \includegraphics[width=\textwidth]{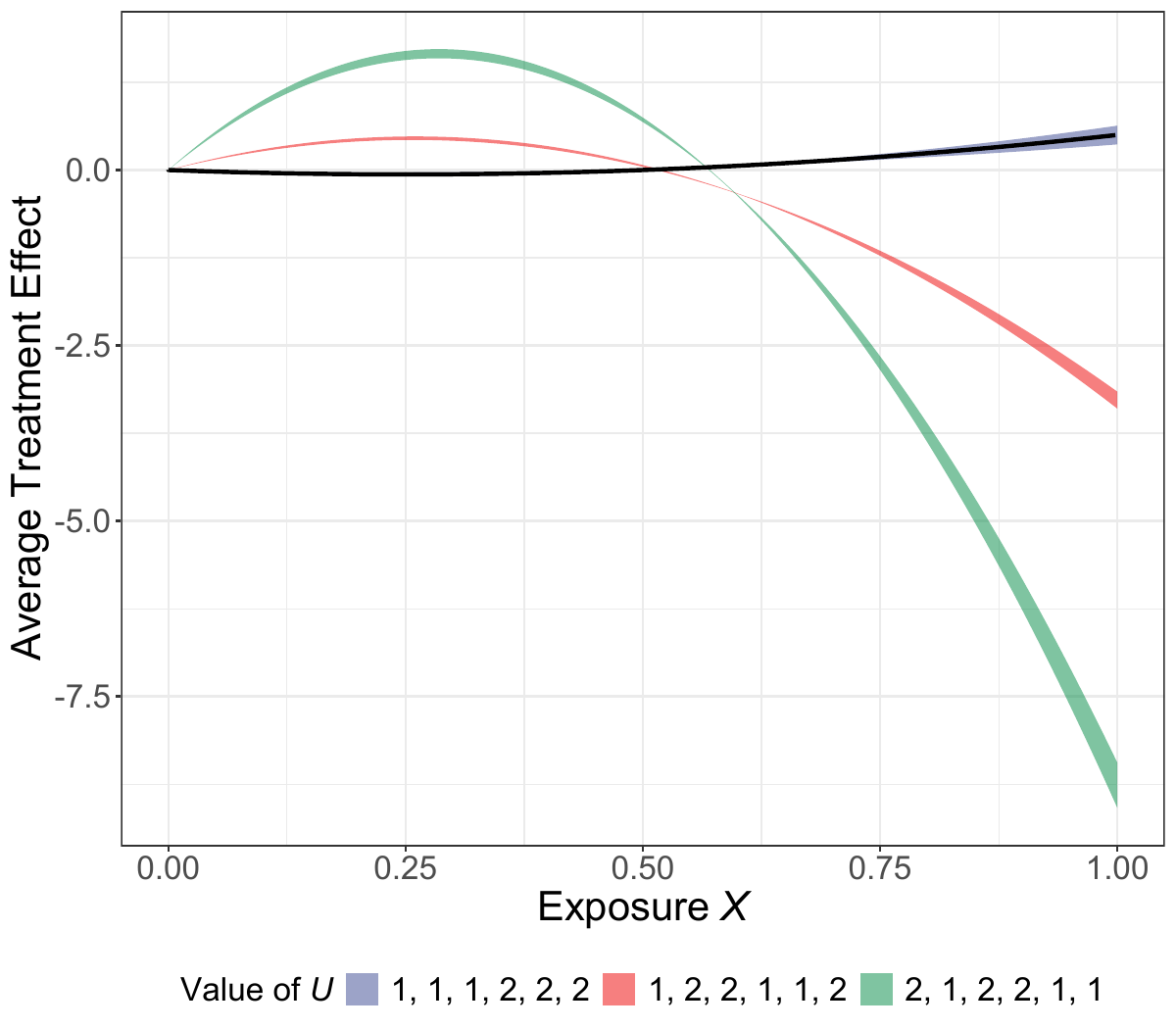}
    \subcaption{$R = 1/2$, $\mathrm{SNR}_y = 1/5$}
    \end{subfigure}
    \begin{subfigure}[t]{0.3\textwidth}
    \centering
    \includegraphics[width=\textwidth]{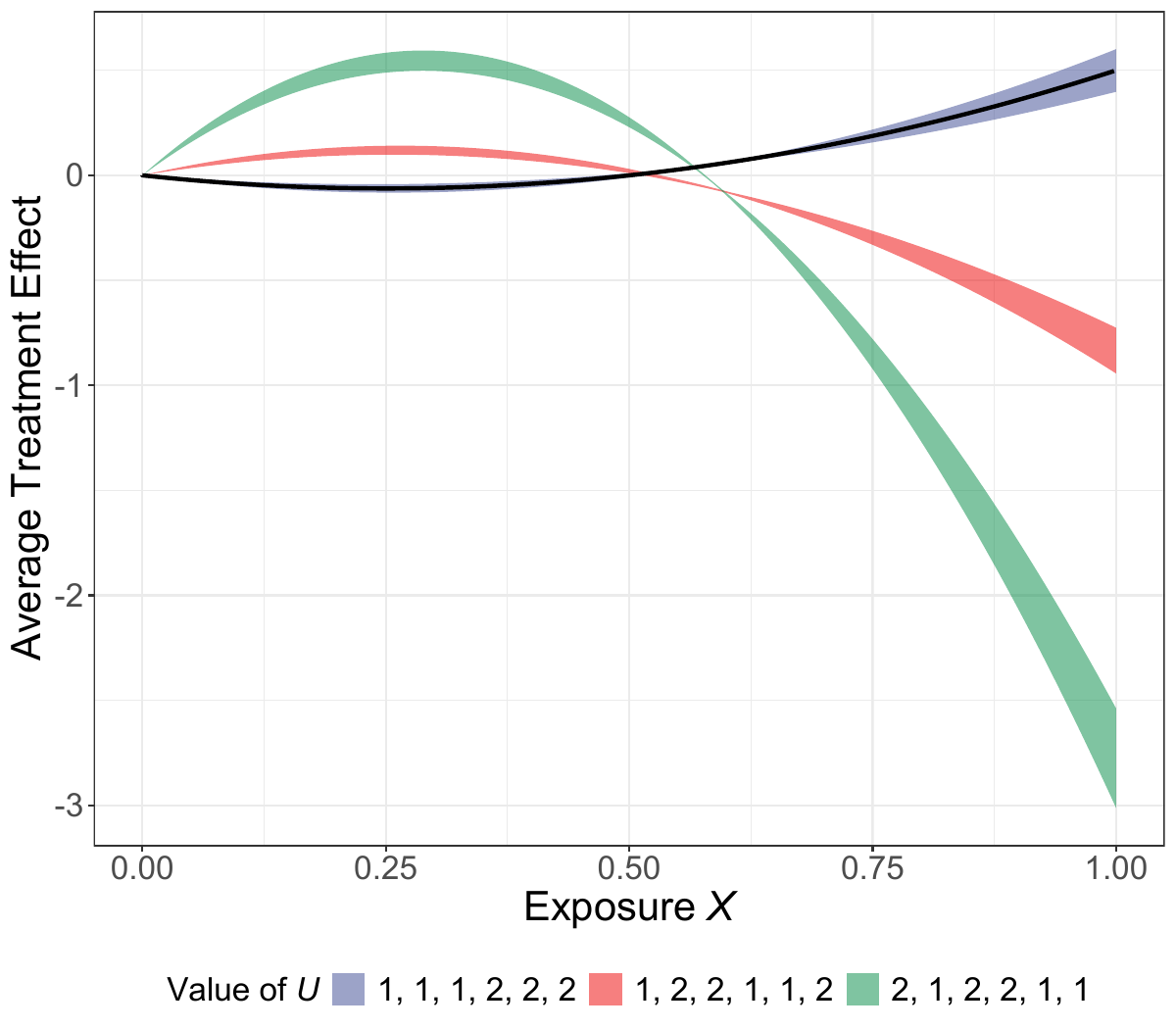}
    \subcaption{$R = 1/2$, $\mathrm{SNR}_y = 1$}
    \end{subfigure}
    \begin{subfigure}[t]{0.3\textwidth}
    \centering
    \includegraphics[width=\textwidth]{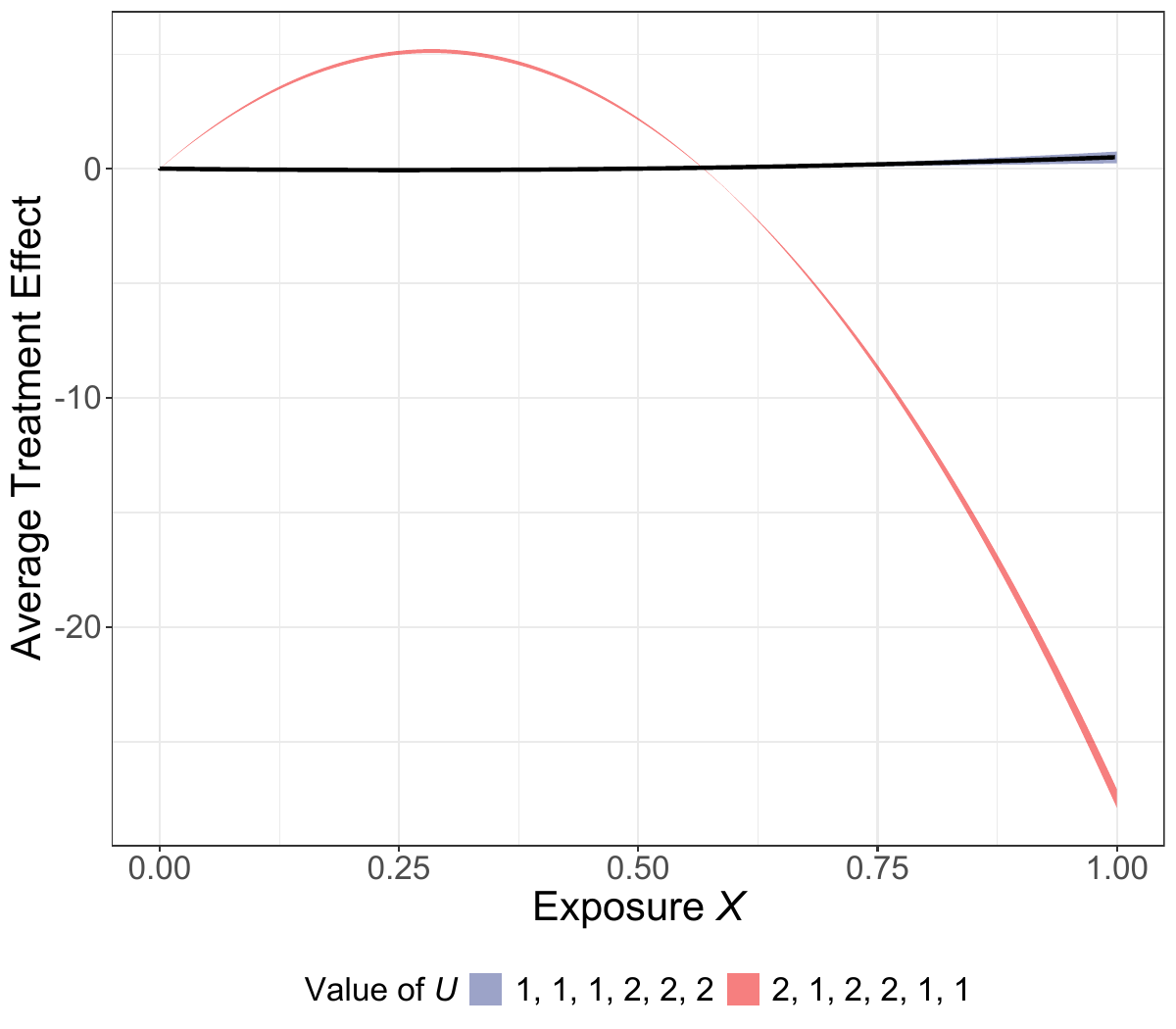}
    \subcaption{$R = 1$, $\mathrm{SNR}_y = 1/10$}
    \end{subfigure}
    \begin{subfigure}[t]{0.3\textwidth}
    \centering
    \includegraphics[width=\textwidth]{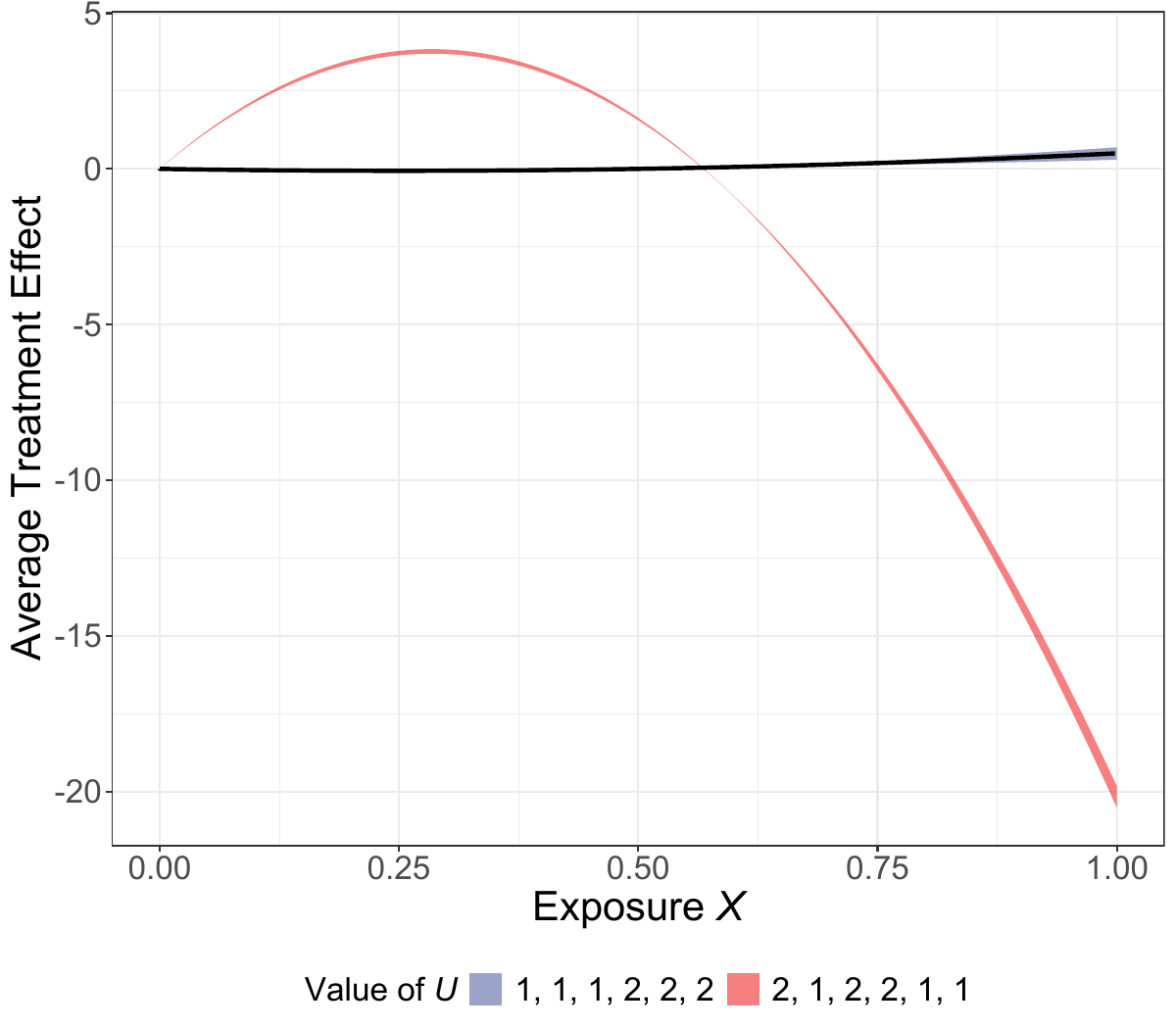}
    \subcaption{$R = 1$, $\mathrm{SNR}_y = 1/5$}
    \end{subfigure}
    \begin{subfigure}[t]{0.3\textwidth}
    \centering
    \includegraphics[width=\textwidth]{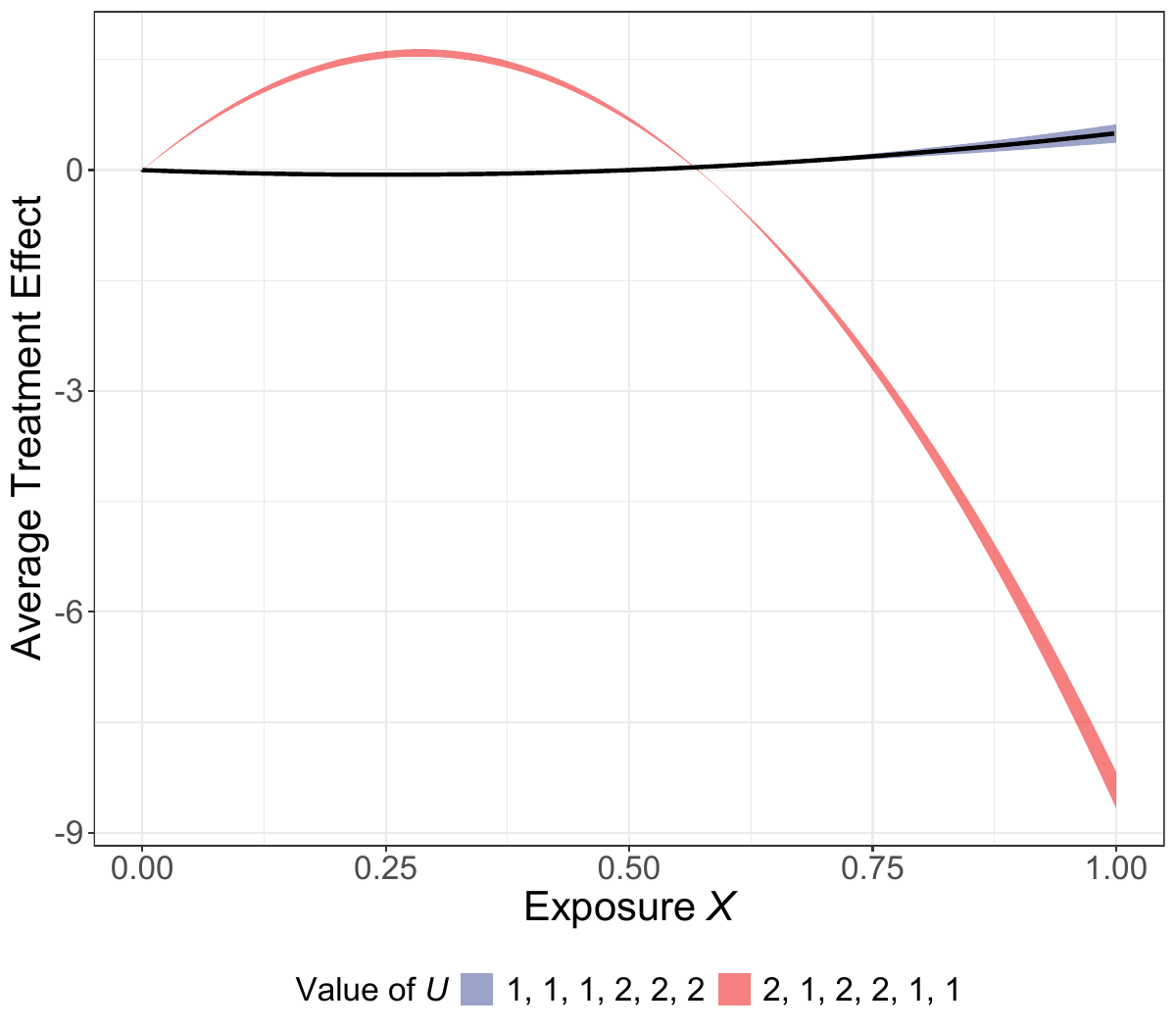}
    \subcaption{$R = 1$, $\mathrm{SNR}_y = 1$}
    \end{subfigure}
    \begin{subfigure}[t]{0.3\textwidth}
    \centering
    \includegraphics[width=\textwidth]{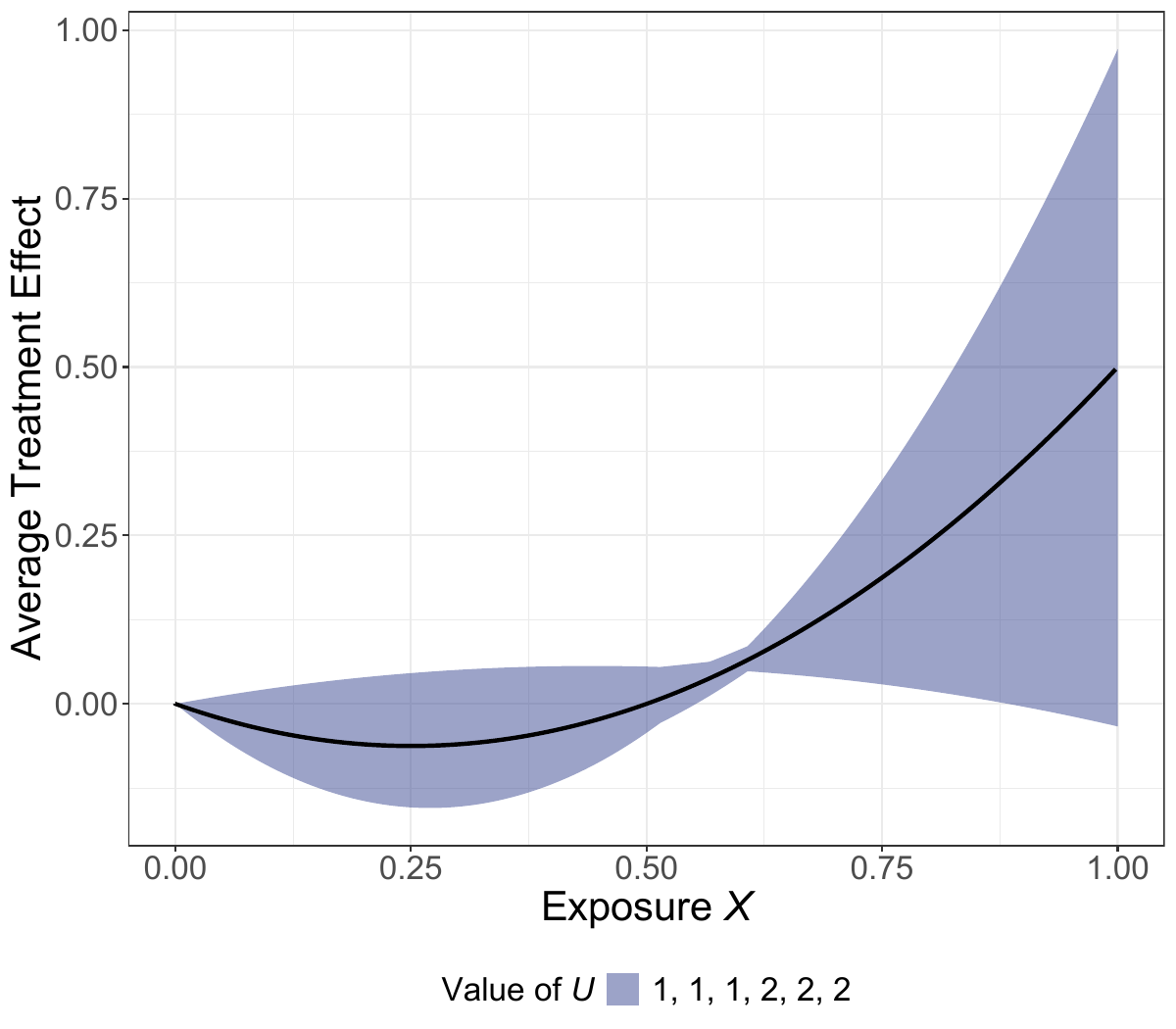}
    \subcaption{$R = 2$, $\mathrm{SNR}_y = 1/10$}
    \end{subfigure}
    \begin{subfigure}[t]{0.3\textwidth}
    \centering
    \includegraphics[width=\textwidth]{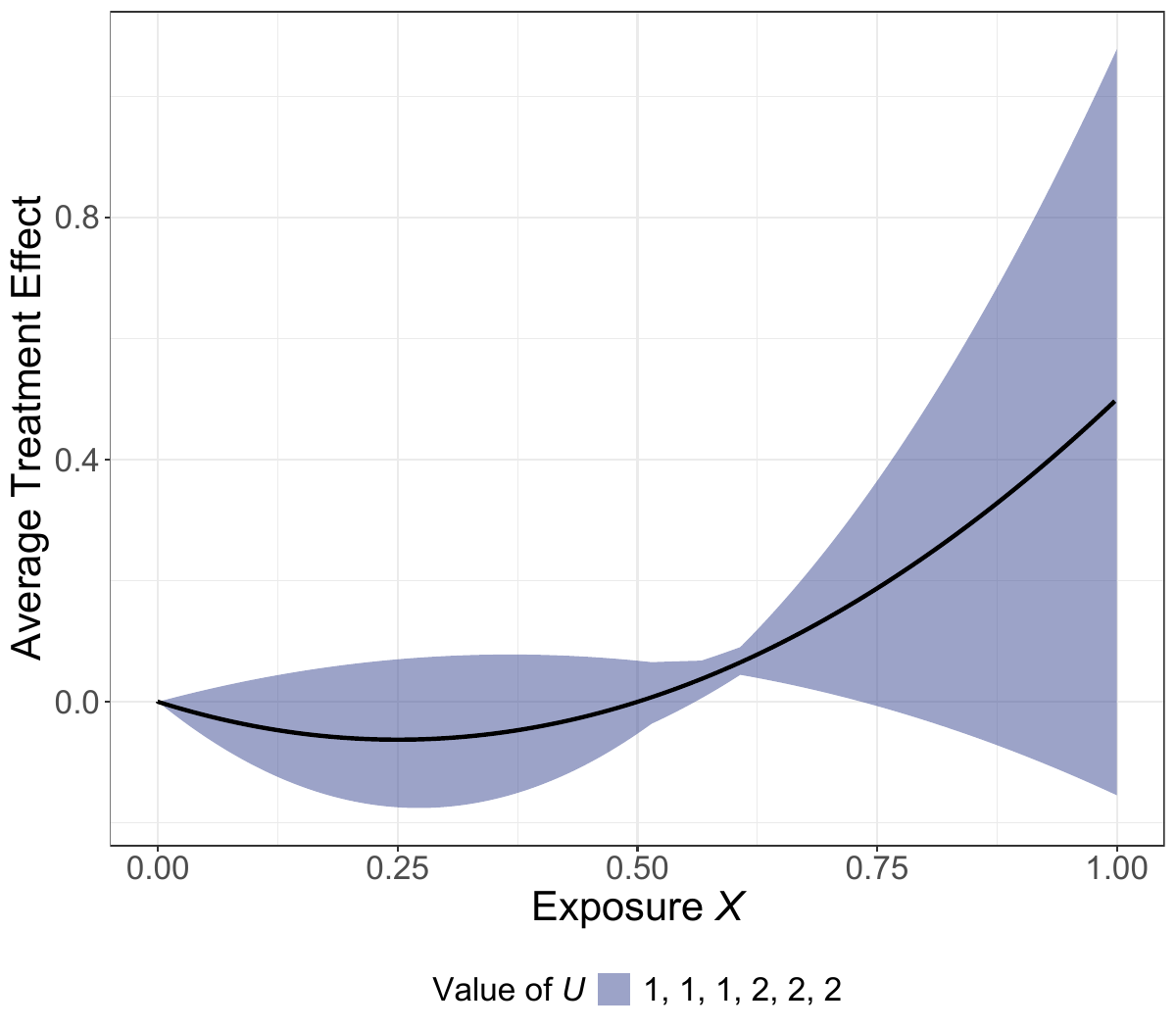}
    \subcaption{$R = 2$, $\mathrm{SNR}_y = 1/5$}
    \end{subfigure}
    \begin{subfigure}[t]{0.3\textwidth}
    \centering
    \includegraphics[width=\textwidth]{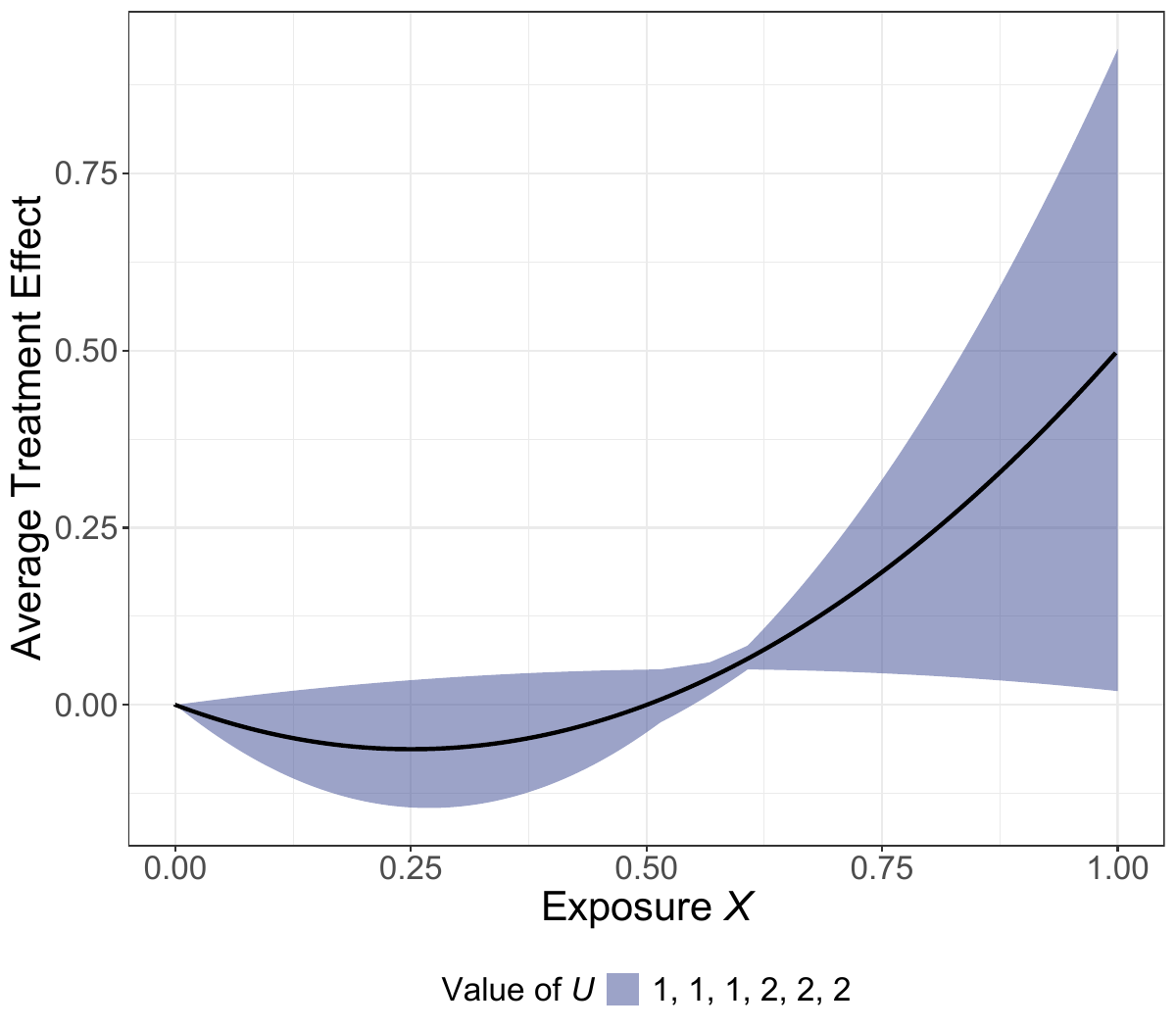}
    \subcaption{$R = 2$, $\mathrm{SNR}_y = 1$}
    \end{subfigure}
    \caption{\textbf{Simulation grid for the experiment discussed in \cref{app:Nonlinear simulation study}}. Some settings (a, b) result in a feasible set in which the ATE is qualitatively very different for each plausible $\Tilde{\bm U}$. In other settings (g--i) $\bm U^*$ is identified exactly. $\Sigma_{\bm b}^{\mathrm{(max)}} = {6 \choose 3} = 20$, so $\texttt{budgetIV}$ significantly reduced the space of plausible $\Tilde{\bm U}$ in this experiment.}
    \label{fig app:Simulation grid}
\end{figure}

\subsubsection{Varying the Simulation Setting} \label{app:nonlinear varying}

The remaining parameters, $(s_{\Phi}, \lambda^{\text{(A$2$)}}, \lambda^{\text{(A$3$)}})$, are set to control the variance $\var (Y)$, signal to noise ratio $\mathrm{SNR}_Y$, and the ratio of effects, defined by:
\begin{align*}
    & R := \frac{\lambda^{\text{(A$3$)}}}{\lambda^{\text{(A$2$)}}} \frac{\lVert \cov (\bm Z \cdot \bm \Lambda \cdot \bm Z, \Tilde{\bm Z}) \rVert_2}{\lVert \cov (\epsilon_y, \Tilde{\bm Z}) \rVert_2},
\end{align*}
where $\Tilde{\bm Z} := (Z_4, \ldots, Z_{6})$.

Using the following shorthand: 
\begin{align*}
    P &:= \varphi (X) := \sqrt{1+(X-0.5)^2}, \\
    \gamma_i &:= \bm z_i^{\top} \cdot \bm \Lambda \cdot \bm z_i\\
    r &:= \frac{\lVert \cov (\bm \gamma, \Tilde{\bm Z}) \rVert_2}{\lVert \cov (\epsilon_y, \Tilde{\bm Z}) \rVert_2}, \\
    U &:= u(\bm Z, \epsilon_y) := \frac{r}{R} \gamma(\bm Z) + \epsilon_y,
\end{align*}
the remaining free parameters $s_{\Phi}, \lambda^{\text{(A$2$)}}, \lambda^{\text{(A$3$)}} \geq 0$ are expressed as:
\begin{align*}
    & s_{\Phi}^2 = \frac{1}{\var P} \, \frac{\var Y}{1+\frac{1}{\mathrm{SNR}_Y}}, \label{eqn:app s phi} \\ 
    & \lambda^{\text{(A$2$)}} = \frac{s_{\Phi} \cov (P,U)}{\var U} \left( -1 + \frac{1}{2} \sqrt{1 + \frac{\var U}{s_{\Phi}^2 (\cov (P,U))^2} \, \frac{\var Y}{1+\mathrm{SNR}_Y}} \right),\\
    & \lambda^{\text{(A$3$)}} = \frac{R}{r} \lambda^{\text{(A$2$)}}.
\end{align*}
We choose to fix $\var (Y) := 10$ (this is simply a choice of the scale with which we choose to measure $Y$).
We produce a three-by-three grid of experiments with the values of $\mathrm{SNR}_Y = 1/10, 1/5, 1$ and $R = 1/2, 1, 2$. 

The remaining parameters are fixed to approximately fit these constraints by a post-hoc method using plug-in empirical estimates: $\widehat{\var} (P)$, $\widehat{\var} (U)$, $\widehat{\cov} (P, U)$ and a plug-in ratio estimate:
\begin{align*}
    \hat{r} := \frac{\lVert \widehat{\cov} (\bm Z \cdot \bm \Lambda \cdot \bm Z, \Tilde{\bm Z}) \rVert_2}{\lVert \widehat{\cov} (\epsilon_y, \Tilde{\bm Z}) \rVert_2},
\end{align*}
where a dataset of $5\times 10^5$ samples is used to construct the estimates. 

\subsubsection{Results from Main Text}

With the same $N = 5 \times 10^5$ sample dataset, we run an implementation of \texttt{budgetIV}, included in the supplement, with feasible search space $\vect{\Gamma} (\tau = 0, b = 3)$. 
\cref{fig app:Simulation grid} shows the results for the full simulation grid. 

\begin{figure}
    \centering 
    \begin{subfigure}[t]{0.3\textwidth}
    \centering
    \includegraphics[width=\textwidth]{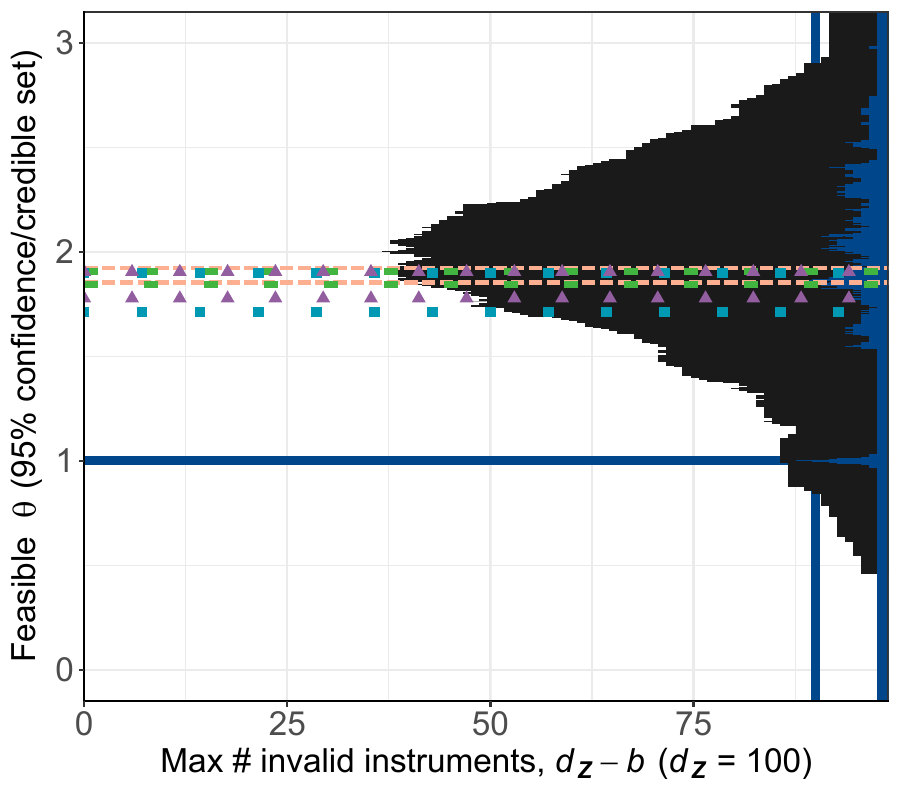}
    \subcaption{$b^* = 10$}
    \end{subfigure}
    \begin{subfigure}[t]{0.3\textwidth}
    \centering
    \includegraphics[width=\textwidth]{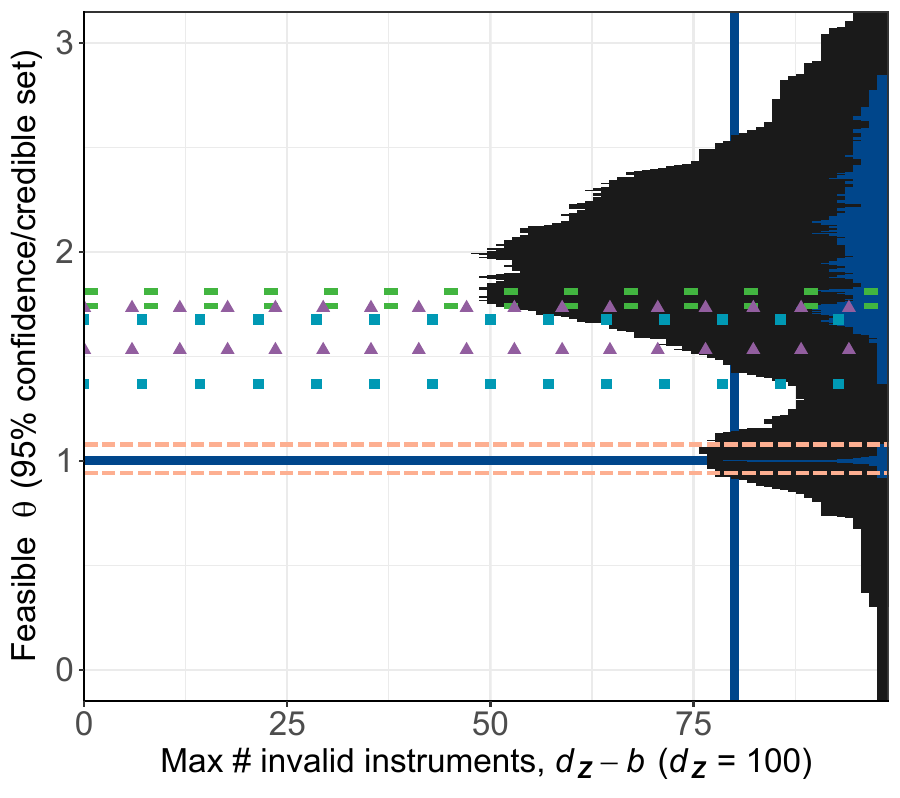}
    \subcaption{$b^* = 20$}
    \end{subfigure}
    \begin{subfigure}[t]{0.3\textwidth}
    \centering
    \includegraphics[width=\textwidth]{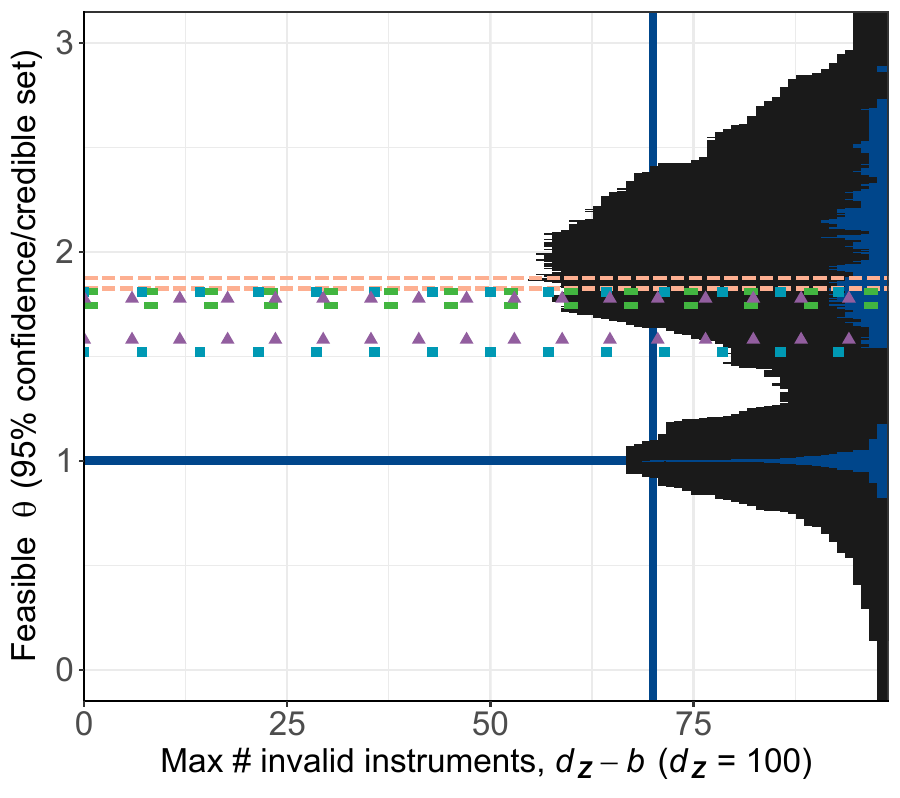}
    \subcaption{$b^* = 30$}
    \end{subfigure}
    \begin{subfigure}[t]{0.3\textwidth}
    \centering
    \includegraphics[width=\textwidth]{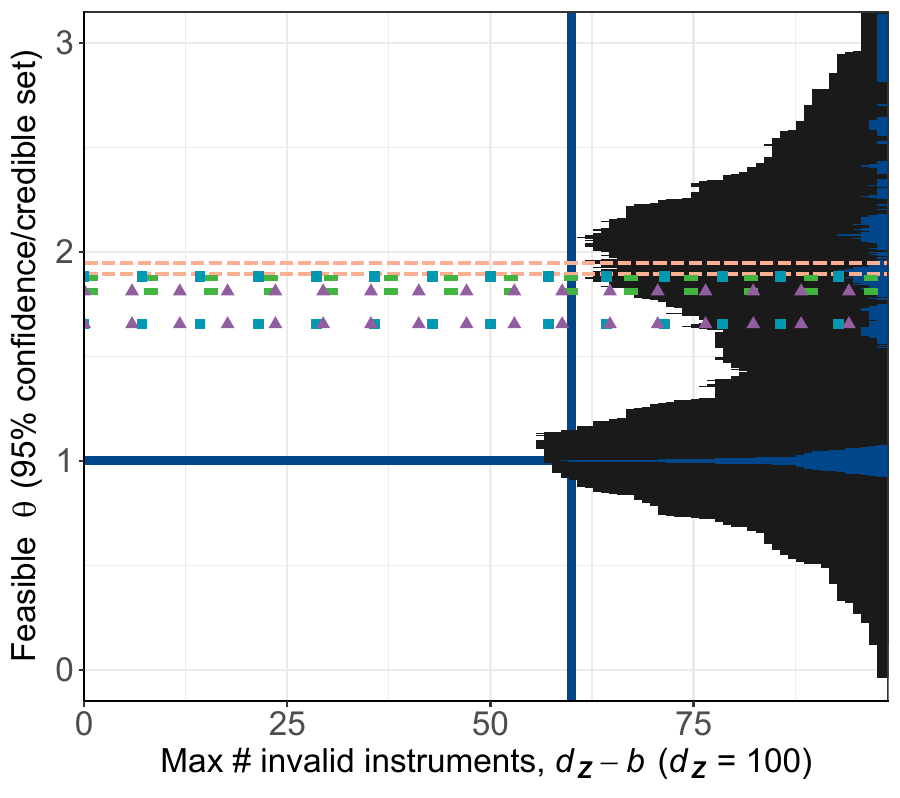}
    \subcaption{$b^* = 40$}
    \end{subfigure}
    \begin{subfigure}[t]{0.3\textwidth}
    \centering
    \includegraphics[width=\textwidth]{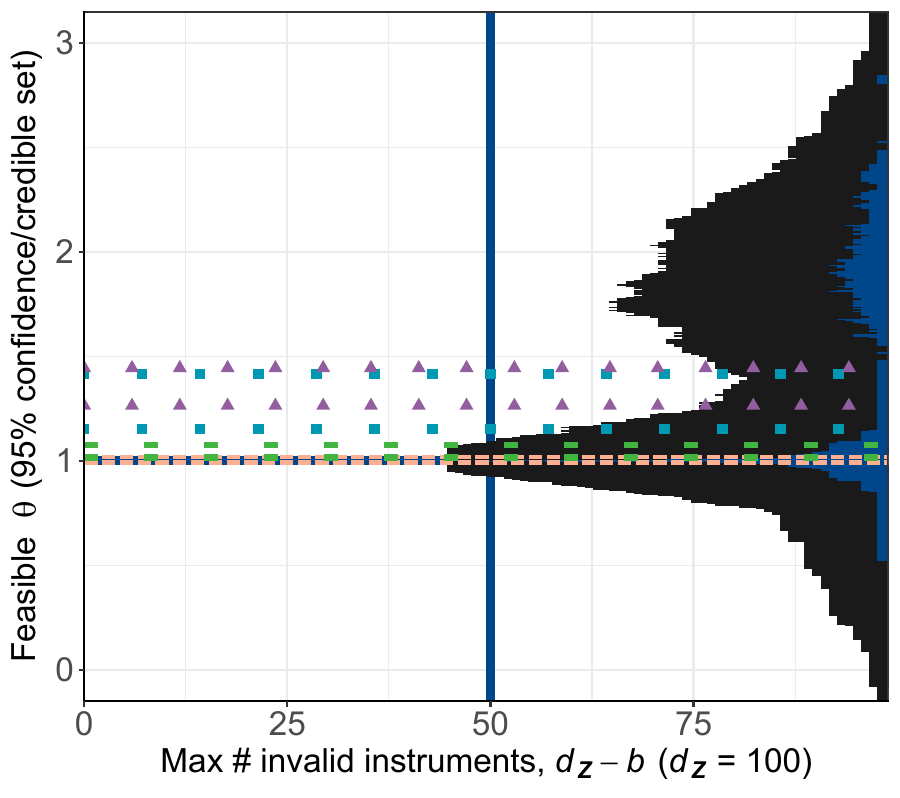}
    \subcaption{$b^* = 50$}
    \end{subfigure}
    \begin{subfigure}[t]{0.3\textwidth}
    \centering
    \includegraphics[width=\textwidth]{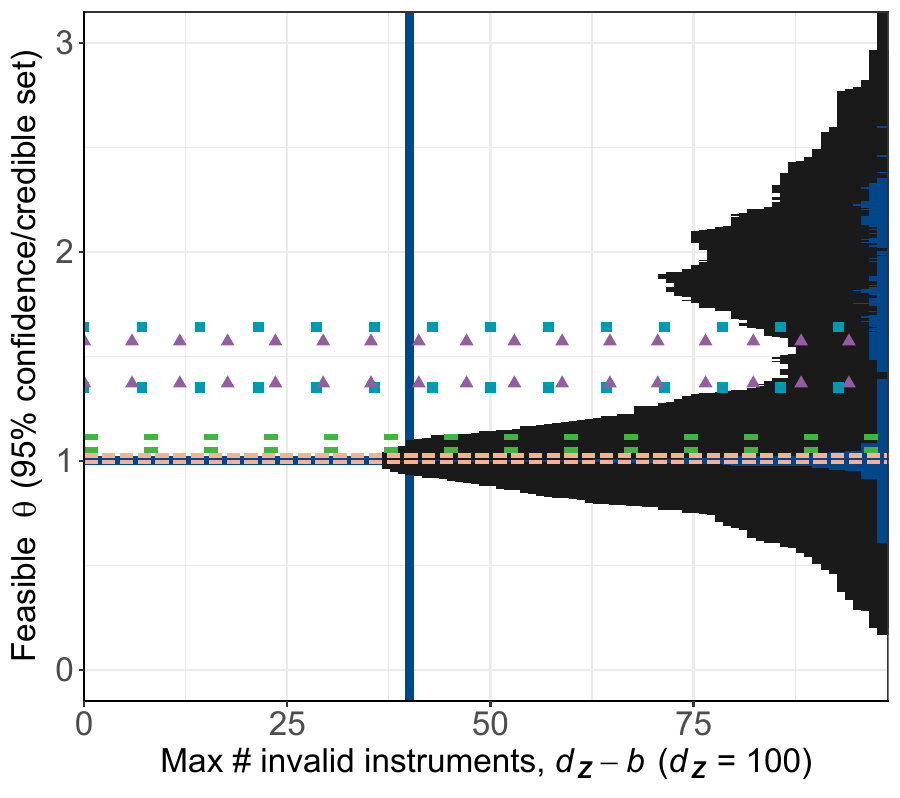}
    \subcaption{$b^* = 60$}
    \end{subfigure}
    \begin{subfigure}[t]{0.3\textwidth}
    \centering
    \includegraphics[width=\textwidth]{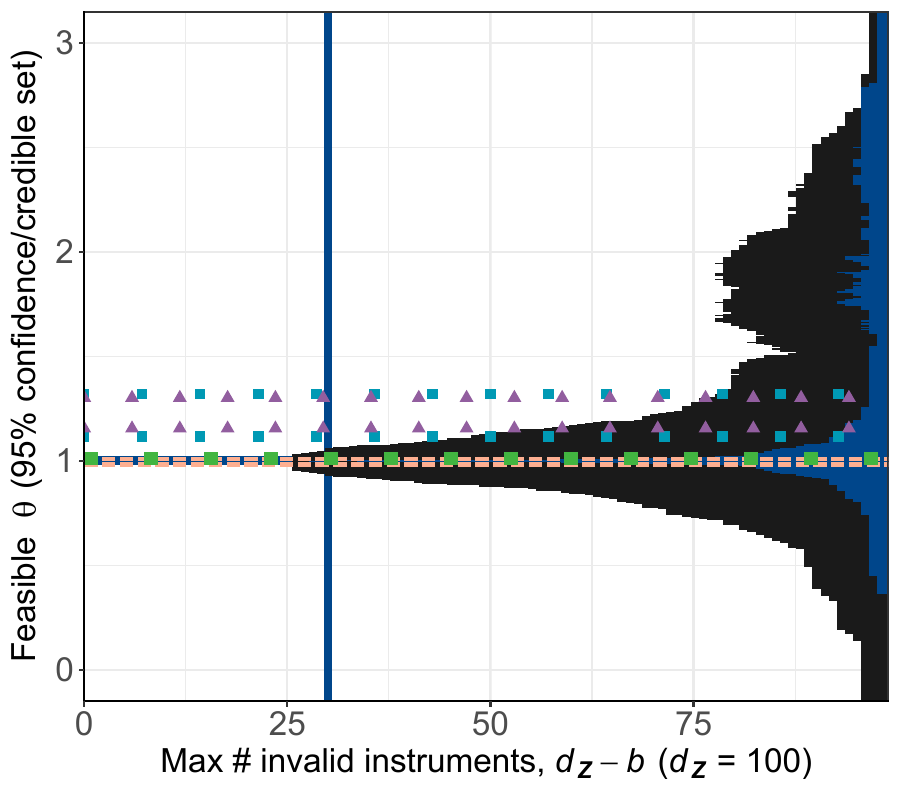}
    \subcaption{$b^* = 70$}
    \end{subfigure}
    \begin{subfigure}[t]{0.3\textwidth}
    \centering
    \includegraphics[width=\textwidth]{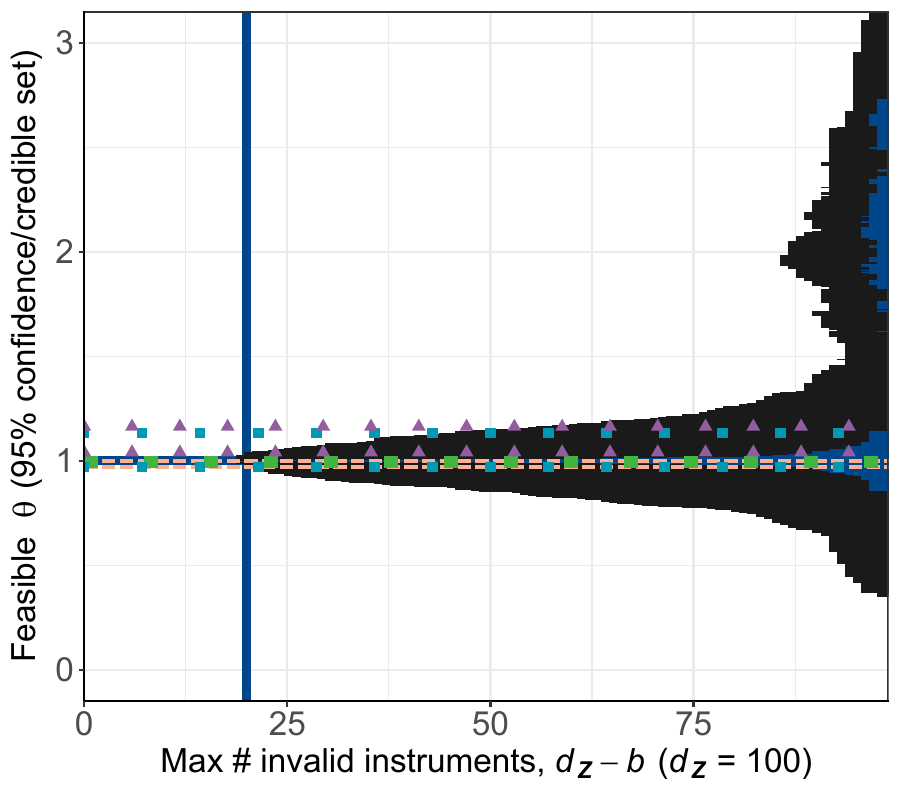}
    \subcaption{$b^* = 80$}
    \end{subfigure}
    \begin{subfigure}[t]{0.3\textwidth}
    \centering
    \includegraphics[width=\textwidth]{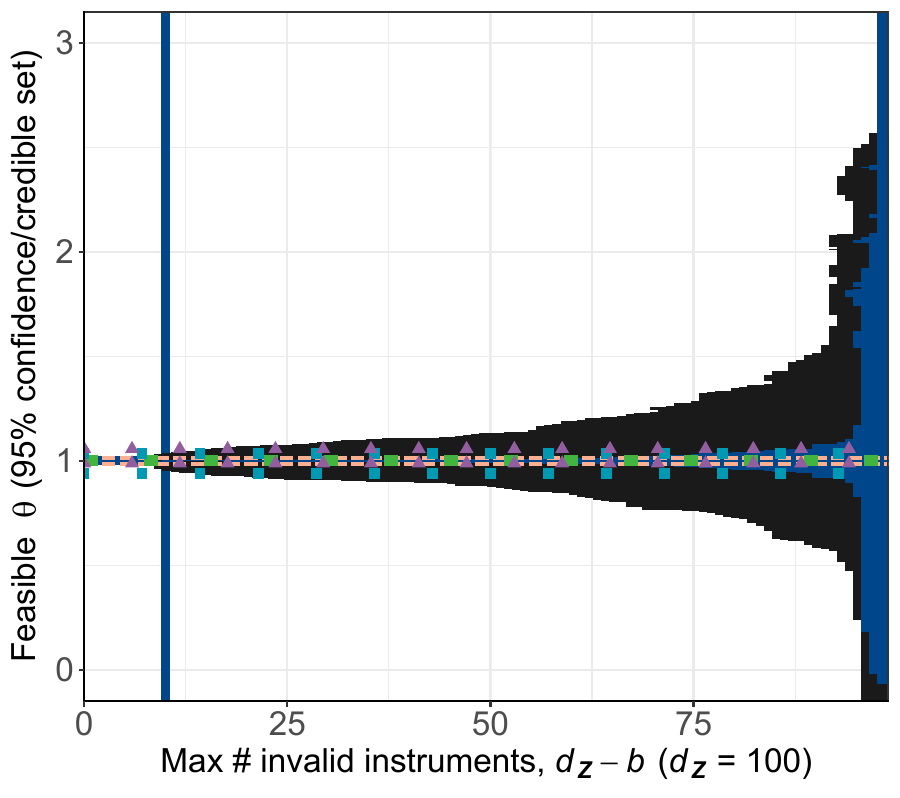}
    \subcaption{$b^* = 90$}
    \end{subfigure}
    \caption{\textbf{Simulation grid for the experiment discussed in \cref{app:Linear high dZ experiment}}. Each subfigure corresponds to a different ground truth number of valid IVs $b^*$. Finite sample confidence sets using \textcolor[HTML]{333333}{\textbf{\texttt{budgetIV\_scalar}}} and \textbf{\textcolor[HTML]{00468B}{Oracle}} are shown for the constraints $\bm \Gamma (\tau = 0, b)$ where $b$ is varied along the horizontal axis. Confidence intervals for the benchmarking methods \textcolor[HTML]{0099B4}{\textbf{MR-Egger}}, \textcolor[HTML]{42B540}{\textbf{MR-Median}}, \textcolor[HTML]{FDAF91}{\textbf{MBE}} and \textcolor[HTML]{925E9F}{\textbf{IVW}}, under which $b$ is not an adjustable parameter, are labeled for each experiment. Sub-figures (a) through (e) show variability in the confidence intervals of \textcolor[HTML]{FDAF91}{\textbf{MBE}} when valid IVs are a minority---despite the required modal assumption of holding for each ground truth model. The approach \textcolor[HTML]{ED0000}{\textbf{MASSIVE}}, included in \cref{fig:benchmark main} in the main text, takes significantly more computational resources than the the methods included in this simulation grid.} 
    \label{fig app:Linear high dZ simulation grid}
\end{figure}

\subsection{Linear Many Candidate IV Study}
\label{app:Linear high dZ experiment}

\paragraph{Simulation setting}

We performed a grid of simulations using the following linear SEM with (A3) violation among a varying proportion of the $100$ candidate IVs:
\begin{align*}
    \bm Z &:= \bm \epsilon_{\bm z}, \\
    X &:= \bm \delta \cdot \bm Z + \epsilon_x. \\
    Y &:= \theta^* X + \bm \gamma \cdot \bm Z + \epsilon_y, \\
    b^* &:= \lVert \bm \gamma_{g}^{\bm *} \rVert_0 \in  \{ 10, 20, 30, 40, 50, 60,  70, 80, 90 \}, \\
    \theta^* &:= 1.
\end{align*}
We set $(\bm \epsilon_{\bm z}, \epsilon_x, \epsilon_y)$ to a multivariate normal centered at $\bm 0$ with no correlation except $\rho := \mathrm{Corr} (\epsilon_x, \epsilon_y)$ drawn uniformly from $[-1, 1]$. The marginal standard deviations are drawn $\mathrm{i.i.d.}$ from $\mathrm{Exp (1)}$. The first $b^*$-many entries of $\bm \gamma$ are set to $0$ and the remaining entries, as well as the entries of $\bm \delta$, are samples $\mathrm{i.i.d.}$ from the uniform distribution $\mathcal{U} [1, 2]$. 

A two-sample approach was applied with $N_x = 1 \times 10^6$ samples used to sample data to generate the summary statistics $\bm{\hat \beta}_{\bm x} := \widehat{\cov} (\bm Z, X)$ and $N_y = 1 \times 10^5$ samples used to generate $\bm{\hat \beta}_{\bm y} := \widehat{\cov} (\bm Z, X)$. This was to model the typical occurrence that $\bm{\hat \beta}_{\bm x} / \mathrm{SE} (\bm{\hat \beta}_{\bm x}) \gg \bm{\hat \beta}_{\bm y} / \mathrm{SE} (\bm{\hat \beta}_{\bm y})$ ($\mathrm{SE}$ is the empirical standard error) because candidate instruments are selected if they have a strong association with $X$ (see \citet{Bowden2016NOME}).

This simulation setting reflects common modeling assumptions that are applied in the original experiments for each of the benchmark methods.

\paragraph{Results} Results for the case $b^* = 30$ were interpreted in \cref{sec:experiments}, including the caption to \cref{fig:benchmark main}.
Further interpretation, including to the stability of the MBE estimator is given in the caption to \cref{fig app:Linear high dZ simulation grid}. 









\section{SEARCHING THROUGH THE BUDGET CONSTRAINTS} \label{app:Further technical details}

\begin{figure}
    \centering 
    \begin{subfigure}[t]{0.4\textwidth}
    \centering
    \begin{tikzpicture} [node distance=10mm and 30mm,>=stealth',sh/.style={shade},line width=1pt, scale=1.4]
    \filldraw[color=olive!70, fill=olive!30, very thick, opacity = 0.5] [draw=none] (1,0) -- (2, 0) -- (2, 1) -- (3, 1) -- (3,2) -- (2, 2) -- (2, 3) -- (1, 3) -- (1, 2) -- (0, 2) -- (0, 1) -- (1, 1) -- (1, 0);
    \filldraw[color=blue!70, fill=blue!30, very thick, opacity = 1][draw=none] (1,2) -- (2,2) -- (2,1) -- (1,1) -- (1,2);
    \draw[] (1.5, 1.5) node {$(1,1)$};
    \draw[] (2.5, 1.5) node {$(2,1)$};
    \draw[] (1.5, 2.5) node {$(1,2)$};
    \draw[] (0.5, 1.5) node {$(2,1)$};
    \draw[] (1.5, 0.5) node {$(1,2)$};
    \end{tikzpicture}
    \subcaption{Values of $\bm U \in \Sigma_{\bm b}$, which partition $\bm \Gamma$ into subspaces, which themselves are disconnected.}
    \label{fig:gamma (b)}
    \end{subfigure}
    \hspace{5em}
    \begin{subfigure}[t]{0.4\textwidth} 
    \centering
    \begin{tikzpicture} [node distance=10mm and 30mm,>=stealth',sh/.style={shade},line width=1pt, scale=1.4]
    \filldraw[color=olive!70, fill=olive!30, very thick, opacity = 0.5, draw=none] (1,0) -- (2, 0) -- (2, 1) -- (3, 1) -- (3,2) -- (2, 2) -- (2, 3) -- (1, 3) -- (1, 2) -- (0, 2) -- (0, 1) -- (1, 1) -- (1, 0);
    \draw[dashed] (1, 3) -- node[above, black, opacity = 1] {$\vect{\tilde{U}}=(1,2)$} (2, 3) -- (2, 0) -- (1,0) -- (1,3);
    \draw[dashed, color=blue] (3, 1) -- node[right] {$\vect{\Tilde{U}}=(2,1)$} (3, 2) -- (0, 2) -- (0,1) -- (3,1);
    \end{tikzpicture}
    \subcaption{Values of $\Tilde{\bm U} \in \Sigma_{\bm b}^{\mathrm{(max)}}$ and their corresponding $\Tilde{\bm \Gamma}_{\Tilde{\bm U}}$, which are connected and convex.}
    \label{fig:one method}
    \end{subfigure}
    \caption{A comparison between representing the search space $\bm \Gamma (\bm \tau = (\tau_1, \tau_2), \bm b = (1,2))$ as a disjoint union over $\bm \Gamma_{\bm U}$ or a union over $\Tilde{\bm \Gamma}_{\Tilde{\bm U}}$.}
    \label{fig:budget constraint approaches}
\end{figure}

The number of unique decision variables $\lvert \Sigma_{\bm b} \rvert$ is given by the number of assignments $\bm U \in [K+1]^{d_{\bm Z}}$ for which at least $b_1$ components of $\bm U$ are equal to $1$, at least $b_2$ components are no greater than $2$ and so on until exactly $b_{K+1} := d_{\bm Z}$ components.

Suppose exactly $d_1$ components are equal to $1$, exactly $d_2 - d_1$ are equal to $2$ and so on. 
This number of combinations with this property is equal to the multinomial coefficient ${d_{\bm Z} \choose d_1, (d_2 - d_1), \ldots, (d_{K+1} - d_K)}$.

Given that we require $d_i \geq b_i$ for all $i \in [K+1]$, we can write the following: 
\begin{align*}
    \lvert \Sigma_{\bm b} \rvert = \sum_{d_1 = b_1}^{d_2} \cdots \sum_{d_K = b_K}^{d_{\vect{Z}}} \frac{d_{\vect{Z}}!}{\prod_{\ell=1}^{K+1} \left(d_\ell - d_{\ell-1} \right)!}.
\end{align*}


By comparison, $\lvert \Sigma_{\bm b}^{\mathrm{(max)}} \rvert$ count $\Tilde{\bm U} \in [K+1]^{d_{\bm Z}}$ for which exactly $b_1$ components equal $1$, $b_2 - b_1$ components equal $2$ and so on. 
Therefore, denoting $b_0 := 0$, we have: 
\begin{align*}
    \lvert \Sigma_{\bm b}^{\mathrm{(max)}} \rvert = \frac{d_{\bm Z}!}{\prod_{\ell = 1}^{K+1} (b_{\ell} - b_{\ell - 1})!}.
\end{align*}

The ratio gap between these sizes is given exactly by: 
\begin{align*}
    \frac{\lvert \Sigma_{\bm b} \rvert}{\lvert \Sigma_{\bm b}^{\mathrm{(max)}} \rvert} = \sum_{d_1 = b_1}^{d_2} \cdots \sum_{d_K = b_K}^{d_{\vect{Z}}} \frac{\prod_{\ell = 1}^{K+1} (b_{\ell} - b_{\ell - 1})!}{\prod_{\ell=1}^{K+1} \left(d_\ell - d_{\ell-1} \right)!} > 1.
\end{align*}
Depending on the scaling of $d_{\bm Z}$ and $\bm b$, asymptotic gap can be $\mathcal{O}(1)$ or as extreme as $\mathcal{O} ({d_{\bm Z}}^{K})$.   

\cref{fig:budget constraint approaches} shows that the subsets $\bm \Gamma_{\bm U} \subset \bm \Gamma (\bm \tau, \bm b)$ for $\bm U \in \Sigma_{\bm b}$ are disconnected, while $\Tilde{\bm \Gamma}_{\Tilde{\bm U}} \subset \bm \Gamma$ for $\Tilde{\bm U} \in \Sigma_{\bm b}^{\mathrm{(max)}}$ are cuboids in general. 
Therefore, it is clear that testing for intersection between $h(\bm \theta)$ and $\Tilde{\bm \Gamma}_{\Tilde{\bm U}}$ is more efficient than searching through the decision variables $\bm U$ directly. 

\newpage
\begin{landscape}
\section{SUMMARY OF RELAXED IV METHODS}\label{appx:lit}
\begin{table}[h!]
\centering
\begin{tabular}{|l|c|c|c|c|c|c||c|c|}
\hline
\textbf{Paper}   & $d_X > 1$ & $d_Z > 1$ & \textbf{Cont.} & \textbf{Nonlin.} & \textbf{2-sample} & \textbf{Inference} & \textbf{Violation} & \textbf{Feasible} $\bm \gamma_g$ \\ 
\hline
\citet{Conley2012} & \cmark & \cmark & \cmark & \xmark & \xmark & \cmark & (A3) & Convex set \\
\citet{nevo_identification_2012} & \cmark & \cmark & \cmark & \xmark & \xmark & \cmark & (A2) & Convex set \\
\citet{ramsahai2012} & \xmark & \xmark & \xmark & \cmark & \cmark & \xmark & (A2) $\lor$ (A3) & N/A\\
\textcolor[HTML]{800080}{\citet{Bowden2015}} & \xmark & \cmark & \cmark & \xmark & \cmark & \cmark & (A3)$^\dagger$ & $d_Z \rightarrow \infty$\\
\citet{Kolesar2015} & \xmark & \cmark & \cmark & \xmark & \xmark & \cmark & (A3)$^\dagger$ & $d_Z \rightarrow \infty$\\
\textcolor[HTML]{FF0000}{\citet{Bowden2016WeightedMedian}} & \xmark & \cmark & \cmark & \xmark & \xmark & \cmark & (A3) & Sparse \\
\citet{kang2016} & \xmark & \cmark & \cmark & \xmark & \xmark & \xmark & (A3)* & Sparse \\
\citet{silva_evans:16} & \xmark & \cmark & \xmark & \cmark & \cmark & \cmark & (A2) \& (A3) & N/A\\
\citet{Hartwig2017} & \xmark & \cmark & \cmark & \xmark & \cmark & \cmark & (A3) & Mode zero\\
\citet{Guo2018} & \xmark & \xmark & \cmark & \cmark & \xmark & \cmark & (A3) & Mode zero\\
\citet{Windmeijer2019} & \xmark & \cmark & \cmark & \xmark & \xmark & \cmark & (A3)* & Sparse\\
\citet{Shapland2019} & \xmark & \cmark & \cmark & \xmark & \xmark & \cmark & (A3)* & $L_0$-norm\\ 
\citet{bucur2020} & \xmark & \cmark & \cmark & \xmark & \cmark & \partialmark & (A3)* & $L_0$-norm\\ 
\citet{kang2020} & \xmark & \cmark & \cmark & \xmark & \xmark & \cmark & (A2) \& (A3) & Point\\
\citet{kuang2020} & \xmark & \cmark & \xmark & \xmark & \xmark & \xmark & (A2)* \& (A3)* & Point \\
\citet{hartford2021} & \xmark & \cmark & \cmark & \cmark & \xmark & \xmark & (A3)* & Mode zero\\
\citet{vancak_2023} & \xmark & \xmark & \cmark & \cmark & \xmark & \cmark & (A2) \& (A3) & Point\\
\citet{Xue2023} & \xmark & \cmark & \cmark & \xmark & \cmark & \partialmark & (A2) \& (A3) & $L_0$-norm\\
\citet{watson2024bounding} & \xmark & \cmark & \cmark & \xmark & \cmark & \cmark & (A3) & Convex set\\
\textbf{\texttt{budgetIV}} & \cmark & \cmark & \cmark & \cmark & \cmark & \cmark & (A2) \& (A3) & Star domain \\
\hline
\end{tabular}
\caption{\textbf{Summary of the constraints and affordances of some notable relaxed IV methods.} We exclude approaches that adhere to classical IV assumptions but provide nonparametric partial identification bounds as out of scope (e.g., \citet{Balke1997, kilbertus2020class, kennedy2023}). We do the same with more generic algorithms not specifically designed for the IV setting (e.g., \citet{hu2021,  duarte2023, padh_stochastic2023}). 
The columns denote, in order, if the method allows for: (1) multidimensional exposures; (2) multiple candidate instruments; (3) continuous data; (4) nonlinear structural equations; (5) summary statistic input; and (6) statistical inference (where \partialmark corresponds to point estimators in partially identifiable settings). The final two columns indicate (7) which IV assumptions (if any) may be violated and (8) the assumed geometry of the feasible region. The * in column (7) indicates that only \textit{some} candidate IVs are allowed to violate starred assumptions (typically fewer than 50\%), while $\dagger$ indicates that the mechanism behind a candidate IV's (A3) violation must be independent to the effect of the candidate on the exposure. The N/A entries in column (8) correspond to fully nonparametric models in which $\bm \gamma_g$ is undefined.}
\label{tab:literature}
\end{table}
\end{landscape}


\end{document}